\documentclass[preprint,12pt]{elsarticle}

\usepackage{amssymb}
\usepackage[fleqn]{amsmath}
\usepackage{amsthm,bm}
\usepackage{amsfonts}
\usepackage{mathtools}
\usepackage{fdsymbol}
\usepackage{mathrsfs}
\usepackage{dutchcal}
\usepackage{algorithm}
\usepackage{algpseudocode}

\usepackage[colorlinks=true, allcolors=blue]{hyperref}
\usepackage[margin=2.5cm,includefoot]{geometry}
\usepackage{caption}
\usepackage{subcaption}
\usepackage{booktabs}
\usepackage{longtable}
\usepackage{multirow}
\usepackage[dvipsnames]{xcolor}
\usepackage{colortbl}
\graphicspath{ {./} }

\hypersetup{pdftitle={An unscented universal filter for joint input-state-parameter estimation of a partially measured system: an application to drive-by structural health monitoring},
            pdfauthor={Zihao Liu, Chul-Woo Kim, Mehrisadat Makki Alamdari}}

\journal{Elsevier}

\usepackage{natbib}
\begin{document}
\begin{frontmatter}

\title{An unscented universal filter for joint input-state-parameter estimation of a partially measured system: an application to drive-by structural health monitoring}

\author[a]{Zihao Liu}
\author[b]{Chul-Woo Kim}
\author[a]{Mehrisadat Makki Alamdari~\corref{cor1}}
\ead{m.makkialamdari@unsw.edu.au}
\cortext[cor1]{Corresponding author}

\address[a]{Centre for Infrastructure Engineering and Safety, School of Civil and Environmental Engineering, University of New South Wales, Sydney, 2052, NSW, Australia}
\address[b]{Department of Civil \& Earth Resources Engineering, Graduate School of Engineering, Kyoto University, Kyoto 615-8540, Japan}

\begin{abstract}

\noindent Drive-by structural health monitoring infers the condition of a bridge from the response of an instrumented vehicle, whose measurements are dominated by the road roughness and the vehicle dynamics. The bridge responses, contact forces, and bridge parameters depend on one another through the vehicle-bridge interaction (VBI), so estimating them jointly is desirable. This paper proposes a novel filter in the Universal Filter family, designated as the Unscented Universal Filter (UUF), for joint input-state-parameter estimation of nonlinear systems with unknown inputs. The unknown input is estimated by weighted least squares from the innovation, without any evolution model or statistics, and the system inversion imposes no rank condition on the feedforward matrix. The nonlinear propagation is linearised using the scaled unscented transform, and the linearisation residuals are propagated into the error covariances. For the acceleration-only configuration, auxiliary integrators form pseudo-measurements from the acceleration measurements themselves. The UUF is then applied to the coupled VBI system, with the contact forces as the unknown input and the local stiffness of the bridge as the tracked parameter. The measurements are the responses of the vehicle body and, at most, a few bridge responses. In numerical case studies on a simply supported beam and the Old Ada Bridge, a steel truss bridge in Japan, the UUF is compared with classical and state-of-the-art joint estimation filters, all tuned by the same practical strategy. Across the damage severities, damage locations, road classes, and noise levels examined, the UUF yields the lowest total error in most scenarios, and its error varies little with damage severity, damage location, and road class. On the Old Ada Bridge, the UUF identifies the stiffness of a damaged member from acceleration-only measurements, without the suspension parameters of the vehicle, and with a rank-deficient feedforward matrix.
\end{abstract}

\begin{keyword}
Drive-by structural health monitoring \sep Vehicle-bridge interaction \sep Joint input-state-parameter estimation \sep Unknown input \sep Unscented transform \sep Universal Filter
\end{keyword}
\end{frontmatter}

\begin{table}[ht]
    \centering
    \small
    \caption*{List of frequently used abbreviations}
    \label{tab1}
    \begin{tabular}{r l}
    \toprule
    AEKF & Augmented extended Kalman filter \\
    DKF-UKF & Combined dual and unscented Kalman filter \\
    EGDF-DF & Extended Gillijns-De Moor filter with direct feedthrough \\
    IPS-UKF & Input-parameter-state unscented Kalman filter \\
    MVU & Minimum-variance unbiased \\
    NRMSE & Normalised root-mean-square error \\
    NUF & Nonlinear Universal Filter \\
    RMS & Root-mean-square \\
    UUF & Unscented Universal Filter \\
    VBI & Vehicle-bridge interaction \\
    \bottomrule
    \end{tabular}
\end{table}

\section{Introduction} \label{sec:introduction}

Many bridges in service are ageing, and their safe operation and lifecycle extension rely on maintenance informed by structural health monitoring~\cite{HaoReview2023}. Sensors installed on the bridge serve this purpose, but their installation is costly and time-consuming, so that monitoring an entire road network in this way is infeasible~\cite{MalekjafarianReview2022}. Drive-by structural health monitoring, also known as vehicle scanning, is an indirect or mobile sensing method that infers bridge condition from the response of an instrumented vehicle, which can pass over several bridges in one journey~\cite{LiReview2026}. However, road roughness and vehicle dynamics dominate the measured response, making it challenging to extract bridge-related information~\cite {LiReview2026}. The methods developed for this purpose are broadly data-driven or model-based. Data-driven methods include artificial neural networks trained on the contact-point response to separate damage from temperature and speed effects~\cite{CorballyMalekjafarian2022}, and generative autoencoders trained only on the reference state of the bridge~\cite{CalderonHurtado2024}. Convolutional neural networks have extracted damage-sensitive features from thousands of laboratory runs~\cite{Peng2025}, and neural operators have been trained as real-time surrogates of the coupled system~\cite{ChenKimWang2025}. These methods learn an implicit mapping from the measurements to the bridge condition, and their generalisation across bridges and vehicle conditions remains an active research topic~\cite{Peng2025}. Model-based methods instead infer the bridge condition through a physical model of the coupled system. The modal properties of the bridge have been identified from the vehicle response, such as the mode shapes by a physics-informed optimisation over crowdsensing trips~\cite{PengLiHaoYang2023} and the frequencies from train accelerations once the contact forces are estimated~\cite{Stoura2025}. The bridge parameters and the axle loads have been identified simultaneously by iterative optimisation~\cite{FengSunFeng2015}, and the bridge response has been reconstructed from contact forces estimated on a train with known properties~\cite{Siu2024}. In these approaches, the contact forces and the bridge properties are obtained either in separate steps or through iterations over the entire record. Since the responses, contact forces, and bridge parameters depend on one another through the bridge-vehicle interaction (VBI), estimating each benefits from estimating them simultaneously. Furthermore, an estimation that proceeds as the vehicle passes follows the bridge condition in time. Recursive Bayesian filtering offers such a treatment, updating estimates and their uncertainties at each timestep.

The Kalman filter (KF)~\cite{Kalman1960} is the most widely used recursive Bayesian filter for the state estimation of linear systems, but it requires the input to be known. To relax this requirement, \citet{LourensReyndersEATAL2012} proposed the augmented Kalman filter (AKF), in which the unknown input is appended to the state vector with a fictitious random walk and estimated jointly with the state. To address the drift of the estimated input and displacements under acceleration-only measurements, \citet{EftekharAzam2015} proposed the dual Kalman filter (DKF), in which the input and the state are estimated by two successive filters. These filters are linear and cannot track the time-varying parameters of a deteriorating structure. The extended Kalman filter (EKF) is a widely used nonlinear approach for parameter identification~\cite{Roffel2014}. The unscented Kalman filter (UKF) is another approach that has been shown to be effective for highly nonlinear structural identification~\cite{ChatziSmyth2009}. Joint input-state-parameter estimation then follows by combining an input treatment with a nonlinear filter. \citet{NaetsCroesDesmet2015} proposed an augmented extended Kalman filter, referred to as the AEKF hereafter, which appends both the input and the parameters to the state vector of an EKF. \citet{Dertimanis2019} combined the DKF and the UKF, referred to as the DKF-UKF hereafter, with the former handling the unknown input and the latter the state and parameter estimation. \citet{ImpraimakisSmyth2022} developed an unscented Kalman filter that estimates the input in two stages within each timestep, first from the predicted states and then from the corrected states and parameters, referred to as the IPS-UKF hereafter. In a Bayesian expectation-maximisation framework, the noise parameters have also been identified along with the input, state, and parameters~\cite{Teymouri2023}. A common feature of Kalman-type methods is that the unknown input is modelled as a fictitious evolution process whose process noise covariance must be calibrated.

The requirement of a fictitious input model motivated the minimum-variance unbiased (MVU) filters, which estimate the input by weighted least squares from the innovation without any prior on its evolution or statistics. The pioneering filter of \citet{Kitanidis1987} estimates only the state under an unknown input. Building on it, \citet{Gillijns2007a} proposed a joint input-state filter for systems without direct feedthrough, referred to as the GDF-NDF hereafter, which in structural dynamics means that no acceleration is measured. A separate filter, referred to as the GDF-DF hereafter, covers systems with direct feedthrough and requires a full-rank feedforward matrix, i.e., at least as many acceleration measurements as unknown inputs~\cite{Gillijns2007b}. Both filters are linear. The GDF-DF was extended to joint input-state-parameter estimation by \citet{Wan2018}, referred to as the EGDF-DF hereafter. This extension retains the rank condition and the input indexing of the GDF-DF, so that systems with a rank-deficient feedforward matrix are not covered. The Universal Filter (UF)~\cite{EbrahimzadehHassanabadi2023} removed this restriction for linear systems, and the Universal Smoothing (US)~\cite{Liu2025} extended the observation window to reduce the ill-conditionedness of the instantaneous system inversion. The Nonlinear Universal Filter (NUF)~\cite{LiuNUF2025} then enabled joint input-state-parameter estimation regardless of the rank condition of the feedforward matrix. The NUF linearises the nonlinear propagation by the Jacobian matrix, as the EKF does. Such a linearisation is reliable only where the error propagation is close to linear, and the Jacobian matrices can be difficult to derive~\cite{Julier2004}.

In drive-by structural health monitoring, Bayesian filters have been applied mainly to identify the road profile or the contact forces, which form the unknown input of the vehicle. On the vehicle alone, the AKF has been used to estimate the dynamic tyre forces from the body responses~\cite{Wang2019}, and the DKF to identify the road roughness before the bridge damage is identified in a second step~\cite{JiantaoLi2020}. The DKF has also been used to identify the contact-point responses of two successive vehicles for the modal identification of the bridge~\cite{JiantaoLi2022}, and the AKF has been combined with Bayesian expectation-maximisation for the contact-point displacements~\cite{Li2024}. On the coupled system, Kalman-type filters with unknown input have identified the track irregularities of railway bridges~\cite{Xiao2020} and the pavement profile of a highway bridge~\cite{HeYang2021}, with the vehicle and the bridge parameters given. MVU filters have been applied to road roughness identification using a quarter-car model, a vehicle parameter identification step, and adaptive correction of the noise statistics~\cite{Kong2025}. The US has identified the road profile from axle accelerations, with a backward-difference approximation of the roughness rate to satisfy the rank condition~\cite{LiuRoughness2026}. The closest approach to joint estimation of the coupled system identifies the bridge frequencies together with the track irregularities using an EKF~\cite{Xiao2021} and a sigma-point filter with quantified uncertainty~\cite{Xiao2022}. In these two studies, the irregularity is modelled as a random-walk state, the vehicle parameters are assumed known, and relative displacement and velocity measurements are required in addition to acceleration measurements. The joint estimation of the responses, the contact forces and a stiffness parameter of the bridge on the coupled VBI system, with the input estimated without an evolution model, has not been established.

Acknowledging the limitations of the existing nonlinear MVU filters and the absence of joint input-state-parameter estimation in drive-by structural health monitoring, this paper proposes a novel filter in the Universal Filter family, designated as the Unscented Universal Filter (UUF). Unlike Kalman-type methods, the UUF requires no evolution model or statistics for the unknown input, which is estimated by weighted least squares from the innovation. Its system inversion has no restriction on the rank condition of the feedforward matrix. The nonlinear propagation is linearised statistically by the scaled unscented transform, so that no Jacobian matrix is required and the residual of the linearisation is carried through the error propagation into the covariances. The derivation is presented in a general form for structural systems with unknown inputs and time-varying parameters. For the acceleration-only configuration, auxiliary integrators are introduced to form pseudo-measurements from the acceleration measurements themselves, so that no fictitious measurement with a prescribed variance is needed. The UUF is then applied to drive-by structural health monitoring on the coupled VBI system, with the contact forces as the unknown input and the bridge parameters tracked, from the responses of the vehicle body and at most a few bridge measurements. Numerical case studies compare the UUF with the classical AEKF and EGDF-DF and with the state-of-the-art DKF-UKF and IPS-UKF, all tuned by the same practical strategy that requires no true response.

The remainder of the paper is organised as follows. Section~\ref{sec:vbi} formulates the coupled VBI system and casts it into a state-space representation. Section~\ref{sec:method} derives the UUF, including the statistical linearisation, the error propagation, the gain matrices and the auxiliary integrators for acceleration-only estimation. Section~\ref{sec:application} applies the UUF to drive-by bridge inspection, where the unknown input is re-parameterised and a practical tuning strategy is presented. Sections~\ref{sec:cs1} and~\ref{sec:cs2} present the numerical case studies on a simply supported beam and the Old Ada Bridge, respectively. Finally, conclusions are drawn in Section~\ref{sec:conclusions}.

\section{Formulation of the coupled VBI system} \label{sec:vbi}
\noindent This section formulates the coupled vehicle-bridge interaction (VBI) system that underlies the estimation problem addressed in this paper. The governing equations are established first, then assembled into a monolithic matrix form, and finally cast into a state-space representation and discretised to give the deterministic forward model. Section~\ref{sec:method} derives the filtering method on the general system class to which this model belongs, and Section~\ref{sec:application} poses the drive-by estimation problem on the model itself.

\subsection{Governing equations of the coupled system} \label{sec:vbi_gov}
\noindent Vehicle models of varying complexity have been used to formulate the dynamic interaction between a moving vehicle and the supporting bridge. The half-car model is a common choice, since its dynamic characteristics represent those of most commercial vehicles well~\cite{Cebon1999}. A two-degree-of-freedom half-car model is adopted herein, with the vehicle crossing the bridge at a constant speed $v$, and the coupled system is formulated following \citet{Alamdari2021} and \citet{YangWu2001}.

\begin{figure}[ht]
    \centering
    \includegraphics[width=0.65\textwidth]{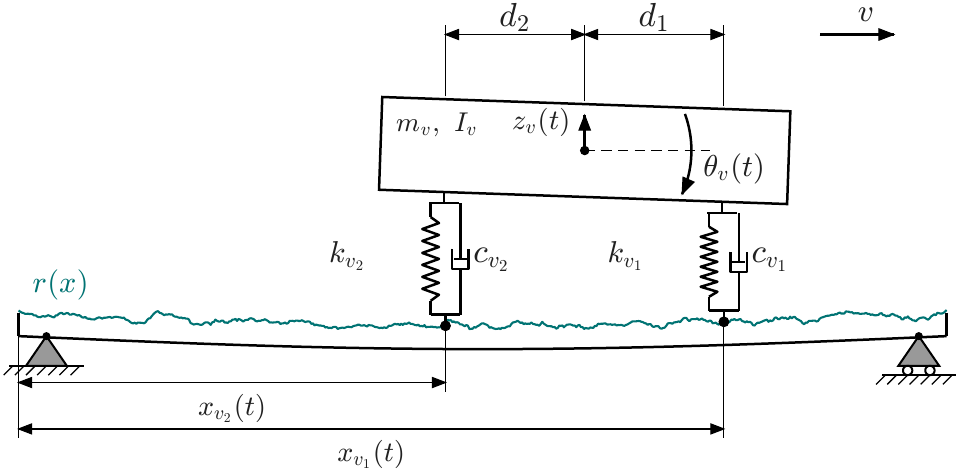}
    \caption{Half-car model crossing a bridge with a road roughness profile.}
    \label{fig:vbi_halfcar}
\end{figure}

As shown in Fig.~\ref{fig:vbi_halfcar}, the half-car model has two independent degrees of freedom, the bounce $z_v(t)$ and the pitch $\theta_v(t)$. Eight parameters define the vehicle. The stiffness $k_{v_i}$ and damping $c_{v_i}$ of the combined tyre-suspension system of axle $i$ describe the two suspensions, with $i=1$ the front axle and $i=2$ the rear. The remaining parameters are the vehicle mass $m_v$, the moment of inertia $I_v$, and the distances $d_1$ and $d_2$ from the centre of gravity to the two axles, giving the axle spacing $d = d_1 + d_2$. It is assumed that the tyres remain in contact with the road surface at all times. The surface of the bridge carries a stochastic roughness profile $r(x)$, described by its power spectral density according to ISO~8608~\cite{ISO8608}. Since the speed is constant, the axle positions are known at all times, $x_{v_1}(t) = vt$ and $x_{v_2}(t) = x_{v_1}(t) - d$, with $t = 0$ at the entry of the front axle to the deck.

The vehicle responds to the deformations of its two tyre-suspension systems. Written about the static equilibrium position, the equation of motion of the vehicle can be written as:
\begin{equation} \label{eq:vbi_vehicle_eom}
    \mathbf{M}_v\ddot{\mathbf{u}}_v(t) + \mathbf{C}_v\dot{\mathbf{u}}_v(t) + \mathbf{K}_v\mathbf{u}_v(t) = \mathbf{f}_v(t),
\end{equation}
\noindent in which $\mathbf{u}_v(t)=\left[\begin{matrix} z_v(t) & \theta_v(t) \end{matrix}\right]^T \in \mathbb{R}^{2}$ collects the two degrees of freedom, $\mathbf{M}_v$, $\mathbf{C}_v$ and $\mathbf{K}_v \in \mathbb{R}^{2\times2}$ are the mass, damping and stiffness matrices of the vehicle, respectively, and $\mathbf{f}_v(t) \in \mathbb{R}^{2}$ is the excitation transmitted to the vehicle by the motion of the two contact points. The mass matrix is diagonal:
\begin{equation} \label{eq:vbi_Mv}
    \mathbf{M}_v = \begin{bmatrix} m_v & 0 \\ 0 & I_v \end{bmatrix}.
\end{equation}

\noindent The stiffness and damping matrices follow from the way the two vehicle degrees of freedom combine at the axles. The vehicle body is taken as rigid, so its displacement at axle $i$ is $z_{a_i} = \mathbf{L}_i^T\mathbf{u}_v$, with
\begin{equation} \label{eq:vbi_axle_disp}
    \mathbf{L}_1 = \begin{bmatrix} 1 \\ d_1 \end{bmatrix}, \qquad
    \mathbf{L}_2 = \begin{bmatrix} 1 \\ -d_2 \end{bmatrix},
\end{equation}
\noindent so that $\mathbf{L}_i \in \mathbb{R}^{2}$ maps the bounce and the pitch to the vehicle body motion at axle $i$. Transposed, it distributes a force acting at that axle onto the two vehicle degrees of freedom. Since $\mathbf{L}_i$ is constant, the same relation holds for velocities and accelerations, and the axle acceleration $\ddot{z}_{a_i} = \mathbf{L}_i^T\ddot{\mathbf{u}}_v$ is the quantity recorded by an accelerometer mounted on the vehicle body above the axle. No unsprung mass is modelled, so axle hop is not represented. Each tyre-suspension system therefore contributes one symmetric rank-one term to the vehicle matrices, giving the half-car stiffness and damping matrices:
\begin{equation} \label{eq:vbi_Kv}
\begin{aligned}
    \mathbf{K}_v &= \sum_{i=1}^{2} k_{v_i}\mathbf{L}_i\mathbf{L}_i^T =
    \begin{bmatrix}
        k_{v_1}+k_{v_2} & k_{v_1}d_1-k_{v_2}d_2 \\
        k_{v_1}d_1-k_{v_2}d_2 & k_{v_1}d_1^2+k_{v_2}d_2^2
    \end{bmatrix}, \\[2mm]
    \mathbf{C}_v &= \sum_{i=1}^{2} c_{v_i}\mathbf{L}_i\mathbf{L}_i^T =
    \begin{bmatrix}
        c_{v_1}+c_{v_2} & c_{v_1}d_1-c_{v_2}d_2 \\
        c_{v_1}d_1-c_{v_2}d_2 & c_{v_1}d_1^2+c_{v_2}d_2^2
    \end{bmatrix}.
\end{aligned}
\end{equation}

\noindent The excitation reaches the vehicle through the two contact points and therefore depends on the bridge response. The bridge is discretised by the finite element method, with $n_b$ degrees of freedom after applying the boundary conditions, and its nodal displacements are collected in $\mathbf{u}_b(t)\in\mathbb{R}^{n_b}$. The excitation can then be written as:
\begin{equation} \label{eq:vbi_fv}
\begin{aligned}
    \mathbf{f}_v(t) = {}& \mathbf{L}_1\left[k_{v_1}\left(\boldsymbol{\iota}_1(t)^T\mathbf{u}_b(t) + r_1(t)\right) + c_{v_1}\left(\boldsymbol{\iota}_1(t)^T\dot{\mathbf{u}}_b(t) + v r'_1(t)\right)\right] \\
    &+ \mathbf{L}_2\left[k_{v_2}\left(\boldsymbol{\iota}_2(t)^T\mathbf{u}_b(t) + r_2(t)\right) + c_{v_2}\left(\boldsymbol{\iota}_2(t)^T\dot{\mathbf{u}}_b(t) + v r'_2(t)\right)\right],
\end{aligned}
\end{equation}
\noindent where $\boldsymbol{\iota}_i(t)\in\mathbb{R}^{n_b}$ is the contact influence vector of axle $i$, so that $\boldsymbol{\iota}_i(t)^T\mathbf{u}_b(t)$ is the deck deflection at the contact point. The roughness height and slope beneath the axle are $r_i(t) = r\!\left(x_{v_i}(t)\right)$ and $r'_i(t) = r'\!\left(x_{v_i}(t)\right)$, the prime denoting the derivative with respect to space, so that $v r'_i(t)$ is the rate at which the roughness passes under the axle. For a deck discretised by beam elements, $\boldsymbol{\iota}_i$ collects the Hermitian interpolation functions of the element containing $x_{v_i}(t)$, assembled in the global coordinates. While axle $i$ is off the deck, $\boldsymbol{\iota}_i(t)=\mathbf{0}$ and the approaches are taken as rigid and smooth. The convective term associated with the slope of the deck is neglected~\cite{Alamdari2021}. The roughness reaches each axle only through the combination $k_{v_i}r_i(t) + c_{v_i}v r'_i(t)$.

Equivalently, Eq.~\eqref{eq:vbi_vehicle_eom} can be written in terms of the forces carried by the tyre-suspension systems as $\mathbf{M}_v\ddot{\mathbf{u}}_v(t) + \mathbf{L}_1F_1(t) + \mathbf{L}_2F_2(t) = \mathbf{0}$, where
\begin{equation} \label{eq:vbi_susp_force}
\begin{aligned}
    F_i(t) &= k_{v_i} \delta_i(t) + c_{v_i} \dot{\delta}_i(t),\\
    \delta_i(t) &= \mathbf{L}_i^T\mathbf{u}_v(t) - \boldsymbol{\iota}_i(t)^T\mathbf{u}_b(t) - r_i(t),\\
    \dot{\delta}_i(t) &= \mathbf{L}_i^T\dot{\mathbf{u}}_v(t) - \boldsymbol{\iota}_i(t)^T\dot{\mathbf{u}}_b(t) - v r'_i(t)
\end{aligned}
\end{equation}
\noindent are the force, the deformation and the deformation rate of the $i$-th tyre-suspension system, respectively. This form is used hereafter to assemble the coupled system. Since the equations are written about static equilibrium, gravity does not appear in Eq.~\eqref{eq:vbi_vehicle_eom}. It enters the coupled system through the static axle loads carried to the bridge.

The bridge carries the reactions of the suspension forces together with the moving static axle loads, and its equation of motion can be written as:
\begin{equation} \label{eq:vbi_bridge_eom}
    \mathbf{M}_b\ddot{\mathbf{u}}_b(t) + \mathbf{C}_b\dot{\mathbf{u}}_b(t) + \mathbf{K}_b\mathbf{u}_b(t)
    = \boldsymbol{\iota}_1(t)\left[F_1(t) - m_{v_1}g\right] + \boldsymbol{\iota}_2(t)\left[F_2(t) - m_{v_2}g\right],
\end{equation}
\noindent in which $\mathbf{M}_b$, $\mathbf{C}_b$ and $\mathbf{K}_b\in\mathbb{R}^{n_b\times n_b}$ are the mass, damping and stiffness matrices of the bridge, $g$ is the gravitational acceleration, and $m_{v_1} = m_v d_2/d$ and $m_{v_2} = m_v d_1/d$ are the portions of the vehicle mass carried by the front and rear axles, respectively. The contact forces in Eq.~\eqref{eq:vbi_bridge_eom} are the reactions of the suspension forces in Eq.~\eqref{eq:vbi_vehicle_eom}, including the pitch contributions $\pm d_i\theta_v$ that $\mathbf{L}_i$ carries into the deformations. Omitting these terms would violate the equilibrium of the massless contact.

\subsection{Coupled matrix form} \label{sec:vbi_coupled}
\noindent The vehicle and bridge equations are coupled through the suspension forces and can be assembled into a single matrix equation. The displacement vectors are stacked as $\mathbf{u}(t)=\left[\begin{matrix} \mathbf{u}_v(t)^T & \mathbf{u}_b(t)^T \end{matrix}\right]^T\in\mathbb{R}^{n_s}$ with $n_s = n_b + 2$, and the combined influence vector of the $i$-th axle is defined as:
\begin{equation} \label{eq:vbi_phi}
    \boldsymbol{\phi}_i(t) = \begin{bmatrix} \mathbf{L}_i \\ -\boldsymbol{\iota}_i(t) \end{bmatrix} \in \mathbb{R}^{n_s},
\end{equation}
\noindent with which the suspension deformation of Eq.~\eqref{eq:vbi_susp_force} takes the compact form $\delta_i(t) = \boldsymbol{\phi}_i(t)^T\mathbf{u}(t) - r_i(t)$. A single influence vector per axle thus carries both the vehicle body motion at the axle and the deck deflection beneath it. Substituting Eq.~\eqref{eq:vbi_susp_force} into the force form of the vehicle equation of motion and into Eq.~\eqref{eq:vbi_bridge_eom} then yields the monolithic coupled system:
\begin{equation} \label{eq:vbi_coupled}
    \mathbf{M}_s\ddot{\mathbf{u}}(t) + \mathbf{C}_s(t)\dot{\mathbf{u}}(t) + \mathbf{K}_s(t)\mathbf{u}(t)
    = \mathbf{\Phi}_k(t) \mathbf{r}(t) + v \mathbf{\Phi}_c(t) \mathbf{r}'(t) + \mathbf{f}_g(t),
\end{equation}

\noindent where $\mathbf{r}(t)=\left[\begin{matrix} r_1(t) & r_2(t) \end{matrix}\right]^T \in \mathbb{R}^{2}$ collects the roughness heights at the two axles, $\mathbf{r}'(t)=\left[\begin{matrix} r'_1(t) & r'_2(t) \end{matrix}\right]^T \in \mathbb{R}^{2}$ the corresponding slopes, and
\begin{equation} \label{eq:vbi_coupled_matrices}
\begin{aligned}
    \mathbf{M}_s &= \begin{bmatrix} \mathbf{M}_v & \mathbf{0} \\ \mathbf{0} & \mathbf{M}_b \end{bmatrix}, \\
    \mathbf{K}_s(t) &= \begin{bmatrix} \mathbf{0} & \mathbf{0} \\ \mathbf{0} & \mathbf{K}_b \end{bmatrix} + \sum_{i=1}^{2} k_{v_i} \boldsymbol{\phi}_i(t)\boldsymbol{\phi}_i(t)^T, \\
    \mathbf{C}_s(t) &= \begin{bmatrix} \mathbf{0} & \mathbf{0} \\ \mathbf{0} & \mathbf{C}_b \end{bmatrix} + \sum_{i=1}^{2} c_{v_i} \boldsymbol{\phi}_i(t)\boldsymbol{\phi}_i(t)^T, \\
    \mathbf{\Phi}_k(t) &= \begin{bmatrix} k_{v_1}\boldsymbol{\phi}_1(t) & k_{v_2}\boldsymbol{\phi}_2(t) \end{bmatrix}, \\
    \mathbf{\Phi}_c(t) &= \begin{bmatrix} c_{v_1}\boldsymbol{\phi}_1(t) & c_{v_2}\boldsymbol{\phi}_2(t) \end{bmatrix}, \\
    \mathbf{f}_g(t) &= -g\begin{bmatrix} \mathbf{0} \\ m_{v_1}\boldsymbol{\iota}_1(t) + m_{v_2}\boldsymbol{\iota}_2(t) \end{bmatrix}.
\end{aligned}
\end{equation}

Each axle contributes one symmetric rank-one term $\boldsymbol{\phi}_i\boldsymbol{\phi}_i^T$ to the stiffness and damping matrices, generating the vehicle blocks, the vehicle-bridge coupling blocks and the local bridge contribution at once. This is the assembled counterpart of the per-axle structure of Eq.~\eqref{eq:vbi_Kv}: restricting $\boldsymbol{\phi}_i$ to its vehicle partition reduces $\boldsymbol{\phi}_i\boldsymbol{\phi}_i^T$ to $\mathbf{L}_i\mathbf{L}_i^T$, so the assembly reproduces $\mathbf{K}_v$ and $\mathbf{C}_v$ exactly rather than adding them separately. The same vectors distribute the contact force caused by the road roughness, so that the coupling and the input structure both follow from one influence vector per axle.

Eq.~\eqref{eq:vbi_coupled} is linear but time-varying. The time dependence enters only through the contact influence vectors $\boldsymbol{\iota}_i(t)$, i.e., through the known axle positions, and the matrices are piecewise smooth, switching only at the instants of axle entry and exit.

\subsection{State-space representation} \label{sec:vbi_ss}
\noindent By defining the state vector $\mathbf{z}(t)=\left[\begin{matrix} \mathbf{u}(t)^T & \dot{\mathbf{u}}(t)^T \end{matrix}\right]^T\in\mathbb{R}^{2n_s}$ and the input vector $\mathbf{p}(t)=\left[\begin{matrix} \mathbf{r}(t)^T & v \mathbf{r}'(t)^T \end{matrix}\right]^T\in\mathbb{R}^{4}$, which collects the roughness heights and their convective rates at the two axles, Eq.~\eqref{eq:vbi_coupled} can be written in the equivalent first-order form:
\begin{equation} \label{eq:vbi_ct_ss}
    \dot{\mathbf{z}}(t) = \mathbcal{A}(t)\mathbf{z}(t) + \mathbcal{B}(t)\mathbf{p}(t) + \mathbcal{b}_g(t),
\end{equation}
\noindent where
\begin{equation} \label{eq:vbi_ct_matrices}
\begin{aligned}
    \mathbcal{A}(t) &= \begin{bmatrix}
        \mathbf{0} & \mathbf{I}_{n_s} \\
        -\mathbf{M}_s^{-1}\mathbf{K}_s(t) & -\mathbf{M}_s^{-1}\mathbf{C}_s(t)
    \end{bmatrix}, \\
    \mathbcal{B}(t) &= \begin{bmatrix}
        \mathbf{0} \\
        \mathbf{M}_s^{-1}\begin{bmatrix} \mathbf{\Phi}_k(t) & \mathbf{\Phi}_c(t) \end{bmatrix}
    \end{bmatrix}, \\
    \mathbcal{b}_g(t) &= \begin{bmatrix}
        \mathbf{0} \\
        \mathbf{M}_s^{-1}\mathbf{f}_g(t)
    \end{bmatrix}.
\end{aligned}
\end{equation}
\noindent The term $\mathbcal{b}_g(t)$ represents the moving static axle loads. Since the vehicle weight and the axle positions are known, it is a known deterministic input.

For the discrete-time form of the forward model, a zero-order hold is assumed on the input over the time interval $\Delta t$, with the input indexed at the current timestep~\cite{EbrahimzadehHassanabadi2023, Liu2025}. The discretised form of Eq.~\eqref{eq:vbi_ct_ss} then reads:
\begin{equation} \label{eq:vbi_dt_ss}
    \mathbf{z}_{k} = \mathbf{A}_{k-1}\mathbf{z}_{k-1} + \mathbf{B}_{k-1}\mathbf{p}_k + \mathbf{b}_{k-1},
\end{equation}
\noindent where $t_k = k\Delta t$, $\mathbf{z}_k = \mathbf{z}(t_k)$, and $\mathbf{A}_{k-1} \in \mathbb{R}^{2n_s \times 2n_s}$, $\mathbf{B}_{k-1} \in \mathbb{R}^{2n_s \times 4}$ and $\mathbf{b}_{k-1} \in \mathbb{R}^{2n_s}$ are the discrete-time system matrix, input matrix and deterministic term, respectively. With the time-varying matrices fixed within each interval, these can be obtained by a time integration scheme over the interval, such as the matrix exponential method~\cite{Brogan1991} or the Newmark-$\beta$ method~\cite{Newmark1959}.

Eq.~\eqref{eq:vbi_dt_ss} constitutes the discrete-time forward model of the coupled VBI system: a linear time-varying system with a known deterministic gravity term, driven by the input $\mathbf{p}$, whose components are generated by a single roughness profile $r(x)$ through the known axle positions. In the setting addressed in this paper, the road profile is not measured, and some of the system parameters are unknown. The estimation method is derived hereafter for a general class of systems, of which this model is one instance.

\section{The Unscented Universal Filter} 
\label{sec:method}

\noindent A new joint input-state-parameter estimation method for a dynamic structural system from a limited set of response measurements, named the Unscented Universal Filter (UUF), is derived in this section. Following the philosophy of the Universal Filter family, the input is estimated by minimum-variance unbiased estimation using weighted least squares, which requires no input model, input statistics or random-walk assumption. The system inversion has no restriction on the rank condition of the feedforward matrix. The parameters are tracked and updated as they vary at each timestep. The nonlinear propagation is linearised statistically by the scaled unscented transform~\cite{Julier2002, Julier2004}, whose constants set the sigma-point spread and weighting, thereby controlling the influence of higher-order terms. The propagated means are correct to second order, and the linearisation residual is carried in the covariances. It should be noted that although this study focuses on indirect drive-by structural health monitoring applications, the UUF itself is applicable to various estimation problems. Its derivation is therefore presented in a general form hereafter.

\subsection{System equations}
\label{sec:met_sys}

\noindent Eq.~\eqref{eq:vbi_coupled} is first rewritten into a more general equation of motion, representing a structural system with $n_s$ degrees of freedom and parameters of interest $\boldsymbol{\theta}(t) \in \mathbb{R}^{n_\theta}$ in the stiffness and damping matrices:
\begin{equation} \label{eq:met_class}
    \mathbf{M}_s\ddot{\mathbf{u}}(t) + \mathbf{C}_s(t;\boldsymbol{\theta}(t)) \dot{\mathbf{u}}(t) + \mathbf{K}_s(t;\boldsymbol{\theta}(t)) \mathbf{u}(t)
    = \mathbf{S}(t) \mathbf{p}(t) + \mathbf{f}_g(t),
\end{equation}

\noindent in which $\mathbf{p}(t) \in \mathbb{R}^{m}$ and $\mathbf{S}(t) \in \mathbb{R}^{n_s \times m}$ are the unknown inputs and their distribution matrix, respectively, and $\mathbf{f}_g(t) \in \mathbb{R}^{n_s}$ is a known load. Eq.~\eqref{eq:met_class} represents a system often encountered in structural health monitoring, in which the mass matrix of a structure can be determined from design information~\cite{Teymouri2023}, while the stiffness and damping matrices carry the parameters to be identified, tracked and updated. The coupled VBI system of Section~\ref{sec:vbi} is one instance.

To form a process equation for joint input-state-parameter estimation, an extended state vector $\mathbf{x}(t) \in \mathbb{R}^{n_x}$ can be defined by appending the unknown parameters to the structural responses, i.e., $\mathbf{x}(t) \triangleq \begin{bmatrix} \mathbf{z}(t)^T & \boldsymbol{\theta}(t)^T \end{bmatrix}^T$, where $\mathbf{z}(t) = \begin{bmatrix} \mathbf{u}(t)^T & \dot{\mathbf{u}}(t)^T \end{bmatrix}^T \in \mathbb{R}^{2n_s}$ and $n_x = 2n_s + n_\theta$. The discrete-time process equation can then be written as:
\begin{equation} \label{eq:met_process}
    \mathbf{x}_k = \mathbf{f}\left(\mathbf{x}_{k-1}\right) + \mathbf{G}_{k-1} \mathbf{p}_k + \mathbf{w}_{k-1},
\end{equation}

\noindent where
\begin{equation} \label{eq:met_fB}
\begin{aligned}
    \mathbf{f}\left(\mathbf{x}_{k}\right) &=
    \begin{bmatrix} \mathbf{A}_{k}(\boldsymbol{\theta}_{k}) \mathbf{z}_{k} + \mathbf{b}_{k}(\boldsymbol{\theta}_{k}) \\ \boldsymbol{\theta}_{k} \end{bmatrix},\\
    \mathbf{G}_{k} &= \begin{bmatrix} \mathbf{B}_{k} \\ \mathbf{0}_{n_\theta \times m} \end{bmatrix}.
\end{aligned}
\end{equation}

\noindent In the above equations, $\mathbf{A}_{k}(\boldsymbol{\theta}_k) \in \mathbb{R}^{2n_s \times 2n_s}$ and $\mathbf{B}_{k} \in \mathbb{R}^{2n_s \times m}$ are the system and input matrices, respectively, and $\mathbf{b}_{k}(\boldsymbol{\theta}_k) \in \mathbb{R}^{2n_s}$ is the known-load term obtained by the discretisation. Although $\mathbf{B}_{k}$ also depends on the parameters through the discretisation, it is evaluated at the current parameter estimate and is therefore written without the argument $\boldsymbol{\theta}_k$. The process noise $\mathbf{w}_{k-1} \in \mathbb{R}^{n_x}$ is assumed to be zero-mean white noise.

The sparse measurements of the system outputs are collected in the output vector $\mathbf{y}_k \in \mathbb{R}^{q}$, and the observation equation of the system with direct feedthrough can be written as:
\begin{equation} \label{eq:met_obs}
    \mathbf{y}_k = \mathbf{c}\left(\mathbf{x}_k\right) + \mathbf{D}_k \mathbf{p}_k + \mathbf{v}_k,
\end{equation}
\noindent where the extended state vector and input vector are related to the output vector through the nonlinear function $\mathbf{c}(\cdot)$ and the feedforward matrix $\mathbf{D}_k\in\mathbb{R}^{q\times m}$, respectively. The measurement noise $\mathbf{v}_k \in \mathbb{R}^{q}$ is assumed to be zero-mean white noise. Based on Eq.~\eqref{eq:met_class}, $\mathbf{c}(\mathbf{x}_k)$ and $\mathbf{D}_k$ can be derived as follows:
\begin{equation} \label{eq:met_obs_c}
    \mathbf{c}(\mathbf{x}_k) = \mathfrak{C} \begin{bmatrix}
        \mathbf{I}_{n_s} & \mathbf{0}_{n_s\times n_s} & \mathbf{0}_{n_s\times n_\theta}\\
        \mathbf{0}_{n_s\times n_s} & \mathbf{I}_{n_s} & \mathbf{0}_{n_s\times n_\theta}\\
        -\mathbf{M}_s^{-1}\mathbf{K}_s(\boldsymbol{\theta}_k) & -\mathbf{M}_s^{-1}\mathbf{C}_s(\boldsymbol{\theta}_k) & \mathbf{0}_{n_s\times n_\theta} \\ \end{bmatrix} \mathbf{x}_k
        +\mathfrak{C} \begin{bmatrix} \mathbf{0}_{2n_s \times 1} \\  \mathbf{M}_s^{-1}\mathbf{f}_{g,k}\end{bmatrix},
\end{equation}
\noindent and
\begin{equation} \label{eq:met_obs_feedforward}
    \mathbf{D}_k = \mathfrak{C} \begin{bmatrix}
        \mathbf{0}_{2n_s \times m} \\
        \mathbf{M}_s^{-1}\mathbf{S}_k \\ \end{bmatrix},
\end{equation}

\noindent where $\mathfrak{C} \in \mathbb{R}^{q\times3n_s}$ is a selection matrix mapping the locations of the measurements. The time-varying matrices are evaluated at $t_k$, with $\mathbf{S}_k = \mathbf{S}(t_k)$, $\mathbf{f}_{g,k} = \mathbf{f}_g(t_k)$, and $\mathbf{K}_s$ and $\mathbf{C}_s$ evaluated at $t_k$ and $\boldsymbol{\theta}_k$. The feedforward matrix $\mathbf{D}_k$ is full-rank, rank-deficient or zero according to the acceleration content of the sensor configuration, and the UUF imposes no restriction on its rank condition. Finally, the covariances of the process noise $\mathbf{w}_k$ and the measurement noise $\mathbf{v}_k$ are defined as:
\begin{equation} \label{eq:met_noise}
\begin{aligned}
    \mathbf{Q}_k &\triangleq \mathbb{E}\left[\mathbf{w}_k\mathbf{w}_k^T\right],\\
    \mathbf{R}_k &\triangleq \mathbb{E}\left[\mathbf{v}_k\mathbf{v}_k^T\right].
\end{aligned}
\end{equation}
\noindent The process noise $\mathbf{w}_k$ and the measurement noise $\mathbf{v}_k$ are assumed to be mutually uncorrelated.

\subsection{Statistical linearisation by the unscented transform} 
\label{sec:met_ut}

\noindent The nonlinear propagation of the developed UUF is linearised by the scaled unscented transform~\cite{Julier2002, Julier2004}, which has been well established in structural system identification~\cite{ChatziSmyth2009, Dertimanis2019, ImpraimakisSmyth2022}. Compared with the Jacobian matrix used in the NUF and the classical extended Kalman filter, the unscented transform requires no derivatives, and it captures the propagated mean and covariance correctly to second order in the Taylor expansion~\cite{Julier2004}. Furthermore, it delivers the residual of the linearisation in closed form, which can be carried in the error propagation. The scaled unscented transform adopted in the developed UUF is introduced hereafter. 

Let $\hat{\mathbf{x}}$ be an estimate of the extended state vector $\mathbf{x}$ with an error covariance $\mathbf{P}$, and let $\sqrt{\mathbf{P}}$ be a matrix square root of $\mathbf{P}$. Its $i$-th column is denoted $(\sqrt{\mathbf{P}})_i \in \mathbb{R}^{n_x}$. The transform draws a set of $2n_x + 1$ sigma points $\mathcal{X}_i \in \mathbb{R}^{n_x}$, $i = 0, \dots, 2n_x$. The point $\mathcal{X}_0$ coincides with the estimate, and the remaining $\mathcal{X}_i$ are placed symmetrically around it along $(\sqrt{\mathbf{P}})_i$ at a scale $\gamma$. The sigma points are given by:
\begin{equation} \label{eq:met_sigma}
\begin{aligned}
    \mathcal{X}_0       &= \hat{\mathbf{x}},\\
    \mathcal{X}_i       &= \hat{\mathbf{x}} + \gamma(\sqrt{\mathbf{P}})_i       &\quad \mathrm{for} \ i = 1, \dots, n_x, \\
    \mathcal{X}_{i+n_x} &= \hat{\mathbf{x}} - \gamma(\sqrt{\mathbf{P}})_i       &\quad \mathrm{for} \ i = 1, \dots, n_x,
\end{aligned}
\end{equation}

\noindent in which the scale is $\gamma = \sqrt{n_x + \lambda}$, with $\lambda = \alpha^2\left(n_x + \kappa\right) - n_x$ the scaling parameter of the transform. Equivalently, $\gamma = \alpha\sqrt{n_x + \kappa}$, so that the constant $\alpha$ sets the spread of the set around the estimate. The mean weights $W_i^{(m)}$ and covariance weights $W_i^{(c)}$ are assigned below~\cite{Dertimanis2019}: 
\begin{equation} \label{eq:met_weights}
\begin{aligned}
    W_0^{(m)} &= \frac{\lambda}{n_x + \lambda},\\
    W_0^{(c)} &= W_0^{(m)} + 1 - \alpha^2 + \beta,\\
    W_i^{(m)} &= W_i^{(c)} = \frac{1}{2\left(n_x + \lambda\right)} \quad \left(i \geq 1\right),
\end{aligned}
\end{equation}

\noindent where $\alpha$, $\beta$ and $\kappa$ are the standard constants of the scaled transform, chosen herein such that all covariance weights are nonnegative.

Furthermore, the propagated statistics of the unscented transform, on which the developed UUF draws, are briefly introduced below. More details can be found in~\cite{Sarkka2023}. Each sigma point is propagated through a nonlinear function $\mathbf{h}(\cdot)$, giving the transformed sigma points $\mathcal{Y}_i = \mathbf{h}\left(\mathcal{X}_i\right)$. The mean of the propagated variable $\mathbf{h}(\mathbf{x})$ is estimated as:
\begin{equation} \label{eq:met_ut_mean}
    \bar{\mathbf{h}} = \sum_{i} W_i^{(m)} \mathcal{Y}_i. 
\end{equation}

The covariance $\mathbf{P}^{\mathbf{hh}}$ of the propagated variable and the cross-covariance $\mathbf{P}^{\mathbf{xh}}$ between the extended state $\mathbf{x}$ and the propagated variable are computed as:
\begin{equation} \label{eq:met_ut}
\begin{aligned}
    \mathbf{P}^{\mathbf{hh}} &= \sum_{i} W_i^{(c)}\left(\mathcal{Y}_i - \bar{\mathbf{h}}\right)\left(\mathcal{Y}_i - \bar{\mathbf{h}}\right)^T,\\
    \mathbf{P}^{\mathbf{xh}} &= \sum_{i} W_i^{(c)}\left(\mathcal{X}_i - \hat{\mathbf{x}}\right)\left(\mathcal{Y}_i - \bar{\mathbf{h}}\right)^T.
\end{aligned}
\end{equation}

The propagated moments further define a statistical linearisation: over the spread of the set, the function $\mathbf{h}(\cdot)$ is approximated by the linear map $\mathbf{h}(\mathbf{x}) \approx \bar{\mathbf{h}} + \bar{\mathbcal{H}}\left(\mathbf{x} - \hat{\mathbf{x}}\right)$, in which the statistically linearised matrix $\bar{\mathbcal{H}}$ is obtained by weighted least squares over the sigma points:
\begin{equation} \label{eq:met_statlin}
    \bar{\mathbcal{H}} \triangleq \left(\mathbf{P}^{\mathbf{xh}}\right)^T\mathbf{P}^{-1},
\end{equation}
\noindent and the covariance of its residual is given by:
\begin{equation} \label{eq:met_res}
    \boldsymbol{\Omega} \triangleq \mathbf{P}^{\mathbf{hh}} - \bar{\mathbcal{H}} \mathbf{P} \bar{\mathbcal{H}}^T,
\end{equation}
\noindent which is positive semi-definite whenever all covariance weights are nonnegative.

It should be noted that Eqs.~\eqref{eq:met_ut_mean} to~\eqref{eq:met_res} are written in a general form. In the application of the developed UUF, the nonlinear function $\mathbf{h}(\cdot)$ can stand for the process function $\mathbf{f}(\cdot)$ or the observation function $\mathbf{c}(\cdot)$ of Section~\ref{sec:met_sys}. Similarly, $\bar{\mathbcal{H}}$ is replaced by $\bar{\mathbcal{F}}$ or $\bar{\mathbcal{C}}$ for the linearised matrices of $\mathbf{f}(\cdot)$ and $\mathbf{c}(\cdot)$, respectively. Superscripts such as $(\cdot)^{(1)}$, $(\cdot)^{(2)}$ and $(\cdot)^{(3)}$ are used hereafter to distinguish the sigma sets when multiple sets are drawn per timestep. In addition, the symbol $\mathbf{h}$ in the superscript of the covariance matrices $\mathbf{P}^{(\cdot)}$ is also replaced by the actual contributing variables.

\subsection{Filter steps}
\label{sec:met_steps}

\noindent The developed UUF comprises four main steps. First, the biased state estimate $\hat{\mathbf{x}}_{k|k-1}$ can be formed from the information at the previous timestep as:
\begin{equation} \label{eq:met_biased}
    \hat{\mathbf{x}}_{k|k-1} = \sum_{i} W_i^{(m)}\mathbf{f}\left(\mathcal{X}_i^{(1)}\right), 
\end{equation}

\noindent where $\mathcal{X}_i^{(1)}$ are the sigma points drawn from the previous timestep estimate $\hat{\mathbf{x}}_{k-1|k-1}$ and its error covariance. By comparing Eqs.~\eqref{eq:met_biased} and~\eqref{eq:met_process}, the bias is caused by the missing contribution of the unknown input $\mathbf{p}_k$. This implies that the unknown input can be estimated implicitly from the innovation, i.e., the difference between the observation and its prediction. The innovation at this stage can be written as:
\begin{equation} \label{eq:met_innovation}
    \tilde{\mathbf{y}}_k = \mathbf{y}_k - \sum_{i} W_i^{(m)} \mathbf{c}\left(\mathcal{X}_i^{(2)}\right),
\end{equation}

\noindent where $\mathcal{X}_i^{(2)}$ are the sigma points drawn from the biased estimate $\hat{\mathbf{x}}_{k|k-1}$ and its error covariance. Second, the unknown input is estimated through the input gain matrix $\mathbf{M}_k \in \mathbb{R}^{m \times q}$:
\begin{equation} \label{eq:met_inputest}
    \hat{\mathbf{p}}_k = \mathbf{M}_k\tilde{\mathbf{y}}_k = \mathbf{M}_k\left[\mathbf{y}_k - \sum_{i} W_i^{(m)} \mathbf{c}\left(\mathcal{X}_i^{(2)}\right)\right].
\end{equation}

\noindent In the above equation, the input gain $\mathbf{M}_k$ is obtained by weighted least squares. Third, the a priori state estimate is formed by adding the contribution of the estimated input:
\begin{equation} \label{eq:met_timeupdate}
    \hat{\mathbcal{x}}_{k|k-1} = \hat{\mathbf{x}}_{k|k-1} + \mathbf{G}_{k-1} \hat{\mathbf{p}}_k.
\end{equation}

Finally, drawing the sigma points $\mathcal{X}_i^{(3)}$ from the a priori state estimate $\hat{\mathbcal{x}}_{k|k-1}$ and its error covariance, the a posteriori state estimate can be written as: 
\begin{equation} \label{eq:met_measupdate}
    \hat{\mathbf{x}}_{k|k} = \hat{\mathbcal{x}}_{k|k-1} + \mathbf{K}_k\left[\mathbf{y}_k - \sum_{i} W_i^{(m)} \mathbf{c}\left(\mathcal{X}_i^{(3)}\right) - \mathbf{D}_k\hat{\mathbf{p}}_k\right],
\end{equation}
\noindent in which $\mathbf{K}_k \in \mathbb{R}^{n_x \times q}$ is the state gain matrix that minimises the trace of the state estimation error covariance. With the main filter steps defined, the error propagation with the proof of unbiasedness, the error covariances and the gain matrices are derived in the following subsections.

\subsection{Error propagation} \label{sec:met_err}
\noindent The estimation errors of the input estimate $\tilde{\mathbf{p}}_k$, the a priori state estimate $\tilde{\mathbcal{x}}_k$ and the a posteriori state estimate $\tilde{\mathbf{x}}_k$ are defined below:
\begin{equation} \label{eq:met_errdefs}
    \tilde{\mathbf{p}}_k \triangleq \mathbf{p}_k - \hat{\mathbf{p}}_k,
    \qquad
    \tilde{\mathbcal{x}}_k \triangleq \mathbf{x}_k - \hat{\mathbcal{x}}_{k|k-1},
    \qquad
    \tilde{\mathbf{x}}_k \triangleq \mathbf{x}_k - \hat{\mathbf{x}}_{k|k}.
\end{equation}

To derive the input estimation error, the innovation defined in Eq.~\eqref{eq:met_innovation} is first rewritten by substituting the observation equation, Eq.~\eqref{eq:met_obs}:
\begin{equation} \label{eq:met_innovobs}
    \tilde{\mathbf{y}}_k = \mathbf{c}\left(\mathbf{x}_k\right) - \sum_{i} W_i^{(m)} \mathbf{c}\left(\mathcal{X}_i^{(2)}\right) + \mathbf{D}_k \mathbf{p}_k + \mathbf{v}_k.
\end{equation}

According to the process equation, Eq.~\eqref{eq:met_process}, the true state $\mathbf{x}_k$ carries the unknown input. To separate it, Eq.~\eqref{eq:met_process} is first substituted into $\mathbf{c}\left(\mathbf{x}_k\right)$. Next, the statistical linearisation of Eq.~\eqref{eq:met_statlin} is applied to $\mathbf{c}(\cdot)$ on the second sigma set, which gives the statistically linearised observation matrix $\bar{\mathbcal{C}}_k^{(2)}$ and separates the input contribution from the input-free part:
\begin{equation} \label{eq:met_clin}
    \mathbf{c}\left(\mathbf{x}_k\right) = \mathbf{c}\left(\mathbf{f}\left(\mathbf{x}_{k-1}\right) + \mathbf{w}_{k-1} + \mathbf{G}_{k-1} \mathbf{p}_k\right) \approx \mathbf{c}\left(\mathbf{f}\left(\mathbf{x}_{k-1}\right) + \mathbf{w}_{k-1}\right) + \bar{\mathbcal{C}}_k^{(2)} \mathbf{G}_{k-1} \mathbf{p}_k.
\end{equation}

The approximation in Eq.~\eqref{eq:met_clin} replaces the exact contribution of the input increment $\mathbf{G}_{k-1}\mathbf{p}_k$ to the observation with the linear term $\bar{\mathbcal{C}}_k^{(2)} \mathbf{G}_{k-1} \mathbf{p}_k$. The omitted difference scales with the parameter error times the input magnitude. Substituting Eq.~\eqref{eq:met_clin} into Eq.~\eqref{eq:met_innovobs}, the innovation $\tilde{\mathbf{y}}_k$ can be expressed with the input $\mathbf{p}_k$ and an error term $\mathbf{e}_k$:
\begin{equation} \label{eq:met_innovdecomp}
    \tilde{\mathbf{y}}_k = \mathbcal{G}_k \mathbf{p}_k + \mathbf{e}_k,
\end{equation}

\noindent where $\mathbcal{G}_k \in \mathbb{R}^{q \times m}$ is given by:
\begin{equation} \label{eq:met_Gk}
    \mathbcal{G}_k = \bar{\mathbcal{C}}_k^{(2)}\mathbf{G}_{k-1} + \mathbf{D}_k,
\end{equation}

\noindent and the error term $\mathbf{e}_k \in \mathbb{R}^{q}$ collects the input-free terms and the measurement noise:
\begin{equation} \label{eq:met_ek}
    \mathbf{e}_k = \mathbf{c}\left(\mathbf{f}\left(\mathbf{x}_{k-1}\right) + \mathbf{w}_{k-1}\right) - \sum_{i} W_i^{(m)} \mathbf{c}\left(\mathcal{X}_i^{(2)}\right) + \mathbf{v}_k.
\end{equation}

With Eqs.~\eqref{eq:met_inputest} and~\eqref{eq:met_innovdecomp}, the expectation of the input estimate $\hat{\mathbf{p}}_k$ can be written as $\mathbb{E}[\hat{\mathbf{p}}_k] = \mathbb{E}\left[\mathbf{M}_k\mathbcal{G}_k \mathbf{p}_k + \mathbf{M}_k\mathbf{e}_k\right]$. For the input estimate to be unbiased, i.e., $\mathbb{E}[\hat{\mathbf{p}}_k] = \mathbb{E}[\mathbf{p}_k]$, it is required that $\mathbb{E}[\mathbf{e}_k] = \mathbf{0}$ and that the input gain satisfies:
\begin{equation} \label{eq:met_unbias}
    \mathbf{M}_k \mathbcal{G}_k = \mathbf{I}_m.
\end{equation}

The condition $\mathbb{E}[\mathbf{e}_k] = \mathbf{0}$ can be established based on Eq.~\eqref{eq:met_ek}. Applying the statistical linearisation of Eq.~\eqref{eq:met_statlin} to $\mathbf{c}(\cdot)$ on the second sigma set, and considering its residual $\boldsymbol{\epsilon}_k^{(2)}$ at the evaluation point, the first term of Eq.~\eqref{eq:met_ek} can be written as:
\begin{equation} \label{eq:met_cblin}
    \mathbf{c}\left(\mathbf{f}\left(\mathbf{x}_{k-1}\right) + \mathbf{w}_{k-1}\right) = \sum_{i} W_i^{(m)} \mathbf{c}\left(\mathcal{X}_i^{(2)}\right) + \bar{\mathbcal{C}}_k^{(2)}\left[\mathbf{f}\left(\mathbf{x}_{k-1}\right) + \mathbf{w}_{k-1} - \hat{\mathbf{x}}_{k|k-1}\right] + \boldsymbol{\epsilon}_k^{(2)}.
\end{equation}

Similarly, applying Eq.~\eqref{eq:met_statlin} to $\mathbf{f}(\cdot)$ on the first sigma set, one can obtain the statistically linearised process matrix $\bar{\mathbcal{F}}_k$ and its residual $\boldsymbol{\epsilon}_k^{(1)}$. Using Eq.~\eqref{eq:met_biased}, the $\mathbf{f}(\cdot)$-related term in Eq.~\eqref{eq:met_cblin} can be written as:
\begin{equation} \label{eq:met_fblin}
    \mathbf{f}\left(\mathbf{x}_{k-1}\right) + \mathbf{w}_{k-1} - \hat{\mathbf{x}}_{k|k-1} = \bar{\mathbcal{F}}_k \tilde{\mathbf{x}}_{k-1} + \mathbf{w}_{k-1} + \boldsymbol{\epsilon}_k^{(1)}.
\end{equation}

Substituting Eqs.~\eqref{eq:met_cblin} and~\eqref{eq:met_fblin} into Eq.~\eqref{eq:met_ek}, the error term becomes:
\begin{equation} \label{eq:met_elin}
    \mathbf{e}_k = \bar{\mathbcal{C}}_k^{(2)}\left(\bar{\mathbcal{F}}_k \tilde{\mathbf{x}}_{k-1} + \mathbf{w}_{k-1} + \boldsymbol{\epsilon}_k^{(1)}\right) + \boldsymbol{\epsilon}_k^{(2)} + \mathbf{v}_k.
\end{equation}

Under the transform weights, the residuals of the statistical linearisation are zero-mean over their sigma spreads and orthogonal to the state deviations, both properties following from the weighted least squares of Eq.~\eqref{eq:met_statlin}. They are hereafter treated as mutually uncorrelated zero-mean noises~\cite{Sarkka2023}. Given zero-mean $\mathbf{w}_{k-1}$ and $\mathbf{v}_k$ and an unbiased previous estimate, i.e., $\mathbb{E}[\tilde{\mathbf{x}}_{k-1}] = \mathbf{0}$, Eq.~\eqref{eq:met_elin} yields $\mathbb{E}[\mathbf{e}_k] = \mathbf{0}$. This completes the proof of unbiasedness for the input estimation.

The input estimation error $\tilde{\mathbf{p}}_k$ defined in Eq.~\eqref{eq:met_errdefs} is then obtained by combining Eqs.~\eqref{eq:met_inputest}, \eqref{eq:met_innovdecomp} and~\eqref{eq:met_unbias}:
\begin{equation} \label{eq:met_ptilde}
    \tilde{\mathbf{p}}_k = -\mathbf{M}_k \mathbf{e}_k. 
\end{equation}

To formulate the estimation error of the a priori state estimate $\tilde{\mathbcal{x}}_k$, one can substitute Eqs.~\eqref{eq:met_process} and~\eqref{eq:met_timeupdate} into the definition of $\tilde{\mathbcal{x}}_k$ and obtain:
\begin{equation} \label{eq:met_erraprioristep}
    \tilde{\mathbcal{x}}_k = \mathbf{f}\left(\mathbf{x}_{k-1}\right) + \mathbf{w}_{k-1} - \hat{\mathbf{x}}_{k|k-1} + \mathbf{G}_{k-1}\left(\mathbf{p}_k - \hat{\mathbf{p}}_k\right).
\end{equation}

\noindent Replacing the first three terms in the above equation with Eq.~\eqref{eq:met_fblin}, the a priori estimation error follows the recursive form:
\begin{equation} \label{eq:met_errapriori}
    \tilde{\mathbcal{x}}_k = \bar{\mathbcal{F}}_k \tilde{\mathbf{x}}_{k-1} + \mathbf{G}_{k-1} \tilde{\mathbf{p}}_k + \mathbf{w}_{k-1} + \boldsymbol{\epsilon}_k^{(1)}.
\end{equation}

Similarly, the estimation error of the a posteriori state estimate $\tilde{\mathbf{x}}_k$ can be derived by substituting Eqs.~\eqref{eq:met_obs} and~\eqref{eq:met_measupdate} into the definition of $\tilde{\mathbf{x}}_k$:
\begin{equation} \label{eq:met_errapoststep}
    \tilde{\mathbf{x}}_k = \tilde{\mathbcal{x}}_k - \mathbf{K}_k\left[\mathbf{c}\left(\mathbf{x}_k\right) - \sum_{i} W_i^{(m)} \mathbf{c}\left(\mathcal{X}_i^{(3)}\right) + \mathbf{D}_k \tilde{\mathbf{p}}_k + \mathbf{v}_k\right].
\end{equation}

\noindent Applying the statistical linearisation of Eq.~\eqref{eq:met_statlin} to $\mathbf{c}(\cdot)$ on the third sigma set gives the statistically linearised observation matrix $\bar{\mathbcal{C}}_k^{(3)}$ and its residual $\boldsymbol{\epsilon}_k^{(3)}$. The first two terms in the brackets then reduce to $\bar{\mathbcal{C}}_k^{(3)} \tilde{\mathbcal{x}}_k + \boldsymbol{\epsilon}_k^{(3)}$, and the a posteriori estimation error follows the recursive form:
\begin{equation} \label{eq:met_errapost}
    \tilde{\mathbf{x}}_k = \tilde{\mathbcal{x}}_k - \mathbf{K}_k\left(\bar{\mathbcal{C}}_k^{(3)} \tilde{\mathbcal{x}}_k + \mathbf{D}_k \tilde{\mathbf{p}}_k + \mathbf{v}_k + \boldsymbol{\epsilon}_k^{(3)}\right).
\end{equation}

Eqs.~\eqref{eq:met_errapriori} and~\eqref{eq:met_errapost} indicate how the estimation error propagates from the previous timestep estimate to the current a posteriori estimate. The error covariances involved are derived next.

\subsection{Estimation error covariances} 
\label{sec:met_cov}

\noindent The biased predicted covariance over the first sigma set is defined and evaluated by the transform as follows:
\begin{equation} \label{eq:met_Pminus}
\begin{aligned}
    \mathbf{P}_k^{-} &\triangleq \mathbb{E}\left\{\left[\mathbf{f}\left(\mathbf{x}_{k-1}\right) + \mathbf{w}_{k-1} - \hat{\mathbf{x}}_{k|k-1}\right]\left[\mathbf{f}\left(\mathbf{x}_{k-1}\right) + \mathbf{w}_{k-1} - \hat{\mathbf{x}}_{k|k-1}\right]^T\right\} \\
    &= \sum_{i} W_i^{(c)}\left[\mathbf{f}\left(\mathcal{X}_i^{(1)}\right) - \hat{\mathbf{x}}_{k|k-1}\right]\left[\mathbf{f}\left(\mathcal{X}_i^{(1)}\right) - \hat{\mathbf{x}}_{k|k-1}\right]^T + \mathbf{Q}_{k-1}.
\end{aligned}
\end{equation}

Before the remaining covariances are derived, the covariance $\bar{\mathbf{Q}}_{k-1}$ of the combined noise $\bar{\mathbf{w}}_{k-1}$ and the residual covariance $\boldsymbol{\Omega}_k^{(1)}$ are introduced:
\begin{equation} \label{eq:met_Qbar}
\begin{aligned}
    \bar{\mathbf{Q}}_{k-1} &\triangleq \mathbb{E}\left[\bar{\mathbf{w}}_{k-1}\bar{\mathbf{w}}_{k-1}^T\right] = \mathbf{Q}_{k-1} + \boldsymbol{\Omega}_k^{(1)}, \\
    \boldsymbol{\Omega}_k^{(1)} &= \mathbf{P}_k^{-} - \mathbf{Q}_{k-1} - \bar{\mathbcal{F}}_k \mathbf{P}_{k-1}^{\mathbf{x}} \bar{\mathbcal{F}}_k^T,
\end{aligned}
\end{equation}

\noindent in which $\bar{\mathbf{w}}_{k-1} \triangleq \mathbf{w}_{k-1} + \boldsymbol{\epsilon}_k^{(1)}$ is the combination of the residual $\boldsymbol{\epsilon}_k^{(1)}$ of the process function $\mathbf{f}(\cdot)$ and the process noise $\mathbf{w}_{k-1}$, and $\boldsymbol{\Omega}_k^{(1)}$ is the residual covariance of the first sigma set according to Eq.~\eqref{eq:met_res}. Substituting Eq.~\eqref{eq:met_Qbar} into Eq.~\eqref{eq:met_Pminus}, the covariance $\mathbf{P}_k^{-}$ can be written as:
\begin{equation}
    \mathbf{P}_k^{-} = \bar{\mathbcal{F}}_k \mathbf{P}_{k-1}^{\mathbf{x}} \bar{\mathbcal{F}}_k^T + \bar{\mathbf{Q}}_{k-1}.
\end{equation}

Note that $\bar{\mathbf{Q}}_{k-1}$ replaces $\mathbf{Q}_{k-1}$ in every cross-covariance hereafter. Next, the covariance of the error term $\mathbf{e}_k$ introduced in Eq.~\eqref{eq:met_elin} can be computed as follows:
\begin{equation} \label{eq:met_Pe}
\begin{aligned}
    \mathbf{P}_k^{\mathbf{e}}  &\triangleq \mathbb{E}\left[\mathbf{e}_k\mathbf{e}_k^T\right] \\
    & = \mathbb{E}\Big\{\Big[\bar{\mathbcal{C}}_k^{(2)}\left(\bar{\mathbcal{F}}_k \tilde{\mathbf{x}}_{k-1} + \mathbf{w}_{k-1} + \boldsymbol{\epsilon}_k^{(1)}\right) + \boldsymbol{\epsilon}_k^{(2)} + \mathbf{v}_k\Big] \\
    & \qquad\qquad\qquad\qquad\qquad\qquad \times \Big[\bar{\mathbcal{C}}_k^{(2)}\left(\bar{\mathbcal{F}}_k \tilde{\mathbf{x}}_{k-1} + \mathbf{w}_{k-1} + \boldsymbol{\epsilon}_k^{(1)}\right) + \boldsymbol{\epsilon}_k^{(2)} + \mathbf{v}_k\Big]^T\Big\} \\
    & = \bar{\mathbcal{C}}_k^{(2)} \mathbf{P}_k^{-}\left(\bar{\mathbcal{C}}_k^{(2)}\right)^T + \boldsymbol{\Omega}_k^{(2)} + \mathbf{R}_k,
\end{aligned}
\end{equation}

\noindent with $\boldsymbol{\Omega}_k^{(2)}$ the residual covariance of the second sigma set according to Eq.~\eqref{eq:met_res}. The cross-covariances related to the input estimate follow from Eqs.~\eqref{eq:met_ptilde} and~\eqref{eq:met_elin}:
\begin{equation} \label{eq:met_Ppx}
    \mathbf{P}_k^{\mathbf{px}} \triangleq \left(\mathbf{P}_k^{\mathbf{xp}}\right)^T \triangleq \mathbb{E}\left[\tilde{\mathbf{p}}_k \tilde{\mathbf{x}}_{k-1}^T\right]
    = -\mathbf{M}_k \bar{\mathbcal{C}}_k^{(2)} \bar{\mathbcal{F}}_k \mathbf{P}_{k-1}^{\mathbf{x}},
\end{equation}
\vspace{-1.5em}
\begin{equation} \label{eq:met_Ppw}
    \mathbf{P}_k^{\mathbf{pw}} \triangleq \left(\mathbf{P}_k^{\mathbf{wp}}\right)^T \triangleq \mathbb{E}\left[\tilde{\mathbf{p}}_k \bar{\mathbf{w}}_{k-1}^T\right]
    = -\mathbf{M}_k \bar{\mathbcal{C}}_k^{(2)} \bar{\mathbf{Q}}_{k-1},
\end{equation}
\vspace{-1.5em}
\begin{equation} \label{eq:met_Ppv}
    \mathbf{P}_k^{\mathbf{pv}} \triangleq \left(\mathbf{P}_k^{\mathbf{vp}}\right)^T \triangleq \mathbb{E}\left[\tilde{\mathbf{p}}_k \mathbf{v}_k^T\right]
    = -\mathbf{M}_k \mathbf{R}_k.
\end{equation}

\noindent The a priori state error covariance and its related cross-covariances then follow from Eq.~\eqref{eq:met_errapriori}:
\begin{equation} \label{eq:met_Pax}
\begin{aligned}
    \mathbf{P}_k^{\mathbcal{x}} &\triangleq \mathbb{E}\left[\tilde{\mathbcal{x}}_k \tilde{\mathbcal{x}}_k^T\right] \\
    & = \mathbb{E} \left[\left(\bar{\mathbcal{F}}_k \tilde{\mathbf{x}}_{k-1} + \mathbf{G}_{k-1} \tilde{\mathbf{p}}_k + \mathbf{w}_{k-1} + \boldsymbol{\epsilon}_k^{(1)}\right)\left(\bar{\mathbcal{F}}_k \tilde{\mathbf{x}}_{k-1} + \mathbf{G}_{k-1} \tilde{\mathbf{p}}_k + \mathbf{w}_{k-1} + \boldsymbol{\epsilon}_k^{(1)}\right)^T\right]\\
    & = \mathbf{P}_k^{-} + \mathbf{G}_{k-1} \mathbf{P}_k^{\mathbf{p}} \mathbf{G}_{k-1}^T
    + \bar{\mathbcal{F}}_k \mathbf{P}_k^{\mathbf{xp}} \mathbf{G}_{k-1}^T
    + \mathbf{G}_{k-1} \mathbf{P}_k^{\mathbf{px}} \bar{\mathbcal{F}}_k^T 
    + \mathbf{G}_{k-1} \mathbf{P}_k^{\mathbf{pw}} + \mathbf{P}_k^{\mathbf{wp}} \mathbf{G}_{k-1}^T,
\end{aligned}
\end{equation}
\vspace{-1em}
\begin{equation} \label{eq:met_Ppax}
\begin{aligned}
    \mathbf{P}_k^{\mathbf{p}\mathbcal{x}} &\triangleq \mathbb{E}\left[\tilde{\mathbf{p}}_k \tilde{\mathbcal{x}}_k^T\right]\\
    &= \mathbb{E}\left[\left(-\mathbf{M}_k\mathbf{e}_k\right)\left(\bar{\mathbcal{F}}_k \tilde{\mathbf{x}}_{k-1} + \mathbf{G}_{k-1} \tilde{\mathbf{p}}_k + \mathbf{w}_{k-1} + \boldsymbol{\epsilon}_k^{(1)}\right)^T\right]\\
    &= \mathbf{P}_k^{\mathbf{px}} \bar{\mathbcal{F}}_k^T + \mathbf{P}_k^{\mathbf{p}} \mathbf{G}_{k-1}^T + \mathbf{P}_k^{\mathbf{pw}},
\end{aligned}
\end{equation}
\vspace{-1em}
\begin{equation}
\begin{aligned}
    \mathbf{P}_k^{\mathbf{v}\mathbcal{x}} &\triangleq \mathbb{E}\left[\mathbf{v}_k \tilde{\mathbcal{x}}_k^T\right]\\
    & = \mathbb{E}\left[\mathbf{v}_k \left(\bar{\mathbcal{F}}_k \tilde{\mathbf{x}}_{k-1} + \mathbf{G}_{k-1} \tilde{\mathbf{p}}_k + \mathbf{w}_{k-1} + \boldsymbol{\epsilon}_k^{(1)}\right)^T\right] \\
    & = \mathbf{P}_k^{\mathbf{vp}} \mathbf{G}_{k-1}^T,
\end{aligned}
\end{equation}

\noindent with $\mathbf{P}_k^{\mathbf{p}} \triangleq \mathbb{E}\left[\tilde{\mathbf{p}}_k \tilde{\mathbf{p}}_k^T\right]$ the input estimation error covariance. From Eq.~\eqref{eq:met_errapost}, the a posteriori state error covariance can be computed as follows:
\begin{equation} \label{eq:met_Px}
\begin{aligned}
    \mathbf{P}_k^{\mathbf{x}} &\triangleq \mathbb{E}\left[\tilde{\mathbf{x}}_k \tilde{\mathbf{x}}_k^T\right] \\
    & = \mathbb{E}\Big\{\Big[\tilde{\mathbcal{x}}_k - \mathbf{K}_k\left(\bar{\mathbcal{C}}_k^{(3)} \tilde{\mathbcal{x}}_k + \mathbf{D}_k \tilde{\mathbf{p}}_k + \mathbf{v}_k + \boldsymbol{\epsilon}_k^{(3)}\right)\Big] \\
    & \qquad\qquad\qquad\qquad\qquad\qquad \times \Big[\tilde{\mathbcal{x}}_k - \mathbf{K}_k\left(\bar{\mathbcal{C}}_k^{(3)} \tilde{\mathbcal{x}}_k + \mathbf{D}_k \tilde{\mathbf{p}}_k + \mathbf{v}_k + \boldsymbol{\epsilon}_k^{(3)}\right)\Big]^T\Big\} \\
    & = \mathbf{P}_k^{\mathbcal{x}} + \mathbf{K}_k \boldsymbol{\Psi}_k \mathbf{K}_k^T - \boldsymbol{\Upsilon}_k^T \mathbf{K}_k^T - \mathbf{K}_k \boldsymbol{\Upsilon}_k,
\end{aligned}
\end{equation}
\noindent in which
\begin{equation} \label{eq:met_Ups}
\begin{aligned}
    \boldsymbol{\Upsilon}_k &\triangleq \mathbb{E}\left[\left(\bar{\mathbcal{C}}_k^{(3)} \tilde{\mathbcal{x}}_k + \mathbf{D}_k \tilde{\mathbf{p}}_k + \mathbf{v}_k + \boldsymbol{\epsilon}_k^{(3)}\right)\tilde{\mathbcal{x}}_k^T\right] \\
    &= \bar{\mathbcal{C}}_k^{(3)} \mathbf{P}_k^{\mathbcal{x}} + \mathbf{D}_k \mathbf{P}_k^{\mathbf{p}\mathbcal{x}} + \mathbf{P}_k^{\mathbf{v}\mathbcal{x}},
\end{aligned}
\end{equation}
\begin{equation} \label{eq:met_Psi}
\begin{aligned}
    \boldsymbol{\Psi}_k &\triangleq \mathbb{E}\left[\left(\bar{\mathbcal{C}}_k^{(3)} \tilde{\mathbcal{x}}_k + \mathbf{D}_k \tilde{\mathbf{p}}_k + \mathbf{v}_k + \boldsymbol{\epsilon}_k^{(3)}\right)\left(\bar{\mathbcal{C}}_k^{(3)} \tilde{\mathbcal{x}}_k + \mathbf{D}_k \tilde{\mathbf{p}}_k + \mathbf{v}_k + \boldsymbol{\epsilon}_k^{(3)}\right)^T\right] \\
    &= \bar{\mathbcal{C}}_k^{(3)} \boldsymbol{\Upsilon}_k^T
    + \mathbf{D}_k\left[\mathbf{P}_k^{\mathbf{p}\mathbcal{x}}\left(\bar{\mathbcal{C}}_k^{(3)}\right)^T + \mathbf{P}_k^{\mathbf{p}} \mathbf{D}_k^T + \mathbf{P}_k^{\mathbf{pv}}\right]
    + \mathbf{P}_k^{\mathbf{v}\mathbcal{x}}\left(\bar{\mathbcal{C}}_k^{(3)}\right)^T + \mathbf{P}_k^{\mathbf{vp}} \mathbf{D}_k^T + \mathbf{R}_k + \boldsymbol{\Omega}_k^{(3)}.
\end{aligned}
\end{equation}
\noindent The residual covariance of the third sigma set $\boldsymbol{\Omega}_k^{(3)}$ in Eq.~\eqref{eq:met_Psi} follows from Eq.~\eqref{eq:met_res}:
{\footnotesize
\begin{equation} \label{eq:met_Om3}
    \boldsymbol{\Omega}_k^{(3)} = \sum_{i} W_i^{(c)}\left[\mathbf{c}\left(\mathcal{X}_i^{(3)}\right) - \sum_{j} W_j^{(m)} \mathbf{c}\left(\mathcal{X}_j^{(3)}\right)\right] \left[\mathbf{c}\left(\mathcal{X}_i^{(3)}\right) -  \sum_{j} W_j^{(m)} \mathbf{c}\left(\mathcal{X}_j^{(3)}\right)\right]^T 
    - \bar{\mathbcal{C}}_k^{(3)} \mathbf{P}_k^{\mathbcal{x}}\left(\bar{\mathbcal{C}}_k^{(3)}\right)^T.
\end{equation}
}

With the error covariances established, the two gain matrices $\mathbf{M}_k$ and $\mathbf{K}_k$ are derived next.

\subsection{Gain matrices} \label{sec:met_gains}
\noindent To derive the input gain, the weighted least squares is herein adopted to achieve a minimum-variance estimation of the unknown input from the innovation decomposition of Eq.~\eqref{eq:met_innovdecomp}, with the weighting being the covariance $\mathbf{P}_k^{\mathbf{e}}$ of the error term $\mathbf{e}_k$. The input estimate is the solution of the following minimisation problem:
\begin{equation} \label{eq:met_wlscost}
    \hat{\mathbf{p}}_k = \operatorname*{arg\,min}_{\mathbf{p}_k} \left(\tilde{\mathbf{y}}_k - \mathbcal{G}_k \mathbf{p}_k\right)^T\left(\mathbf{P}_k^{\mathbf{e}}\right)^{-1}\left(\tilde{\mathbf{y}}_k - \mathbcal{G}_k \mathbf{p}_k\right).
\end{equation}

\noindent In Eq.~\eqref{eq:met_wlscost}, the matrix inverse $(\cdot)^{-1}$ is replaced with the Moore-Penrose inverse (pseudoinverse), denoted by $(\cdot)^{\dagger}$, to prevent numerical issues caused by rank deficiency and ill-conditionedness. Solving Eq.~\eqref{eq:met_wlscost} with $\hat{\mathbf{p}}_k = \mathbf{M}_k \tilde{\mathbf{y}}_k$ according to Eq.~\eqref{eq:met_inputest}, and using Eq.~\eqref{eq:met_ptilde} for the covariance, the input gain and the input estimation error covariance follow as:
\begin{equation} \label{eq:met_Mk}
\begin{aligned}
    \mathbf{M}_k &= \left[\mathbcal{G}_k^T\left(\mathbf{P}_k^{\mathbf{e}}\right)^{\dagger}\mathbcal{G}_k\right]^{\dagger}\mathbcal{G}_k^T\left(\mathbf{P}_k^{\mathbf{e}}\right)^{\dagger},\\
    \mathbf{P}_k^{\mathbf{p}}
    &= \left[\mathbcal{G}_k^T\left(\mathbf{P}_k^{\mathbf{e}}\right)^{\dagger}\mathbcal{G}_k\right]^{\dagger}.
\end{aligned}
\end{equation}

\noindent Note that the input gain satisfies the unbiasedness constraint of Eq.~\eqref{eq:met_unbias} whenever $\mathbcal{G}_k^T\left(\mathbf{P}_k^{\mathbf{e}}\right)^{\dagger}\mathbcal{G}_k$ is nonsingular.

The state gain can be derived by minimising the trace of the state estimation error covariance $\mathbf{P}_k^{\mathbf{x}}$, i.e., Eq.~\eqref{eq:met_Px}. To avert numerical issues caused by rank deficiency or negligible singular values of $\boldsymbol{\Psi}_k$, a decomposition of $\mathbf{K}_k$ is assumed:
\begin{equation} \label{eq:met_Kdecomp}
    \mathbf{K}_k = \bar{\mathbcal{K}}_k \mathbf{U}_k^T,
\end{equation}

\noindent in which the columns of $\mathbf{U}_k$ are the singular vectors of $\boldsymbol{\Psi}_k$ whose singular values are non-negligible, so that $\mathbf{U}_k^T \boldsymbol{\Psi}_k \mathbf{U}_k$ is the diagonal matrix of these singular values and is nonsingular. Next, $\bar{\mathbcal{K}}_k$ can be derived by substituting Eq.~\eqref{eq:met_Kdecomp} into Eq.~\eqref{eq:met_Px} and minimising the trace of $\mathbf{P}_k^{\mathbf{x}}$:
\begin{equation} \label{eq:met_Kmin}
\begin{aligned}
    \bar{\mathbcal{K}}_k &= \operatorname*{arg\,min}_{\bar{\mathbcal{K}}_k} \operatorname{trace}\left(\mathbf{P}_k^{\mathbcal{x}} + \bar{\mathbcal{K}}_k \mathbf{U}_k^T \boldsymbol{\Psi}_k \mathbf{U}_k \bar{\mathbcal{K}}_k^T - \boldsymbol{\Upsilon}_k^T \mathbf{U}_k \bar{\mathbcal{K}}_k^T - \bar{\mathbcal{K}}_k \mathbf{U}_k^T \boldsymbol{\Upsilon}_k\right) \\
    &= \boldsymbol{\Upsilon}_k^T \mathbf{U}_k \left(\mathbf{U}_k^T \boldsymbol{\Psi}_k \mathbf{U}_k\right)^{-1}.
\end{aligned}
\end{equation}

\noindent Finally, the state gain is obtained by substituting Eq.~\eqref{eq:met_Kmin} into Eq.~\eqref{eq:met_Kdecomp}:
\begin{equation} \label{eq:met_Kk}
    \mathbf{K}_k = \boldsymbol{\Upsilon}_k^T \mathbf{U}_k\left(\mathbf{U}_k^T \boldsymbol{\Psi}_k \mathbf{U}_k\right)^{-1}\mathbf{U}_k^T.
\end{equation}

The unbiasedness constraint of Eq.~\eqref{eq:met_unbias} imposes $m$ independent rows on $\mathbf{M}_k\mathbcal{G}_k$ and is therefore satisfiable only if $\mathbcal{G}_k \in \mathbb{R}^{q \times m}$ has full column rank, i.e., $\operatorname{rank}\mathbcal{G}_k = m$. This rank condition is the only requirement the formulation places on the sensor network, and it requires $q \geq m$: the number of measurements, of any type among displacement, velocity and acceleration, must be no less than the number of unknown inputs. Therefore, the developed UUF applies alike to systems with direct feedthrough, without feedthrough and with a rank-deficient feedforward matrix, in contrast to the specific requirement on the feedforward matrix in other minimum-variance unbiased filters~\cite{Gillijns2007a, Gillijns2007b}.

\subsection{Auxiliary integrators for acceleration-only estimation} \label{sec:met_aux}

\noindent Kalman-type and minimum-variance unbiased approaches typically require displacement or strain measurements alongside accelerations to achieve stable state estimation, and acceleration-only measurements lead to a drift of the estimates~\cite{Yashar2017, Teymouri2022, Liu2026Exp}. However, the acceleration-only configuration is often preferred in practice, as accelerometers are easier to install than displacement sensors, which require a reference point. Therefore, a treatment of the drift is required. One common treatment is to append zero-valued dummy displacement measurements, with their uncertainty modelled as zero-mean white noise of a prescribed variance, thereby suppressing the drift in the estimated displacements~\cite{Naets2015}. Another is to append a dummy measurement of zero input, with a covariance updated from the data, which regularises the input estimation~\cite{Teymouri2022}. Both treatments supply the missing information through zero-valued dummy measurements.

The treatment adopted herein instead derives pseudo-measurements from the acceleration measurements themselves. A causal state-variable filter, as used in continuous-time system identification~\cite{Garnier2003}, is applied to both sides of the observation equation for selected acceleration outputs. The state-variable filter is of second order, with the transfer function $1/\left(s + \omega_c\right)^2$~\cite{Garnier2003}, in which $s$ is the Laplace variable and $\omega_c$ is the cut-off frequency. For ease of reference, it is hereafter referred to as the auxiliary integrator. Above $\omega_c$, the auxiliary integrator acts as a double integration, and its two poles at $-\omega_c$ make it stable. The formulation is presented below.

Let $n_a$ of the measured accelerations be selected, and let $\ddot{u}_{a,j}(t)$, $j = 1, \ldots, n_a$, denote the $j$-th selected acceleration, i.e., an entry of $\ddot{\mathbf{u}}(t)$ in Eq.~\eqref{eq:met_class}. Driven by $\ddot{u}_{a,j}(t)$, the auxiliary integrator in state-space form defines the auxiliary state vector $\boldsymbol{\eta}_j(t) \in \mathbb{R}^{2}$, which holds the filtered output $\zeta_j(t)$ and its time derivative:
\begin{equation} \label{eq:met_auxstate}
\begin{aligned}
    \boldsymbol{\eta}_j(t) &\triangleq \begin{bmatrix}\zeta_j(t) \\ \dot{\zeta}_j(t)\end{bmatrix}, \\
    \dot{\boldsymbol{\eta}}_j(t) &= \mathbcal{A}_a \boldsymbol{\eta}_j(t) + \begin{bmatrix}0 \\ 1\end{bmatrix}\ddot{u}_{a,j}(t), \\
    \mathbcal{A}_a &\triangleq \begin{bmatrix}0 & 1 \\ -\omega_c^2 & -2\omega_c\end{bmatrix}.
\end{aligned}
\end{equation}
\noindent The $n_a$ auxiliary state vectors are stacked into $\boldsymbol{\eta}(t) \in \mathbb{R}^{2n_a}$ and appended to the extended state vector, as is standard in recursive filtering~\cite{Sarkka2023}, i.e., $\mathbf{x}(t) \triangleq \begin{bmatrix}\mathbf{z}(t)^T & \boldsymbol{\eta}(t)^T & \boldsymbol{\theta}(t)^T\end{bmatrix}^T$ with $n_x = 2n_s + 2n_a + n_\theta$. Discretising Eq.~\eqref{eq:met_auxstate} together with the structural dynamics as in Section~\ref{sec:vbi} gives a process equation of the form of Eq.~\eqref{eq:met_process}. In this equation, the auxiliary states are driven by the structural states, the unknown input and the known load through $\ddot{u}_{a,j}(t)$, but do not act back on the structural states. Since the auxiliary integrator is a known deterministic system, the auxiliary block of $\mathbf{Q}_k$ is set to zero, and the initial auxiliary states are set to zero as well, i.e., $\boldsymbol{\eta}_j(0) = \mathbf{0}$.

The same auxiliary integrator is applied to the measurements. Let $y_{a,j,k}$ denote the entry of $\mathbf{y}_k$ measuring the $j$-th selected acceleration at timestep $k$, and let $v_{a,j,k}$ and $\sigma_{j,k}^2$ be its measurement noise and the corresponding diagonal entry of $\mathbf{R}_k$, respectively. In addition, let $\mathbf{A}_a \in \mathbb{R}^{2 \times 2}$ and $\mathbf{B}_a \in \mathbb{R}^{2 \times 1}$ be the matrices of the zero-order-hold discretisation of Eq.~\eqref{eq:met_auxstate} over the time interval $\Delta t$. The pseudo-measurement $y_{\zeta,j,k}$ and the variance $R_{\zeta,j,k}$ of its noise are generated recursively from the measurements:
\begin{equation} \label{eq:met_auxdata}
\begin{aligned}
    \boldsymbol{\eta}_{j,k}^{\mathrm{d}} &= \mathbf{A}_a \boldsymbol{\eta}_{j,k-1}^{\mathrm{d}} + \mathbf{B}_a y_{a,j,k}, &\qquad y_{\zeta,j,k} &\triangleq \begin{bmatrix}1 & 0\end{bmatrix}\boldsymbol{\eta}_{j,k}^{\mathrm{d}}, &\qquad \boldsymbol{\eta}_{j,0}^{\mathrm{d}} &= \mathbf{0},\\
    \boldsymbol{\Pi}_{j,k} &= \mathbf{A}_a \boldsymbol{\Pi}_{j,k-1} \mathbf{A}_a^T + \mathbf{B}_a \sigma_{j,k}^2 \mathbf{B}_a^T, &\qquad R_{\zeta,j,k} &\triangleq \left[\boldsymbol{\Pi}_{j,k}\right]_{11}, &\qquad \boldsymbol{\Pi}_{j,0} &= \mathbf{0},
\end{aligned}
\end{equation}
\noindent in which the subscript $\zeta$ marks the quantities of the pseudo-measurement, and the superscript $\mathrm{d}$ marks the quantities computed from the measurements. $\boldsymbol{\Pi}_{j,k} \in \mathbb{R}^{2 \times 2}$ is the covariance of the part of $\boldsymbol{\eta}_{j,k}^{\mathrm{d}}$ due to the measurement noise $v_{a,j,k}$, and $[\cdot]_{11}$ denotes the $(1,1)$ entry of a matrix. By substituting $y_{a,j,k} = \ddot{u}_{a,j}(t_k) + v_{a,j,k}$ into Eq.~\eqref{eq:met_auxdata}, the pseudo-measurement can be related to the filtered output:
\begin{equation} \label{eq:met_auxerr}
    y_{\zeta,j,k} = \zeta_{j}(t_k) + \xi_{j,k},
\end{equation}
\noindent in which $\xi_{j,k}$ is the measurement noise $v_{a,j,k}$ passed through the auxiliary integrator, with variance $R_{\zeta,j,k}$. It should be noted that $\xi_{j,k}$ is correlated in time and with the current measurement noise $v_{a,j,k}$ of the same output. In the developed UUF, however, $\xi_{j,k}$ is treated as zero-mean white noise uncorrelated with the remaining measurement noise and the past estimation errors. To allow for the neglected correlations, its variance is multiplied by a constant factor $\beta_R \geq 1$, i.e., taken as $\beta_R R_{\zeta,j,k}$, which reduces the weight of the pseudo-measurements. The rigorous alternative of appending the noise to the state vector~\cite{Simon2006} is not pursued herein.

Finally, the pseudo-measurements are appended to the output vector $\mathbf{y}_k$. Their rows of the observation function $\mathbf{c}(\cdot)$ read $\zeta_j$ from $\boldsymbol{\eta}$, their rows of the feedforward matrix $\mathbf{D}_k$ are zero, and their diagonal entries of $\mathbf{R}_k$ are $\beta_R R_{\zeta,j,k}$, giving an observation equation of the form of Eq.~\eqref{eq:met_obs}. Therefore, the developed UUF applies to the extended problem without modification, and setting $n_a = 0$ recovers the configuration without auxiliary integrators. Since the pseudo-measurements are formed from the measurements, they are not counted in the number of measurements for the rank condition of Section~\ref{sec:met_gains}: the measured outputs alone must be no fewer than the unknown inputs.

\subsection{Summary of the algorithm} \label{sec:met_alg}
\noindent Given the three sigma sets of the unscented transform, the error propagation and the computation of the error covariances and cross-covariances, the developed UUF is summarised in Table~\ref{tab:met_alg} for ease of implementation. The table covers the general sensor configuration and the acceleration-only configuration with the auxiliary integrators of Section~\ref{sec:met_aux}, which share the filtering loop.
\clearpage

\begin{table}[htbp]
\footnotesize
\centering
\renewcommand{\arraystretch}{1.08}
\caption{Summary of the UUF for the general and the acceleration-only sensor configurations}
\label{tab:met_alg}
\begin{tabular}{p{0.97\textwidth}}
\toprule
\textbf{Initialisation}\\
\hspace{2mm}\textbullet\ Define the process and observation equations; assign $\hat{\mathbf{x}}_{0|0}$, $\mathbf{P}_0^{\mathbf{x}}$, $\mathbf{Q}_k$, $\mathbf{R}_k$, $\alpha$, $\beta$ and $\kappa$\\
\hspace{2mm}\textbullet\ $\lambda = \alpha^2\left(n_x + \kappa\right) - n_x$, $\gamma = \sqrt{n_x + \lambda}$;\\
\qquad \qquad $W_0^{(m)} = \lambda/\left(n_x + \lambda\right)$, $W_0^{(c)} = W_0^{(m)} + 1 - \alpha^2 + \beta$, $W_i^{(m)} = W_i^{(c)} = 1/\left[2\left(n_x + \lambda\right)\right]$ for $i \geq 1$\\
\hspace{2mm}\textit{For the general sensor configuration}\\
\hspace{5mm}\textopenbullet\ $\mathbf{x} = \begin{bmatrix}\mathbf{z}^T & \boldsymbol{\theta}^T\end{bmatrix}^T$, $n_x = 2n_s + n_\theta$; $\mathbf{y}_k$: the measured outputs\\
\hspace{2mm}\textit{For the acceleration-only configuration}\\
\hspace{5mm}\textopenbullet\ Select $n_a$ measured accelerations, $\omega_c$ and $\beta_R \geq 1$; $\mathbcal{A}_a = \begin{bmatrix}0 & 1 \\ -\omega_c^2 & -2\omega_c\end{bmatrix}$\\
\hspace{5mm}\textopenbullet\ $\mathbf{x} = \begin{bmatrix}\mathbf{z}^T & \boldsymbol{\eta}^T & \boldsymbol{\theta}^T\end{bmatrix}^T$, $n_x = 2n_s + 2n_a + n_\theta$; $\boldsymbol{\eta}_j(0) = \mathbf{0}$, $\boldsymbol{\eta}_{j,0}^{\mathrm{d}} = \mathbf{0}$, $\boldsymbol{\Pi}_{j,0} = \mathbf{0}$; auxiliary block of $\mathbf{Q}_k$ zero\\
\hspace{5mm}\textopenbullet\ Rows of $\mathbf{c}(\cdot)$ for the pseudo-measurements read $\zeta_j$; their rows of $\mathbf{D}_k$ are zero\\
\midrule
\textbf{Unscented transform} of $\left(\hat{\mathbf{x}}, \mathbf{P}\right)$ through $\mathbf{h}(\cdot)$, applied to each sigma set below\\
\hspace{2mm}\textbullet\ $\mathcal{X}_0 = \hat{\mathbf{x}}$, $\mathcal{X}_i = \hat{\mathbf{x}} + \gamma(\sqrt{\mathbf{P}})_i$, $\mathcal{X}_{i+n_x} = \hat{\mathbf{x}} - \gamma(\sqrt{\mathbf{P}})_i$ for $i = 1, \ldots, n_x$; $\bar{\mathbf{h}} = \sum_i W_i^{(m)}\mathbf{h}\left(\mathcal{X}_i\right)$\\
\hspace{2mm}\textbullet\ $\bar{\mathbcal{H}} = \left[\sum_i W_i^{(c)}\left(\mathcal{X}_i - \hat{\mathbf{x}}\right)\left(\mathbf{h}\left(\mathcal{X}_i\right) - \bar{\mathbf{h}}\right)^T\right]^T\mathbf{P}^{-1}$, \quad $\boldsymbol{\Omega} = \sum_i W_i^{(c)}\left(\mathbf{h}\left(\mathcal{X}_i\right) - \bar{\mathbf{h}}\right)\left(\mathbf{h}\left(\mathcal{X}_i\right) - \bar{\mathbf{h}}\right)^T - \bar{\mathbcal{H}}\mathbf{P}\bar{\mathbcal{H}}^T$\\
\midrule
\textbf{Filtering loop}, $k = 1, 2, \ldots$\\
\textbf{Step 1: Read the measurements}\\
\hspace{2mm}\textit{For the general sensor configuration}\\
\hspace{5mm}\textopenbullet\ Read $\mathbf{y}_k$\\
\hspace{2mm}\textit{For the acceleration-only configuration}\\
\hspace{5mm}\textopenbullet\ Read $\mathbf{y}_k$; $\boldsymbol{\eta}_{j,k}^{\mathrm{d}} = \mathbf{A}_a \boldsymbol{\eta}_{j,k-1}^{\mathrm{d}} + \mathbf{B}_a y_{a,j,k}$, \quad $\boldsymbol{\Pi}_{j,k} = \mathbf{A}_a \boldsymbol{\Pi}_{j,k-1} \mathbf{A}_a^T + \mathbf{B}_a \sigma_{j,k}^2 \mathbf{B}_a^T$\\
\hspace{5mm}\textopenbullet\ Append $y_{\zeta,j,k} = \begin{bmatrix}1 & 0\end{bmatrix}\boldsymbol{\eta}_{j,k}^{\mathrm{d}}$ to $\mathbf{y}_k$ and $\beta_R\left[\boldsymbol{\Pi}_{j,k}\right]_{11}$ to the diagonal of $\mathbf{R}_k$\\
\textbf{Step 2: Input estimation}\\
\hspace{2mm}1. $\mathcal{X}_i^{(1)}$ from $\left(\hat{\mathbf{x}}_{k-1|k-1}, \mathbf{P}_{k-1}^{\mathbf{x}}\right)$ through $\mathbf{f}(\cdot)$: $\hat{\mathbf{x}}_{k|k-1} = \sum_i W_i^{(m)}\mathbf{f}\left(\mathcal{X}_i^{(1)}\right)$, $\bar{\mathbcal{F}}_k = \bar{\mathbcal{H}}$, $\boldsymbol{\Omega}_k^{(1)} = \boldsymbol{\Omega}$\\
\hspace{2mm}2. $\mathbf{P}_k^{-} = \sum_i W_i^{(c)}\left[\mathbf{f}\left(\mathcal{X}_i^{(1)}\right) - \hat{\mathbf{x}}_{k|k-1}\right]\left[\mathbf{f}\left(\mathcal{X}_i^{(1)}\right) - \hat{\mathbf{x}}_{k|k-1}\right]^T + \mathbf{Q}_{k-1}$, \quad $\bar{\mathbf{Q}}_{k-1} = \mathbf{Q}_{k-1} + \boldsymbol{\Omega}_k^{(1)}$\\
\hspace{2mm}3. $\mathcal{X}_i^{(2)}$ from $\left(\hat{\mathbf{x}}_{k|k-1}, \mathbf{P}_k^{-}\right)$ through $\mathbf{c}(\cdot)$: $\tilde{\mathbf{y}}_k = \mathbf{y}_k - \sum_i W_i^{(m)}\mathbf{c}\left(\mathcal{X}_i^{(2)}\right)$, $\bar{\mathbcal{C}}_k^{(2)} = \bar{\mathbcal{H}}$, $\boldsymbol{\Omega}_k^{(2)} = \boldsymbol{\Omega}$\\
\hspace{2mm}4. $\mathbf{P}_k^{\mathbf{e}} = \bar{\mathbcal{C}}_k^{(2)} \mathbf{P}_k^{-}\left(\bar{\mathbcal{C}}_k^{(2)}\right)^T + \boldsymbol{\Omega}_k^{(2)} + \mathbf{R}_k$, \quad $\mathbcal{G}_k = \bar{\mathbcal{C}}_k^{(2)}\mathbf{G}_{k-1} + \mathbf{D}_k$\\
\hspace{2mm}5. $\mathbf{P}_k^{\mathbf{p}} = \left[\mathbcal{G}_k^T\left(\mathbf{P}_k^{\mathbf{e}}\right)^{\dagger}\mathbcal{G}_k\right]^{\dagger}$, \quad $\mathbf{M}_k = \mathbf{P}_k^{\mathbf{p}}\mathbcal{G}_k^T\left(\mathbf{P}_k^{\mathbf{e}}\right)^{\dagger}$, \quad $\hat{\mathbf{p}}_k = \mathbf{M}_k\tilde{\mathbf{y}}_k$\\
\textbf{Step 3: State estimation}\\
\hspace{2mm}1. $\hat{\mathbcal{x}}_{k|k-1} = \hat{\mathbf{x}}_{k|k-1} + \mathbf{G}_{k-1} \hat{\mathbf{p}}_k$\\
\hspace{2mm}2. $\mathbf{P}_k^{\mathbf{px}} = \left(\mathbf{P}_k^{\mathbf{xp}}\right)^T = -\mathbf{M}_k \bar{\mathbcal{C}}_k^{(2)} \bar{\mathbcal{F}}_k \mathbf{P}_{k-1}^{\mathbf{x}}$, \quad $\mathbf{P}_k^{\mathbf{pw}} = \left(\mathbf{P}_k^{\mathbf{wp}}\right)^T = -\mathbf{M}_k \bar{\mathbcal{C}}_k^{(2)} \bar{\mathbf{Q}}_{k-1}$, \quad $\mathbf{P}_k^{\mathbf{pv}} = \left(\mathbf{P}_k^{\mathbf{vp}}\right)^T = -\mathbf{M}_k \mathbf{R}_k$\\
\hspace{2mm}3. $\mathbf{P}_k^{\mathbcal{x}} = \mathbf{P}_k^{-} + \mathbf{G}_{k-1} \mathbf{P}_k^{\mathbf{p}} \mathbf{G}_{k-1}^T + \bar{\mathbcal{F}}_k \mathbf{P}_k^{\mathbf{xp}} \mathbf{G}_{k-1}^T + \mathbf{G}_{k-1} \mathbf{P}_k^{\mathbf{px}} \bar{\mathbcal{F}}_k^T + \mathbf{G}_{k-1} \mathbf{P}_k^{\mathbf{pw}} + \mathbf{P}_k^{\mathbf{wp}} \mathbf{G}_{k-1}^T$\\
\hspace{2mm}4. $\mathbf{P}_k^{\mathbf{p}\mathbcal{x}} = \mathbf{P}_k^{\mathbf{px}} \bar{\mathbcal{F}}_k^T + \mathbf{P}_k^{\mathbf{p}} \mathbf{G}_{k-1}^T + \mathbf{P}_k^{\mathbf{pw}}$, \quad $\mathbf{P}_k^{\mathbf{v}\mathbcal{x}} = \mathbf{P}_k^{\mathbf{vp}} \mathbf{G}_{k-1}^T$\\
\hspace{2mm}5. $\mathcal{X}_i^{(3)}$ from $\left(\hat{\mathbcal{x}}_{k|k-1}, \mathbf{P}_k^{\mathbcal{x}}\right)$ through $\mathbf{c}(\cdot)$: $\bar{\mathbcal{C}}_k^{(3)} = \bar{\mathbcal{H}}$, $\boldsymbol{\Omega}_k^{(3)} = \boldsymbol{\Omega}$\\
\hspace{2mm}6. $\boldsymbol{\Upsilon}_k = \bar{\mathbcal{C}}_k^{(3)} \mathbf{P}_k^{\mathbcal{x}} + \mathbf{D}_k \mathbf{P}_k^{\mathbf{p}\mathbcal{x}} + \mathbf{P}_k^{\mathbf{v}\mathbcal{x}}$\\
\hspace{2mm}7. $\boldsymbol{\Psi}_k = \bar{\mathbcal{C}}_k^{(3)} \boldsymbol{\Upsilon}_k^T + \mathbf{D}_k\left[\mathbf{P}_k^{\mathbf{p}\mathbcal{x}}\left(\bar{\mathbcal{C}}_k^{(3)}\right)^T + \mathbf{P}_k^{\mathbf{p}} \mathbf{D}_k^T + \mathbf{P}_k^{\mathbf{pv}}\right] + \mathbf{P}_k^{\mathbf{v}\mathbcal{x}}\left(\bar{\mathbcal{C}}_k^{(3)}\right)^T + \mathbf{P}_k^{\mathbf{vp}} \mathbf{D}_k^T + \mathbf{R}_k + \boldsymbol{\Omega}_k^{(3)}$\\
\hspace{2mm}8. $\mathbf{K}_k = \boldsymbol{\Upsilon}_k^T \mathbf{U}_k\left(\mathbf{U}_k^T \boldsymbol{\Psi}_k \mathbf{U}_k\right)^{-1}\mathbf{U}_k^T$\\
\hspace{2mm}9. $\hat{\mathbf{x}}_{k|k} = \hat{\mathbcal{x}}_{k|k-1} + \mathbf{K}_k\left[\mathbf{y}_k - \sum_i W_i^{(m)}\mathbf{c}\left(\mathcal{X}_i^{(3)}\right) - \mathbf{D}_k\hat{\mathbf{p}}_k\right]$\\
\textbf{Step 4: Error propagation for the next timestep}\\
\hspace{2mm}\textbullet\ $\mathbf{P}_k^{\mathbf{x}} = \mathbf{P}_k^{\mathbcal{x}} + \mathbf{K}_k \boldsymbol{\Psi}_k \mathbf{K}_k^T - \boldsymbol{\Upsilon}_k^T \mathbf{K}_k^T - \mathbf{K}_k \boldsymbol{\Upsilon}_k$; \quad output $\hat{\mathbf{p}}_k$, $\hat{\mathbf{x}}_{k|k}$ and $\mathbf{P}_k^{\mathbf{x}}$, and proceed to timestep $k+1$\\
\bottomrule
\end{tabular}
\end{table}

\clearpage

\section{Application to drive-by bridge inspection} \label{sec:application}

\noindent The developed UUF is applied herein to indirect drive-by structural health monitoring on the coupled VBI system of Section~\ref{sec:vbi}. The target is to identify the parameters of interest of the bridge and to track their changes, such as a reduction in the stiffness of a structural member, while the states and the contact forces are estimated jointly. The road profile is not measured, and the measurements are taken on the vehicle and, at most, at a few locations on the bridge. Given the difficulty of measuring displacement-type quantities on a moving vehicle~\cite{Li2024}, only the two axle accelerations are measured on the vehicle in this study, except in Section~\ref{sec:cs2_rank}, where a single sensor unit at the centre of gravity is considered.

Adopting the road profile as the unknown input, with a dimension of four, as introduced in Section~\ref{sec:vbi_ss}, is not pursued for two reasons. First, as noted after Eq.~\eqref{eq:vbi_fv}, the road roughness profile excites the system only through the term $k_{v_i} r_i + c_{v_i} v r'_i$ at each axle, so that the response depends on the road profile only through the suspension stiffness and damping constants. Hence, an accurate identification of the road profile depends on accurately known suspension constants, which are more difficult to determine in practice than the mass properties of the vehicle~\cite{Hasegawa2023}.

Second, the same term shows that the four components of $\mathbf{p}(t)=\left[\begin{matrix} \mathbf{r}(t)^T & v \mathbf{r}'(t)^T \end{matrix}\right]^T$ act through two quantities, i.e., the two tyre-suspension contact forces driven by the road roughness. Therefore, the distribution matrices $\mathbf{\Phi}_k(t)$ and $\mathbf{\Phi}_c(t)$ of Eq.~\eqref{eq:vbi_coupled_matrices} together have rank at most two at every time instant. Consequently, the matrix $\mathbcal{G}_k$ of Section~\ref{sec:met_gains} inherits this bound, so that $\operatorname{rank}\mathbcal{G}_k \leq 2 < m = 4$ for every sensor configuration, on the vehicle or on the bridge, and the rank condition cannot be met. For instance, the two axle accelerations together with two bridge measurements of any type give $q = m = 4$ and meet the count that the rank condition of Section~\ref{sec:met_gains} requires, yet $\operatorname{rank}\mathbcal{G}_k \leq 2 < m$. This is a constraint imposed by the VBI system formulation rather than a limitation of the developed UUF. A workaround can be adopted to relax this constraint, such as replacing $\mathbf{r}'_k$ by a numerical differentiation with $\mathbf{r}_k$ and $\mathbf{r}_{k-1}$~\cite{LiuRoughness2026, KimConference2025}. However, the error of this numerical differentiation has to be propagated properly during the estimation~\cite{LiuRoughness2026}. Nonetheless, the ultimate goal of drive-by structural health monitoring in this study is to assess the bridge condition. The unknown input is therefore re-parameterised as the two contact forces caused by road roughness, which the UUF estimates as quantities in their own right. The road profile itself is not a quantity of interest in this application, but it can be reconstructed from the estimated contact forces through the same term if the vehicle suspension parameters are known. The re-parameterised system and its state-space representation for the UUF are presented below, followed by the error metric for evaluating the estimation performance and a practical tuning strategy.

\subsection{Re-parameterisation of the unknown input} \label{sec:app_input}

\noindent Following Eqs.~\eqref{eq:vbi_coupled_matrices} and~\eqref{eq:vbi_ct_matrices}, the road roughness profile acts on the VBI system through the distribution matrix $\left[\begin{matrix} \mathbf{\Phi}_k(t) & \mathbf{\Phi}_c(t) \end{matrix}\right] \in \mathbb{R}^{n_s \times 4}$, so that $m = 4$. The contact force it produces under each axle can be written as: 
\begin{equation} \label{eq:app_collapse}
    \begin{bmatrix} \mathbf{\Phi}_k(t) & \mathbf{\Phi}_c(t) \end{bmatrix}
    \begin{bmatrix} \mathbf{r}(t) \\ v \mathbf{r}'(t) \end{bmatrix}
    = \begin{bmatrix} \boldsymbol{\phi}_1(t) & \boldsymbol{\phi}_2(t) \end{bmatrix}\mathbf{T}
    \begin{bmatrix} \mathbf{r}(t) \\ v \mathbf{r}'(t) \end{bmatrix}
    = \boldsymbol{\phi}_1(t) f_1(t) + \boldsymbol{\phi}_2(t) f_2(t),
\end{equation}

\noindent where
\begin{equation}
\begin{aligned}
    \mathbf{T} &= \begin{bmatrix} k_{v_1} & 0 & c_{v_1} & 0 \\ 0 & k_{v_2} & 0 & c_{v_2} \end{bmatrix},\\
    f_i(t) &\triangleq k_{v_i} r_i(t) + c_{v_i} v r'_i(t).
\end{aligned}
\end{equation}

\noindent In the above equations, $\mathbf{T} \in \mathbb{R}^{2\times4}$ is a transform matrix formed from the suspension parameters of the vehicle, and $f_i(t)$ is the contact force at axle $i$ caused by the road roughness profile. The force $f_i(t)$ is taken as positive when transmitted through the tyre-suspension system to the vehicle, with its reaction $-f_i(t)$ carried by the bridge deck, consistent with the sign convention of Eq.~\eqref{eq:vbi_phi}. The unknown input is therefore re-parameterised as the two contact forces. The input vector $\mathbf{p}(t)$ and its distribution matrix $\mathbf{S}(t)$ can be defined as:
\begin{equation} \label{eq:app_input}
\begin{aligned}
    \mathbf{p}(t) &\triangleq \begin{bmatrix} f_1(t) \\ f_2(t) \end{bmatrix} = \mathbf{T}\begin{bmatrix} \mathbf{r}(t) \\ v \mathbf{r}'(t) \end{bmatrix},\\
    \mathbf{S}(t) &\triangleq \begin{bmatrix} \boldsymbol{\phi}_1(t) & \boldsymbol{\phi}_2(t) \end{bmatrix}.
\end{aligned}
\end{equation}

\noindent Hereafter, the input of the VBI system is that of Eq.~\eqref{eq:app_input}, with $m = 2$.

\subsection{Evaluation of the estimation performance} \label{sec:app_metric}

\noindent The estimation performance is evaluated on the three quantities of the joint estimation problem: the state vector, the two contact-force components $f_1$ and $f_2$ of Eq.~\eqref{eq:app_collapse}, and the parameter history $\boldsymbol{\theta}_k$. The state vector carries the displacements and velocities at every degree of freedom of the coupled system and hence the response at the unmeasured locations. In the numerical studies, each estimate is compared with the true time history of the simulation. The comparison is restricted to the timesteps at which both axles are on the bridge deck, from the entry of the rear axle to the exit of the front axle. This window is set by the geometry of the crossing and introduces no free constant.

Every error is made dimensionless by the root-mean-square value of the true history over the window. For a scalar quantity $a$ with true history $a_k$ and estimate $\hat{a}_k$, the normalised root-mean-square error (NRMSE) is defined as:
\begin{equation} \label{eq:app_nrmse}
    \operatorname{NRMSE}(a) \triangleq \sqrt{\frac{\sum_{k}\left(a_k - \hat{a}_k\right)^2}{\sum_{k} a_k^2}},
\end{equation}
\noindent in which the sums run over the timesteps of the window, and $a$ stands for a displacement $u_i$, a velocity $\dot{u}_i$, a contact-force component $f_i$ or a parameter $\theta_j$ of the estimation problem. The errors of the three estimated quantities and the total estimation error are then composed as follows:
\begin{equation} \label{eq:app_metric}
\begin{aligned}
    E_{\mathrm{state}} &\triangleq \frac{1}{n_s}\sum_{i=1}^{n_s} \operatorname{NRMSE}\left(u_i\right) + \frac{1}{n_s}\sum_{i=1}^{n_s} \operatorname{NRMSE}\left(\dot{u}_i\right), \\
    E_{\mathrm{input}} &\triangleq \operatorname{NRMSE}\left(f_1\right) + \operatorname{NRMSE}\left(f_2\right), \\
    E_{\mathrm{param}} &\triangleq \frac{1}{n_\theta}\sum_{j=1}^{n_\theta} \operatorname{NRMSE}\left(\theta_j\right),\\
    E_{\mathrm{total}} &\triangleq E_{\mathrm{state}} + E_{\mathrm{input}} + E_{\mathrm{param}},
\end{aligned}
\end{equation}
\noindent in which the averages over the degrees of freedom and over the parameters give every entry equal weight regardless of its amplitude. The same metric is applied to every method compared in the numerical studies.

Every method compared in the numerical studies requires tuning, of the noise covariances in particular. The true histories are not available in practice, so the tuning cannot be selected by the errors of Eq.~\eqref{eq:app_metric} and has to be selected from the measurements and the model alone. A strategy for this practical tuning is presented below, and the errors of Eq.~\eqref{eq:app_metric} serve to report the performance obtained at the selected values.

\subsection{Tuning strategy in practice} \label{sec:app_tuning}

\noindent The process noise covariance $\mathbf{Q}_k$ is the quantity to be tuned. For every method compared, $\mathbf{Q}_k$ is reduced to one scalar per block of the extended state vector:
\begin{equation} \label{eq:app_Q}
    \mathbf{Q}_k = \begin{bmatrix} Q_s \mathbf{I}_{2n_s} & \mathbf{0} \\ \mathbf{0} & Q_\theta \mathbf{I}_{n_\theta} \end{bmatrix},
\end{equation}
\noindent in which the scalars $Q_s$ and $Q_\theta$ are the quantities to be tuned. It should be noted that methods which append the unknown input to the state vector as a random walk, such as the augmented Kalman filter family, carry a third block for the input in $\mathbf{Q}_k$, and Eq.~\eqref{eq:app_Q} is extended accordingly.

The result of a grid search improves with the grid resolution, whereas the computational cost increases with the number of grid points. A coarse-to-fine search is therefore adopted herein. The objective, introduced below, is first evaluated on a coarse grid over the declared range of each scalar $Q_s$ and $Q_\theta$. The fine grid is then centred on the point with the smallest mean objective among its neighbours rather than on the minimum, to avoid isolated low values of the objective. The values of the scalars are those of the smallest objective on the fine grid. A centre on the boundary of the coarse grid indicates that the range is too narrow, and the range is extended.

As mentioned previously, the true histories of the structural responses, the input and the parameters are not available in practice. Therefore, the tuning is performed by minimising an objective function that requires no true history. The error estimate proposed in~\cite{Vettori2023} is adopted as the objective, in which the inaccuracy of the estimates is quantified separately for the measured and the unmeasured quantities. Let $\mathbf{y}_l$ and $\hat{\mathbf{y}}_l$ collect the $l$-th measured output and its reconstruction from the a posteriori estimates, the pseudo-measurements of Section~\ref{sec:met_aux} being excluded. The error of the measured quantities is defined as:
\begin{equation} \label{eq:app_Eo}
    E_o \triangleq \frac{1}{q}\sqrt{\sum_{l=1}^{q} \left[\frac{\mathbf{y}_l^T\left(\mathbf{y}_l - \hat{\mathbf{y}}_l\right)}{\mathbf{y}_l^T\mathbf{y}_l}\right]^2},
\end{equation}
\noindent where $q$ is the number of measured outputs.

For the unmeasured quantities, a reference is constructed by modal expansion of the measurements of one response type to the unmeasured degrees of freedom~\cite{Vettori2023}. Let $\bar{\mathbf{y}}_l$ and $\hat{\mathbf{y}}^e_l$ collect the reference and the estimate at the $l$-th unmeasured degree of freedom. The error of the unmeasured quantities takes the same form:
\begin{equation} \label{eq:app_Ep}
    E_p \triangleq \frac{1}{n_e}\sqrt{\sum_{l=1}^{n_e} \left[\frac{\bar{\mathbf{y}}_l^T\left(\bar{\mathbf{y}}_l - \hat{\mathbf{y}}^e_l\right)}{\bar{\mathbf{y}}_l^T\bar{\mathbf{y}}_l}\right]^2},
\end{equation}
\noindent where $n_e$ is the number of unmeasured degrees of freedom. The practical objective combines the two terms:
\begin{equation} \label{eq:app_Jprac}
    E_{\mathrm{prac}} \triangleq E_o + E_p.
\end{equation}
\noindent The term of the error estimate formed from the input error covariance is not included, since not every method compared returns one.

It should be noted that Eq.~\eqref{eq:app_Q} is a general form, and its adaptation to the system at hand is recommended. Depending on the dynamic system, the entries of a block of $\mathbf{Q}_k$ can vary widely in magnitude, in which case the identity matrix in Eq.~\eqref{eq:app_Q} can be replaced by a diagonal matrix carrying a suitable scale for each entry. Furthermore, $\mathbf{Q}_k$ can be divided into finer blocks where necessary. Finally, the measurement noise covariance $\mathbf{R}_k$ is not tuned, since sensor noise estimates are often available from manufacturers or can be obtained using standard techniques. 

\section{Case Study I: simply supported beam} 
\label{sec:cs1}

\noindent The performance of the developed UUF is demonstrated through two numerical case studies. The first is a simply supported beam with an idealised sensor configuration, which is used for a systematic assessment across damage severities, damage locations, and levels of road roughness and measurement noise. The second is the Old Ada Bridge, a steel truss bridge in Japan, whose model, inspection vehicle and sensor positions follow the field test of~\cite{KimAdaData2021}, so that the UUF is assessed under conditions close to practice. In both case studies, the UUF is compared with the classical AEKF~\cite{NaetsCroesDesmet2015} and EGDF-DF~\cite{Wan2018}, and with the state-of-the-art DKF-UKF~\cite{Dertimanis2019} and IPS-UKF~\cite{ImpraimakisSmyth2022}. The set-up of the first case study is presented below.

\subsection{Set-up of the case study} \label{sec:cs1_setup}

\begin{figure}[ht]
    \centering
    \begin{subfigure}[c]{0.54\textwidth}
        \caption{} \label{fig:cs1_layout_a}
        \includegraphics[width=\textwidth]{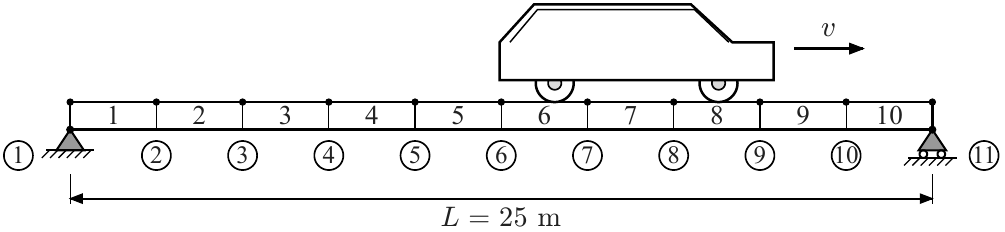}
    \end{subfigure}
    \hfill
    \begin{subfigure}[c]{0.44\textwidth}
        \caption{} \label{fig:cs1_layout_b}
        \includegraphics[width=\textwidth]{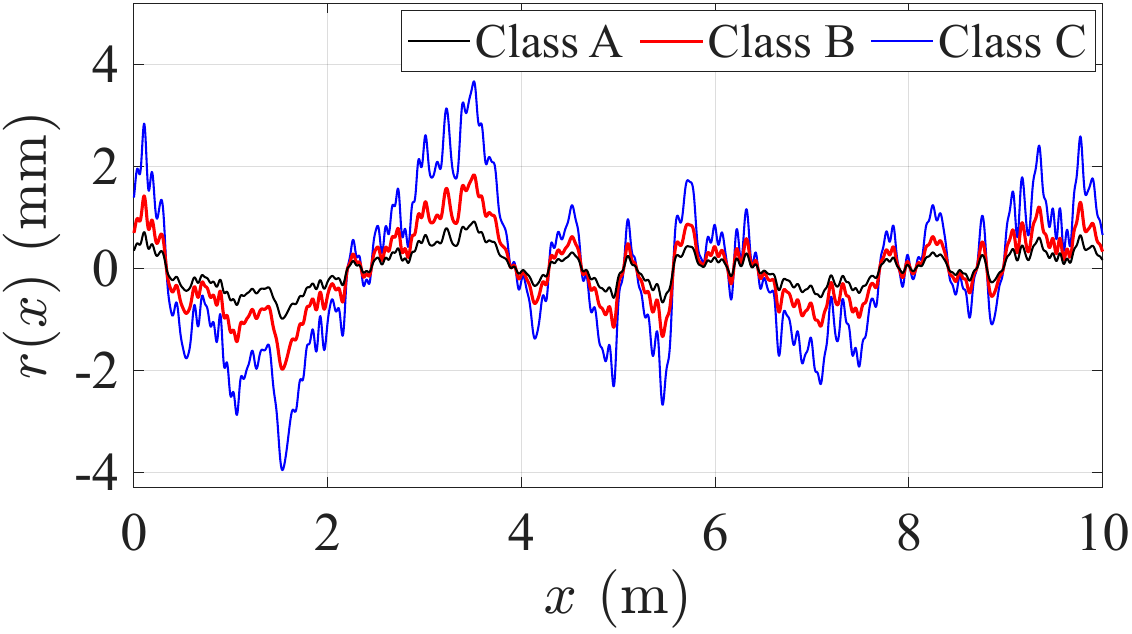}
    \end{subfigure}
    \caption{Set-up of the case study: (a) the vehicle and the beam, and (b) the road roughness profiles.}
    \label{fig:cs1_layout}
\end{figure}

\noindent Fig.~\ref{fig:cs1_layout_a} shows the drive-by structural health monitoring set-up adopted from~\cite{CalderonHurtado2026}, in which an instrumented vehicle crosses a simply supported beam, with sensors on the vehicle and on the beam. The beam has a span of $L = 25$~m and is modelled by ten Euler-Bernoulli beam elements of equal length, with the nodes numbered from $1$ at the entry support to $11$ at the exit support. The flexural stiffness is $EI = 4.865 \times 10^{10}$~$\mathrm{N \cdot m^2}$ and the mass per unit length is $18~358$~kg/m, giving the first three natural frequencies of $4.09$, $16.37$ and $36.84$~Hz. Rayleigh damping is assumed with a damping ratio of $3\%$ for the first two modes. The road roughness profiles are generated according to ISO~8608~\cite{ISO8608}. Classes A, B and C are considered in this case study, as shown in Fig.~\ref{fig:cs1_layout_b}, and the class B profile is used unless stated otherwise. The half-car model of Section~\ref{sec:vbi_gov}, with the properties listed in Table~\ref{tab:cs1_vehicle}, crosses the beam at $v = 30$~km/h.

\begin{table}[ht]
    \centering
    \caption{Properties of the half-car model.}
    \label{tab:cs1_vehicle}
    \begin{tabular}{c c}
    \toprule
    Properties & Values \\
    \midrule
    Mass $m_v$ & $16~200$~kg \\
    Moment of inertia $I_v$ & $93~234$~$\mathrm{kg \cdot m^2}$ \\
    Distances to the axles $d_1$, $d_2$ & $2.375$~m \\
    Suspension stiffness $k_{v_1}$, $k_{v_2}$ & $4 \times 10^{5}$~N/m \\
    Suspension damping $c_{v_1}$, $c_{v_2}$ & $1 \times 10^{4}$~$\mathrm{N \cdot s/m}$ \\
    \bottomrule
    \end{tabular}
\end{table}

The measurements are simulated by adding artificial noise to the responses generated by the Newmark-$\beta$ scheme~\cite{Newmark1959} with a time interval of $\Delta t = 0.001$~s. The added noise is zero-mean white noise with a standard deviation of $1\%$ of the root-mean-square (RMS) value of the corresponding response.

For every method compared, the discrete-time state-space representation is obtained by the matrix exponential method~\cite{Brogan1991} with the same $\Delta t$ as the measurements. It should be noted that different time integration schemes are therefore used for generating the measurements and for the estimation. This discrepancy is intentional and is to be accounted for by the process noise. The initial state estimate $\hat{\mathbf{x}}_{0|0}$ is zero for the responses and the nominal stiffness for the parameter. The initial error covariance $\mathbf{P}^{\mathbf{x}}_{0}$ is diagonal, with a relative standard deviation of $10^{-4}$ for each entry. The measurement noise covariance $\mathbf{R}_k$ is set to the actual covariance of the added noise. For the methods based on the unscented transform, i.e., the developed UUF, the DKF-UKF and the IPS-UKF, the constants are set to $\alpha = 1$, $\beta = 2$ and $\kappa = 0$, in line with~\cite{ImpraimakisSmyth2022, Julier2002}. The matrix of the fictitious input model of the DKF-UKF is set to $0.95\mathbf{I}_2$, within the range recommended in~\cite{Dertimanis2019}.

The process noise covariance $\mathbf{Q}_k$ of every method is tuned by the strategy of Section~\ref{sec:app_tuning}. The scalars $Q_s$ and $Q_\theta$ are searched from $10^{-18}$ to $10^{-1}$. For the AEKF and the DKF-UKF, the scalar of the additional input block is searched from $10^{-14}$ to $10^{0}$ and from $10^{-16}$ to $10^{0}$, respectively.

Three studies are presented hereafter, which examine the damage severity, the damage location, and the road class together with the noise level.

\subsection{Performance under different damage severities} \label{sec:cs1_severity}

\noindent The first study examines the estimation performance against the damage severity. The monitored parameter is the flexural stiffness of element 4, which is reduced by $5\%$, $10\%$, $20\%$, $30\%$ and $50\%$. The damage is present before the vehicle enters the beam, whereas the remaining elements keep their nominal stiffness. The measurements are the two axle accelerations, the vertical acceleration at node 4 and the vertical displacement at node 5, i.e., the bridge sensors are placed at the two nodes of the monitored element.
\begin{figure}[ht]
    \centering
    \includegraphics[trim={0cm 0cm 0cm 0cm}, clip, width=0.45\textwidth]{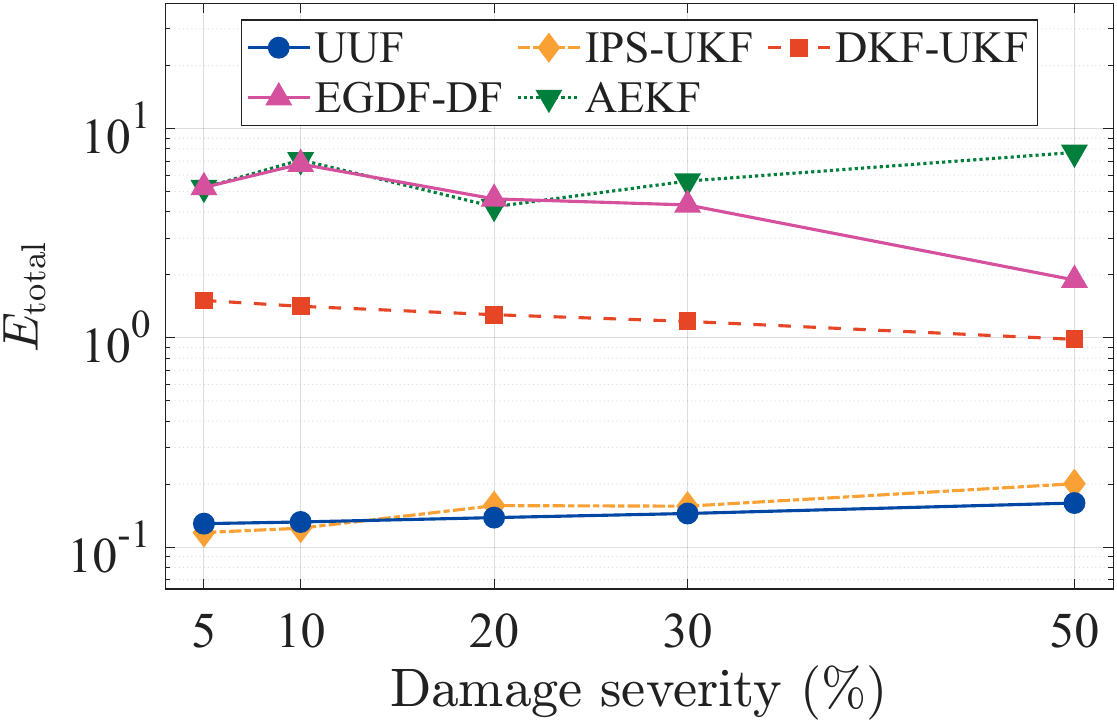}
    \caption{Total estimation error against the damage severity.}
    \label{fig:cs1_sev_errors}
\end{figure}

Fig.~\ref{fig:cs1_sev_errors} shows the total error $E_{\mathrm{total}}$ of each method against the damage severity. The classical AEKF and EGDF-DF, which both linearise by the Jacobian matrix, give total errors between $1.89$ and $7.70$. The DKF-UKF performs better, with total errors between $0.98$ and $1.51$. In contrast, the developed UUF shows a total error between $0.13$ and $0.16$ over the five severities, which is clearly lower than those of the three methods above. The IPS-UKF is the only method of comparable accuracy, with total errors between $0.12$ and $0.20$. It is marginally lower than the UUF at stiffness reductions of $5\%$ and $10\%$. However, its total error grows with the damage severity, whereas that of the UUF remains nearly unchanged. Therefore, the UUF gives the lowest total error at stiffness reductions of $20\%$, $30\%$ and $50\%$.

\begin{figure}[ht]
    \centering
    \begin{subfigure}{0.4\textwidth}
        \caption{} \label{fig:cs1_sev_param_a}
        \includegraphics[width=\textwidth]{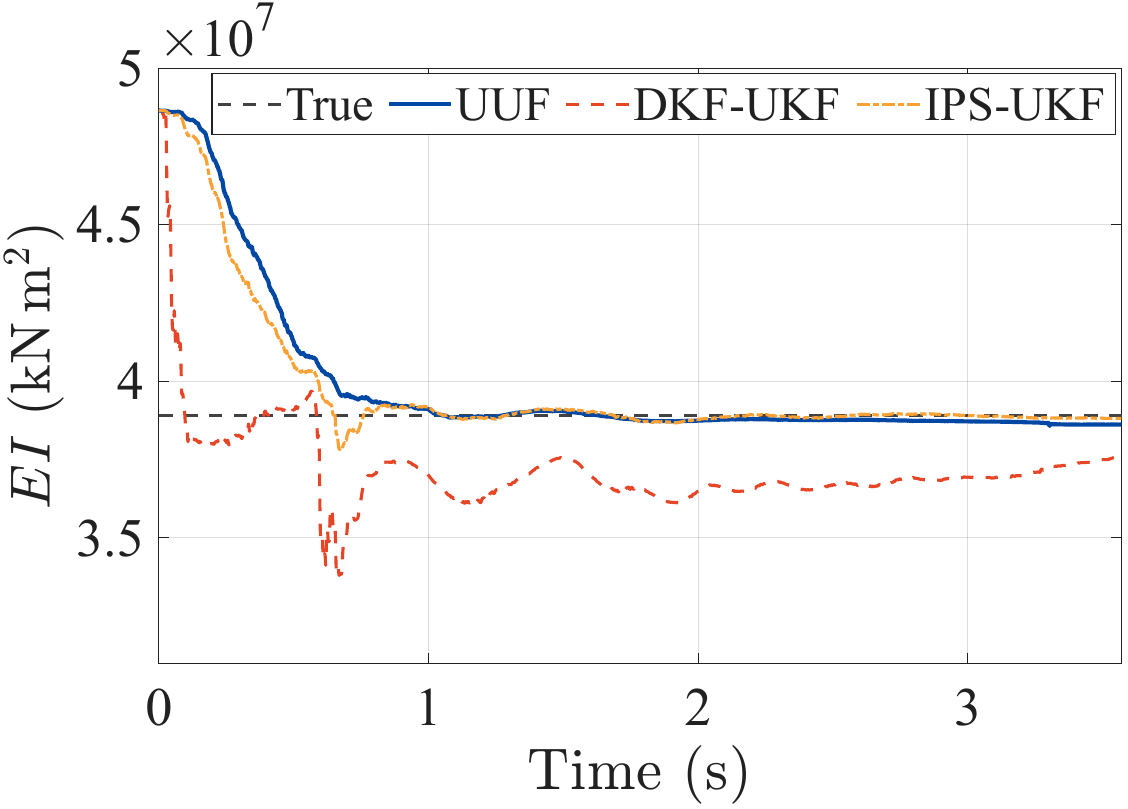}
    \end{subfigure}
    \hspace{1em}
    \begin{subfigure}{0.4\textwidth}
        \caption{} \label{fig:cs1_sev_param_b}
        \includegraphics[width=\textwidth]{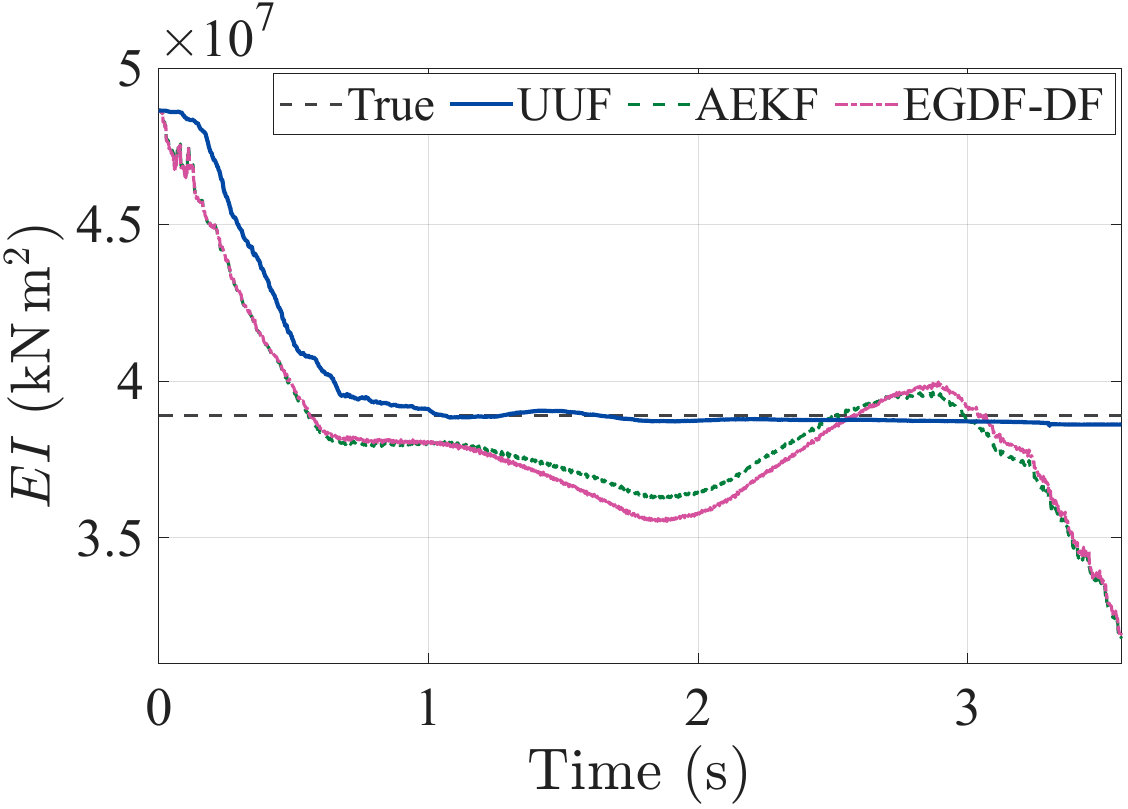}
    \end{subfigure}
    \caption{Flexural stiffness estimates of element 4 by the UUF against (a) the state-of-the-art and (b) the classical methods.}
    \label{fig:cs1_sev_param}
\end{figure}

\begin{figure}[ht]
    \centering
    \begin{subfigure}{0.49\textwidth}
        \caption{} \label{fig:cs1_sev_state_a}
        \includegraphics[width=\textwidth]{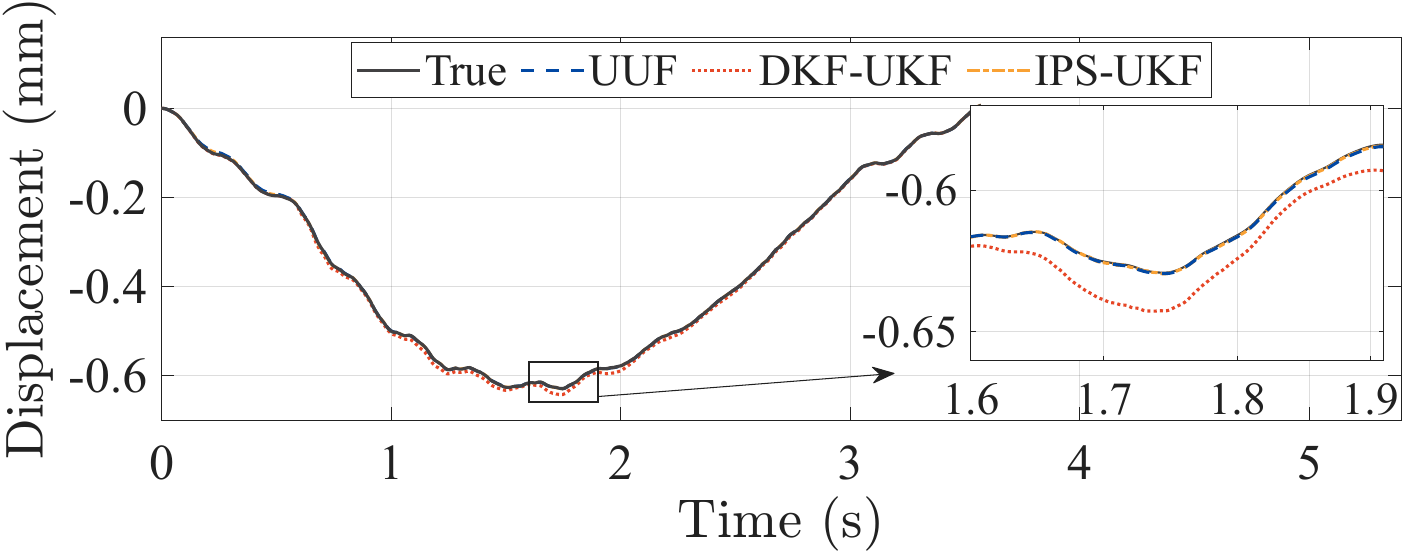}
    \end{subfigure}
    \begin{subfigure}{0.49\textwidth}
        \caption{} \label{fig:cs1_sev_state_b}
        \includegraphics[width=\textwidth]{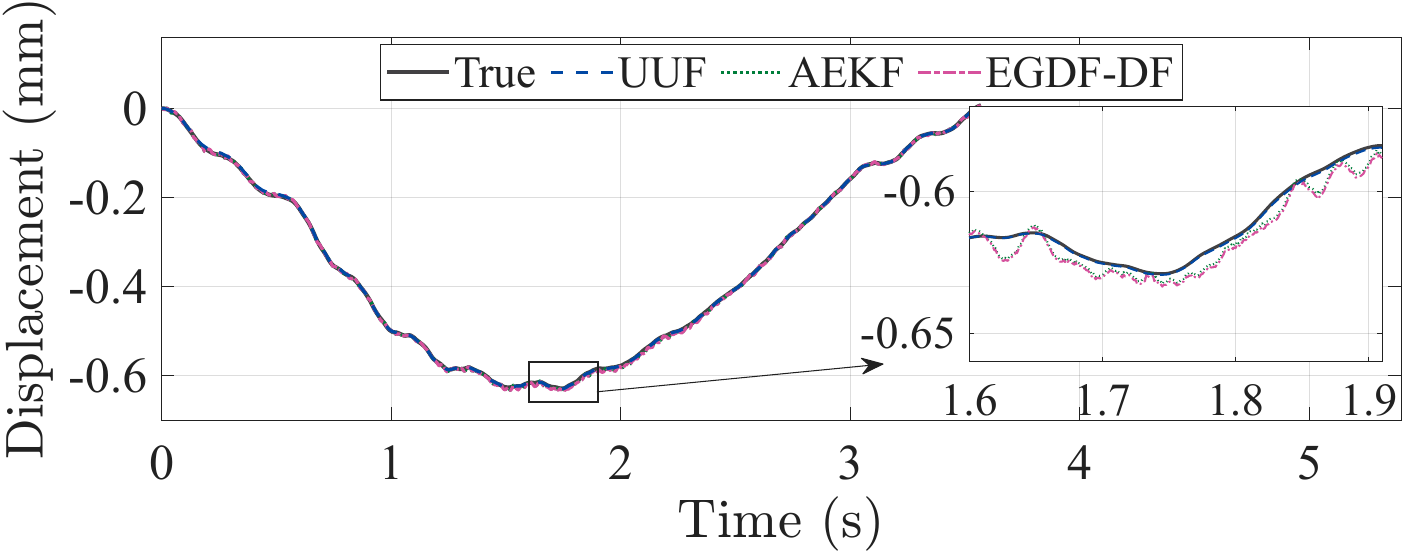}
    \end{subfigure}
    \begin{subfigure}{0.5\textwidth}
        \caption{} \label{fig:cs1_sev_state_c}
        \includegraphics[width=\textwidth]{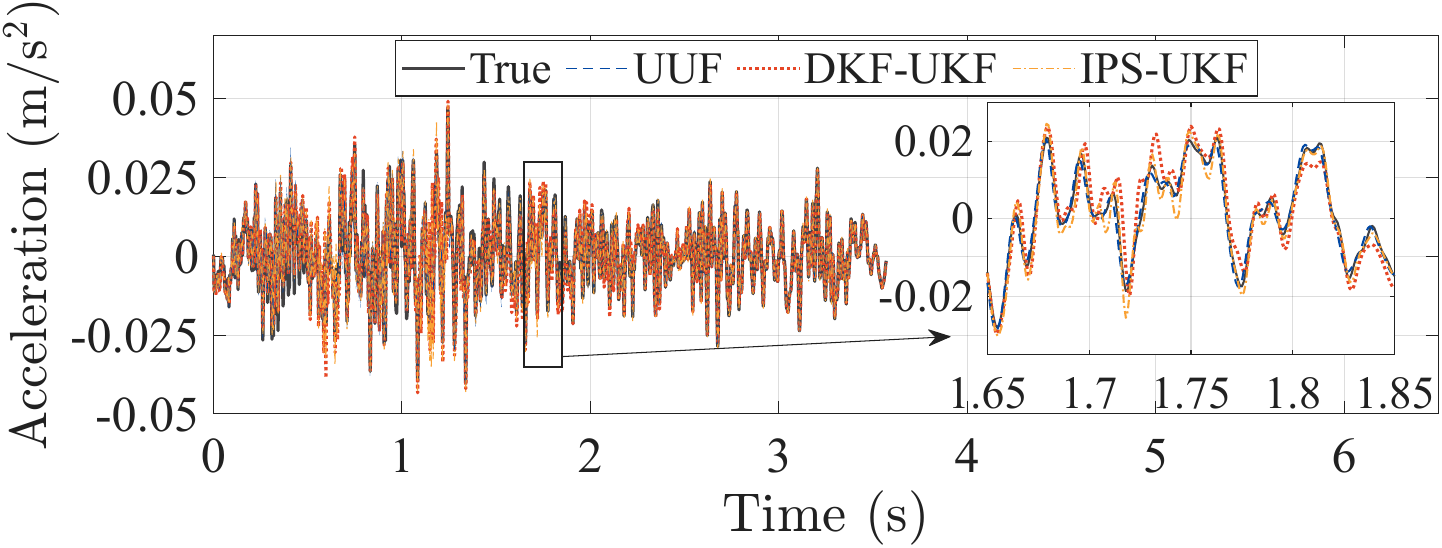}
    \end{subfigure}
    \begin{subfigure}{0.48\textwidth}
        \caption{} \label{fig:cs1_sev_state_d}
        \includegraphics[width=\textwidth]{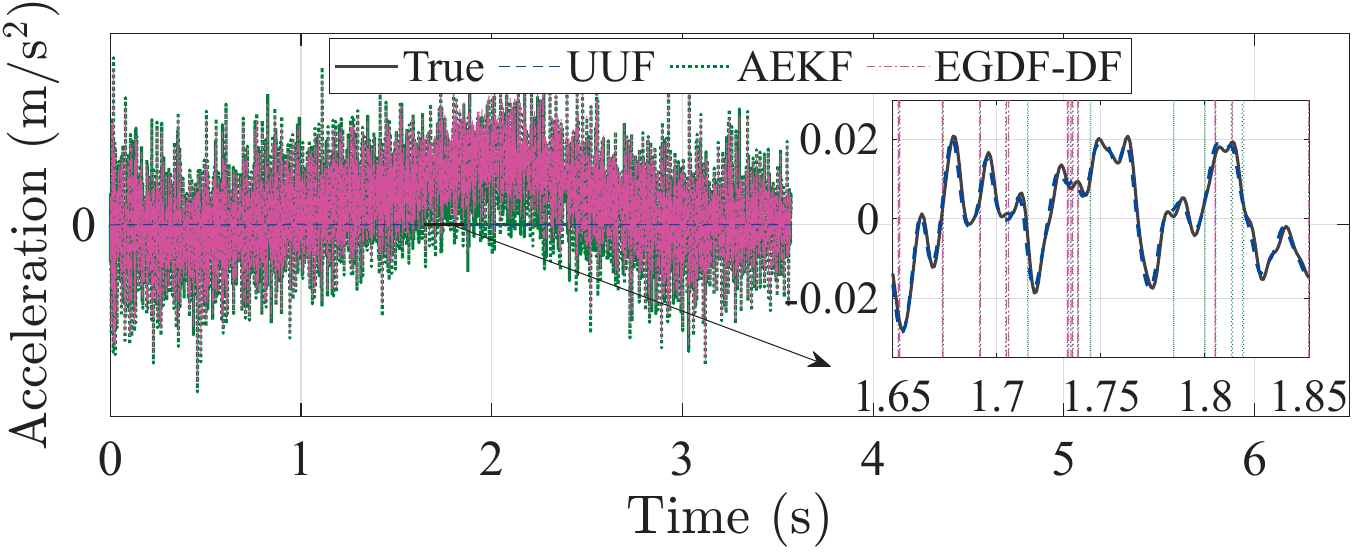}
    \end{subfigure}
    \caption{Estimates at node 3: (a) displacement against the state-of-the-art methods, (b) displacement against the classical methods, (c) acceleration against the state-of-the-art methods, and (d) acceleration against the classical methods.}
    \label{fig:cs1_sev_state}
\end{figure}

\begin{figure}[ht]
    \centering
    \begin{subfigure}{0.4\textwidth}
        \caption{} \label{fig:cs1_sev_input_a}
        \includegraphics[width=\textwidth]{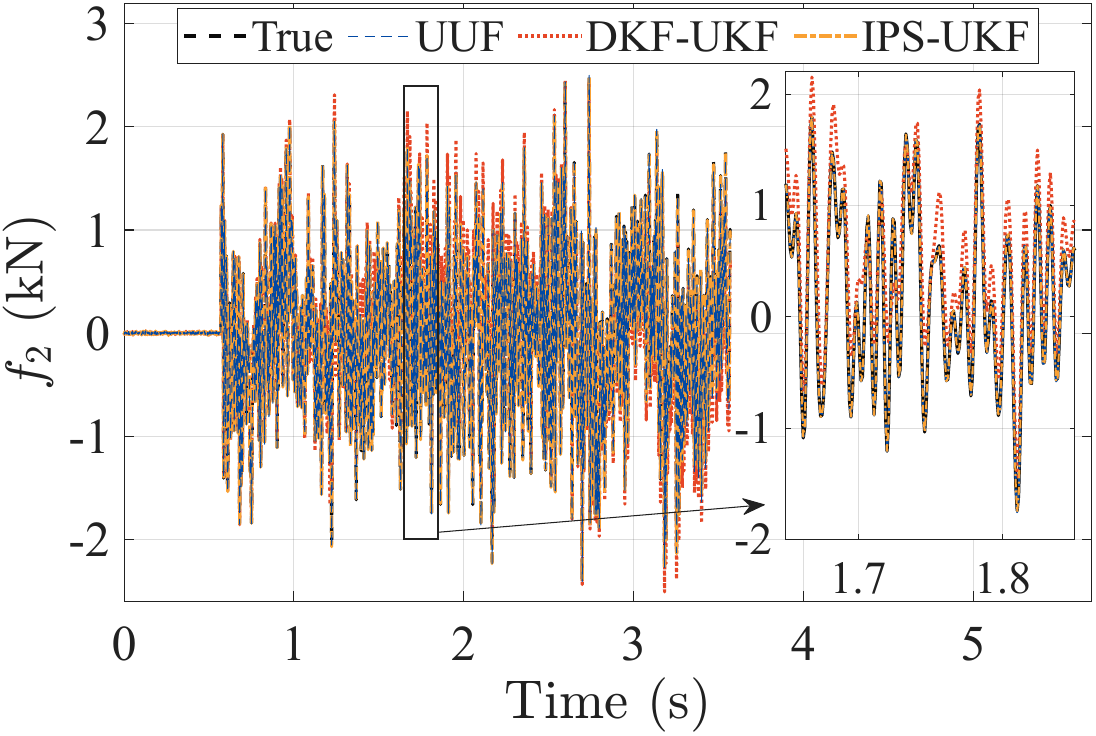}
    \end{subfigure}
    \hspace{1em}
    \begin{subfigure}{0.4\textwidth}
        \caption{} \label{fig:cs1_sev_input_b}
        \includegraphics[width=\textwidth]{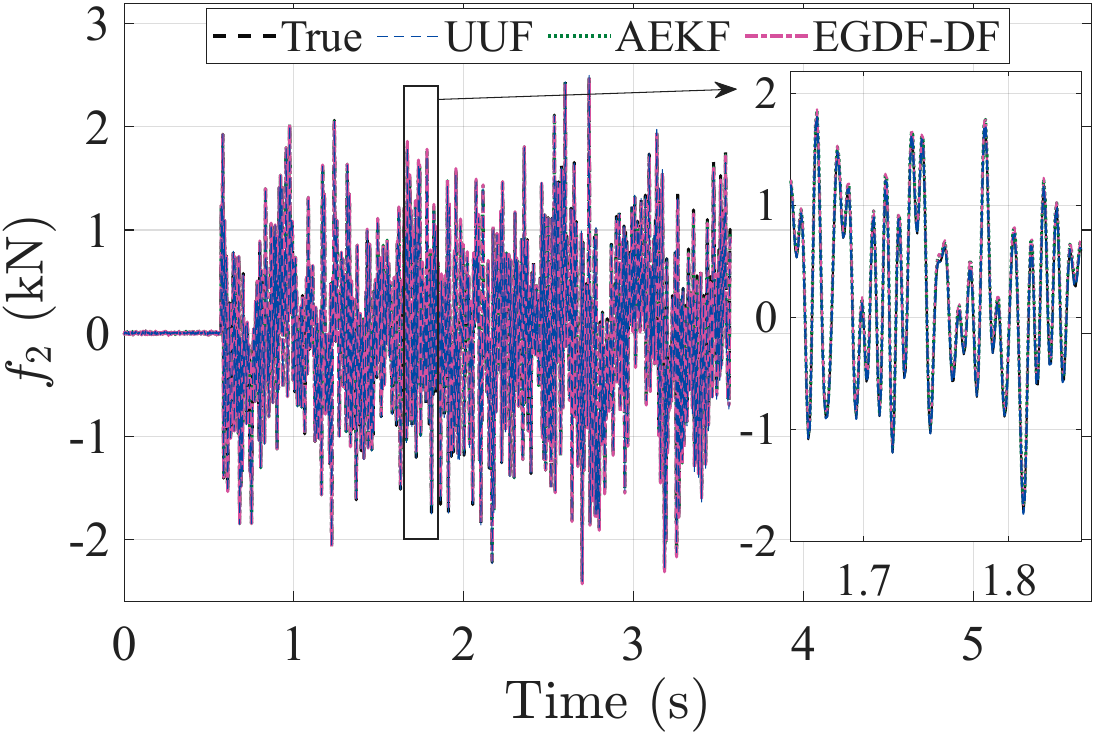}
    \end{subfigure}
    \caption{Rear axle contact force estimates by the UUF against (a) the state-of-the-art and (b) the classical methods, with element 4 damaged.}
    \label{fig:cs1_sev_input}
\end{figure}

Figs.~\ref{fig:cs1_sev_param} to~\ref{fig:cs1_sev_input} show the time histories for the stiffness reduction of $20\%$. As shown in Fig.~\ref{fig:cs1_sev_param_a}, the developed UUF converges from the nominal stiffness to the true value within about $1$~s and remains close to it until the rear axle leaves the beam at $3.57$~s. The IPS-UKF shows a similar behaviour. In contrast, the DKF-UKF underestimates the stiffness for most of the time history. In Fig.~\ref{fig:cs1_sev_param_b}, the estimates of the AEKF and the EGDF-DF do not settle at the true value, whereas the estimate of the UUF remains steady.

Fig.~\ref{fig:cs1_sev_state} shows the displacement and the acceleration estimates at node 3, where neither quantity is measured. In Fig.~\ref{fig:cs1_sev_state_a}, the UUF and the IPS-UKF match the true displacement, whereas the DKF-UKF slightly overestimates the displacement near its peak. In Fig.~\ref{fig:cs1_sev_state_b}, the AEKF and the EGDF-DF match the true displacement but with visible oscillations, as shown in the zoom window. The difference between the methods is larger for the acceleration. In Fig.~\ref{fig:cs1_sev_state_c}, the UUF matches the true acceleration, whereas the DKF-UKF and the IPS-UKF deviate from it at the peaks. In Fig.~\ref{fig:cs1_sev_state_d}, the estimates of the AEKF and the EGDF-DF are dominated by noise.

Fig.~\ref{fig:cs1_sev_input} shows the contact force at the rear axle, which enters the beam at $0.57$~s. In Fig.~\ref{fig:cs1_sev_input_a}, the UUF and the IPS-UKF match the true force, whereas the DKF-UKF deviates from it at the peaks, as shown in the zoom window. In Fig.~\ref{fig:cs1_sev_input_b}, the AEKF and the EGDF-DF also match the true force. Their large total errors in Fig.~\ref{fig:cs1_sev_errors} therefore stem from the parameter and the unmeasured responses rather than from the input.

\clearpage
\subsection{Performance under different damage locations} \label{sec:cs1_location}

\noindent The second study examines the estimation performance against the damage location. The flexural stiffness of one element is reduced by $30\%$, and the damaged element is moved from element 5, next to the mid-span, to element 1, next to the entry support. Since the beam is symmetric, only one half of the span is considered. The stiffness of the damaged element is the monitored parameter, and the damage is present before the vehicle enters the beam. The measurements are the two axle accelerations, together with the vertical acceleration and displacement at the node of the monitored element on the mid-span side, i.e., nodes 6 to 2 for elements 5 to 1, respectively.

\begin{figure}[ht]
    \centering
    \includegraphics[trim={0cm 0cm 0cm 0cm}, clip, width=0.45\textwidth]{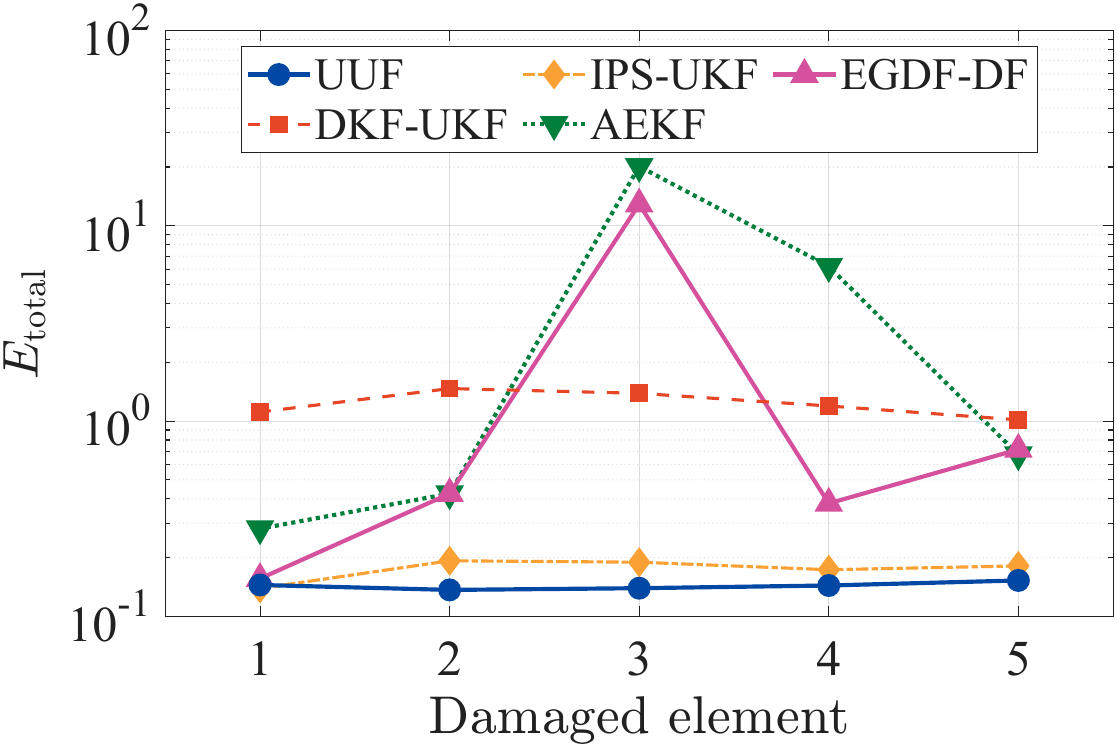}
    \caption{Total estimation error against the damaged element.}
    \label{fig:cs1_loc_errors}
\end{figure}

Fig.~\ref{fig:cs1_loc_errors} shows the total error $E_{\mathrm{total}}$ of each method against the damaged element. The total errors of the classical AEKF and EGDF-DF depend strongly on the damage location, ranging from $0.16$ to $20.13$, and both exceed $12$ when element 3 is damaged. The DKF-UKF gives total errors between $1.01$ and $1.47$. In contrast, the developed UUF shows a total error between $0.14$ and $0.15$ at every location, which is lower than those of the DKF-UKF and both classical methods. The IPS-UKF is again the only method of comparable accuracy, with total errors between $0.14$ and $0.19$. It is marginally lower than the UUF when element 1 is damaged, whereas the UUF gives the lowest total error for elements 2 to 5. It should be noted that the total error of the UUF varies by less than $0.02$ over the five locations, i.e., its performance is nearly independent of the damage location.

\begin{figure}[ht]
    \centering
    \begin{subfigure}{0.4\textwidth}
        \caption{} \label{fig:cs1_loc_param_a}
        \includegraphics[width=\textwidth]{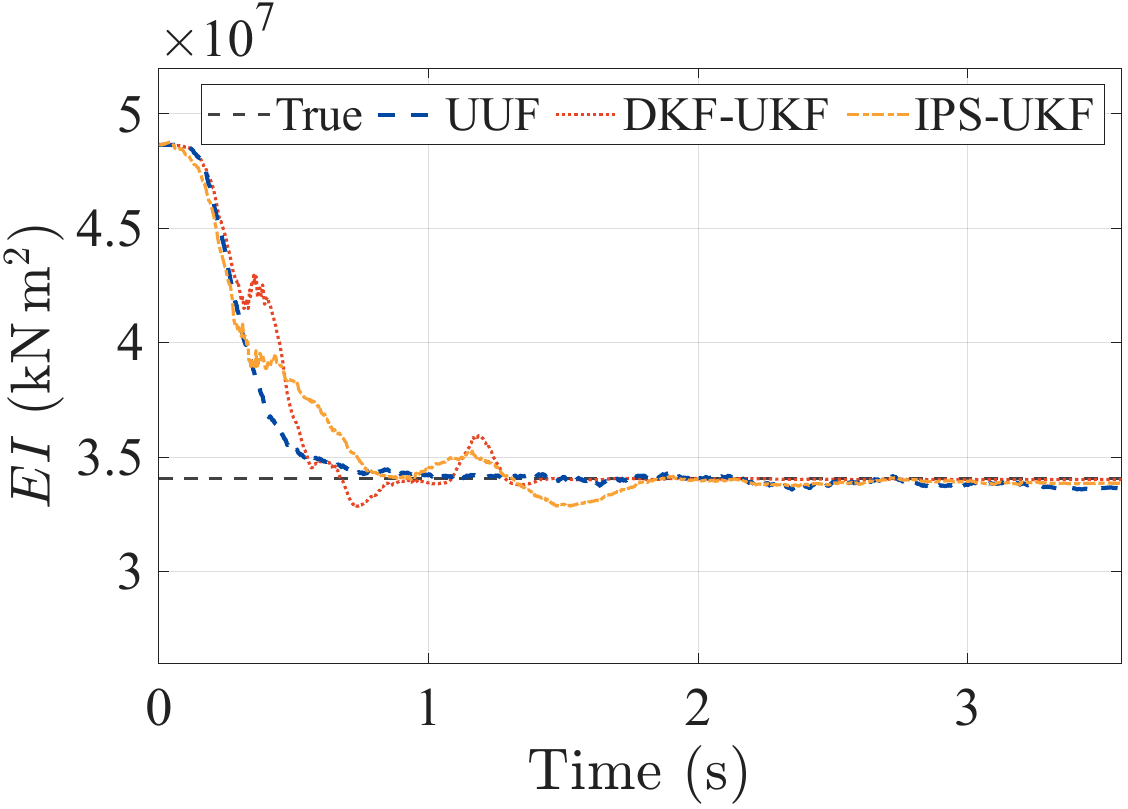}
    \end{subfigure}
    \hspace{1em}
    \begin{subfigure}{0.4\textwidth}
        \caption{} \label{fig:cs1_loc_param_b}
        \includegraphics[width=\textwidth]{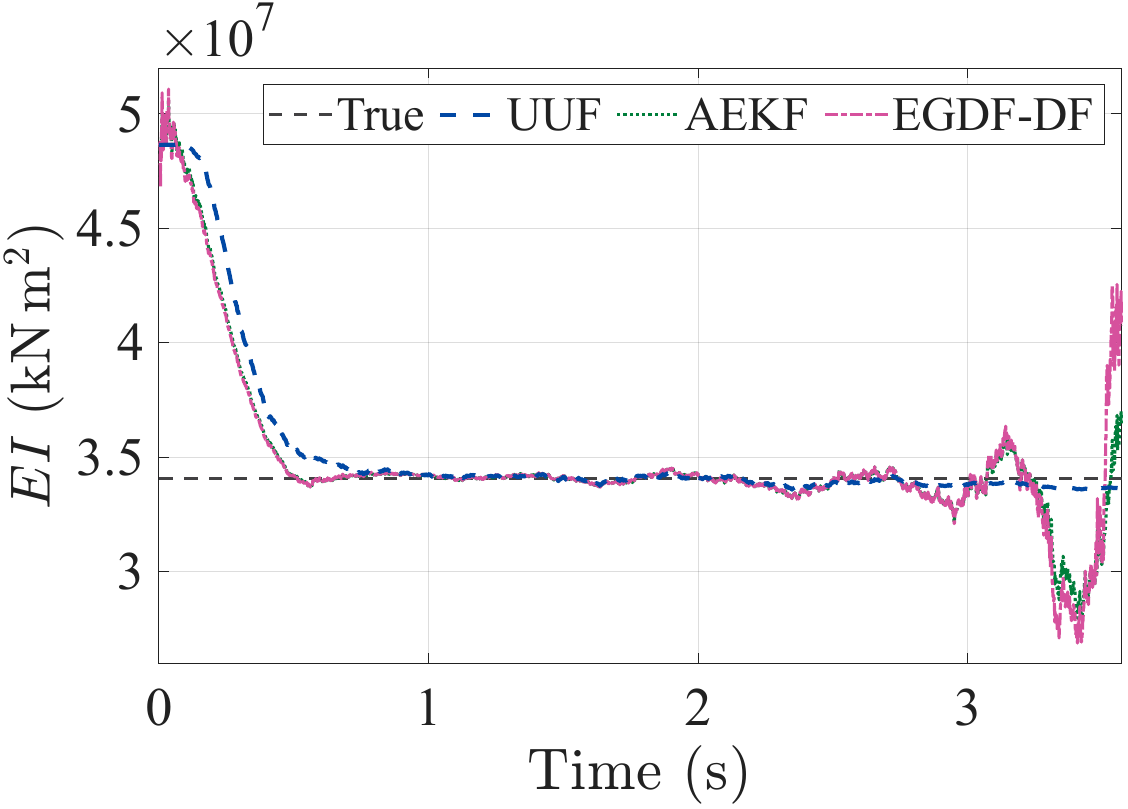}
    \end{subfigure}
    \caption{Flexural stiffness estimates of element 2 by the UUF against (a) the state-of-the-art and (b) the classical methods.}
    \label{fig:cs1_loc_param}
\end{figure}

\begin{figure}[ht]
    \centering
    \begin{subfigure}{0.45\textwidth}
        \caption{} \label{fig:cs1_loc_state_a}
        \includegraphics[width=\textwidth]{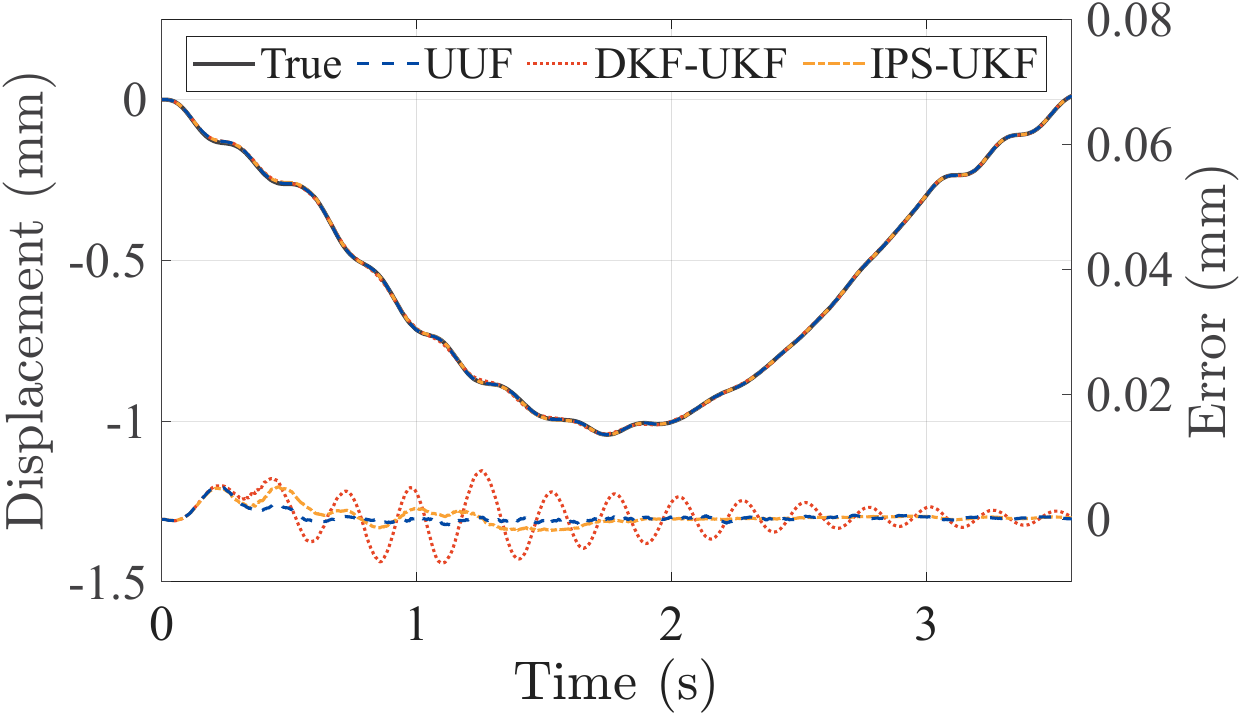}
    \end{subfigure}
    \hspace{1em}
    \begin{subfigure}{0.45\textwidth}
        \caption{} \label{fig:cs1_loc_state_b}
        \includegraphics[width=\textwidth]{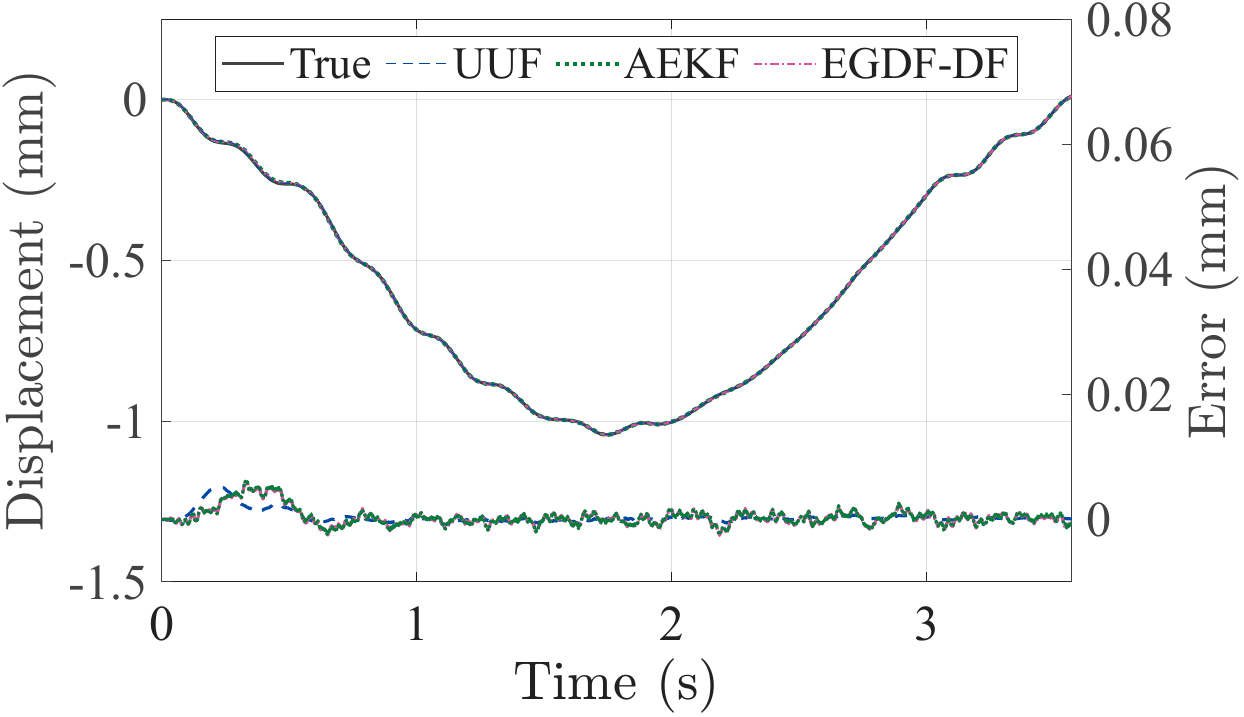}
    \end{subfigure}
    \begin{subfigure}{0.49\textwidth}
        \caption{} \label{fig:cs1_loc_state_c}
        \includegraphics[width=\textwidth]{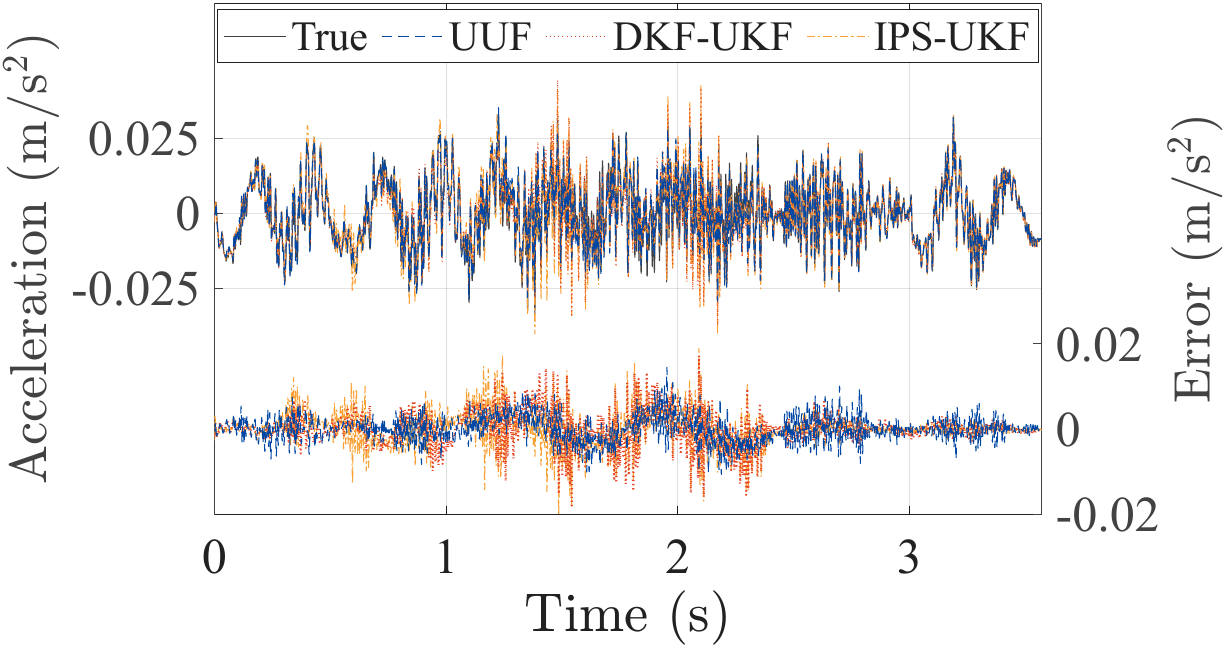}
    \end{subfigure}
    \hspace{1em}
    \begin{subfigure}{0.45\textwidth}
        \caption{} \label{fig:cs1_loc_state_d}
        \includegraphics[width=\textwidth]{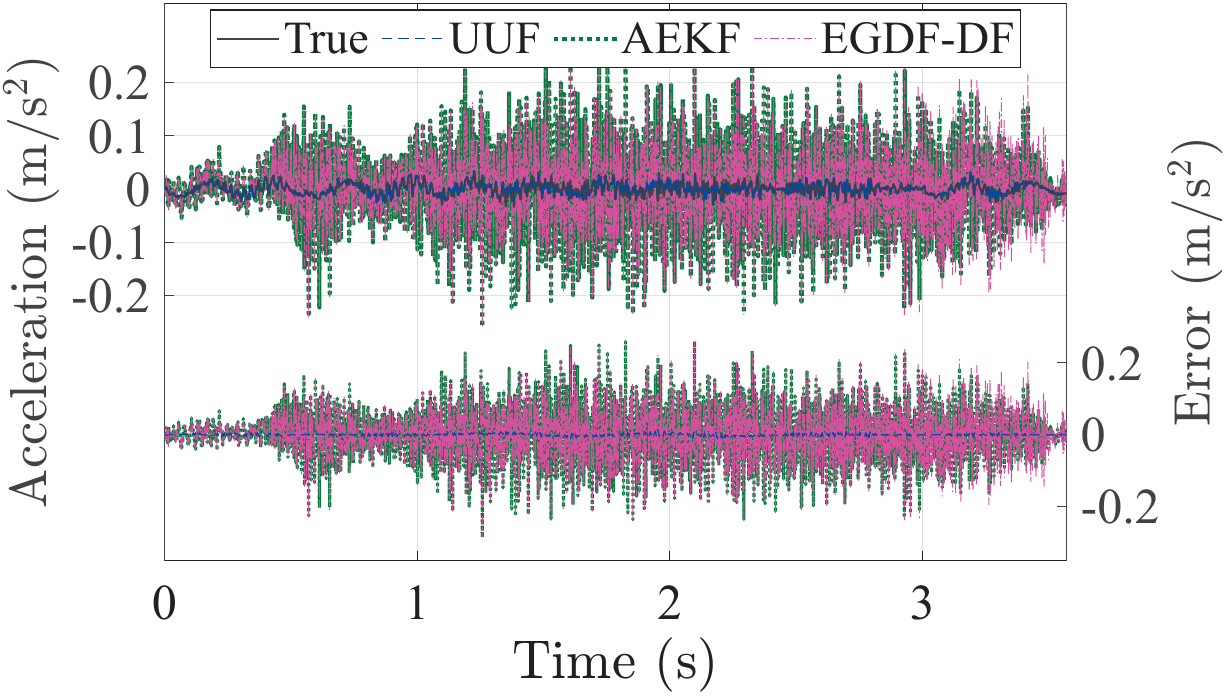}
    \end{subfigure}
    \caption{Estimates at node 6: (a) displacement against the state-of-the-art methods, (b) displacement against the classical methods, (c) acceleration against the state-of-the-art methods, and (d) acceleration against the classical methods.}
    \label{fig:cs1_loc_state}
\end{figure}

\begin{figure}[ht]
    \centering
    \begin{subfigure}{0.4\textwidth}
        \caption{} \label{fig:cs1_loc_input_a}
        \includegraphics[width=\textwidth]{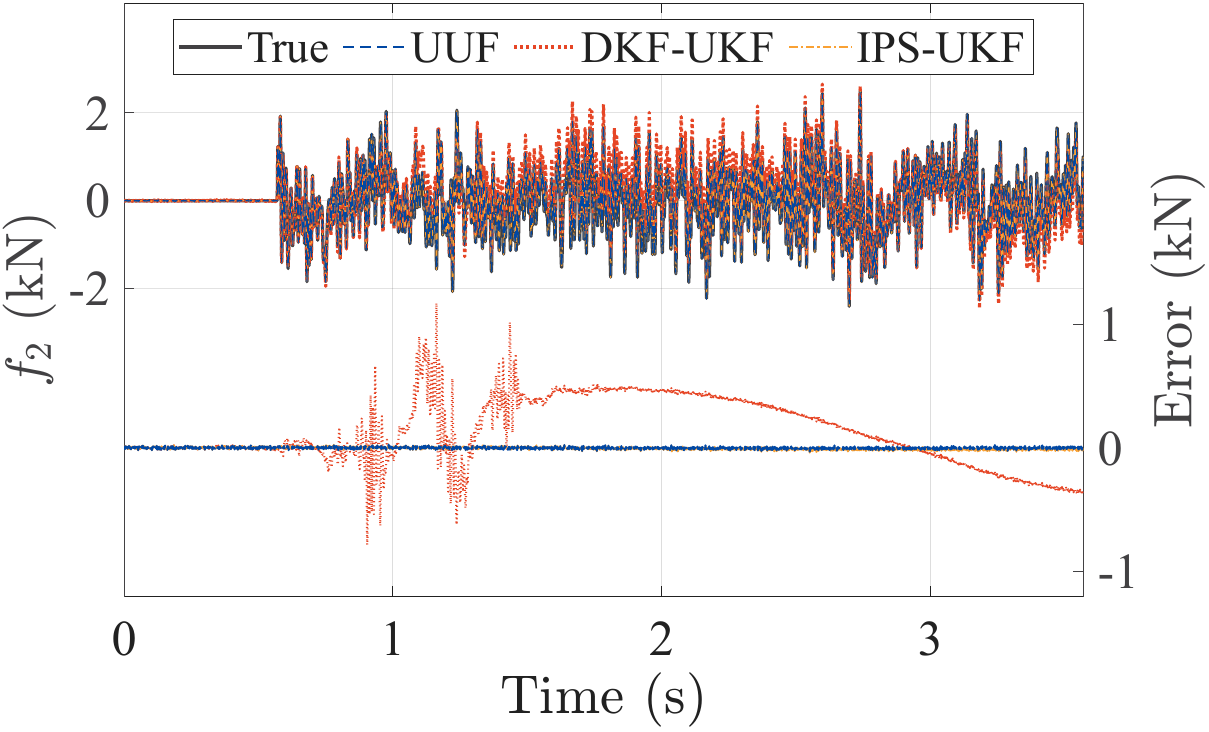}
    \end{subfigure}
    \hspace{1em}
    \begin{subfigure}{0.4\textwidth}
        \caption{} \label{fig:cs1_loc_input_b}
        \includegraphics[width=\textwidth]{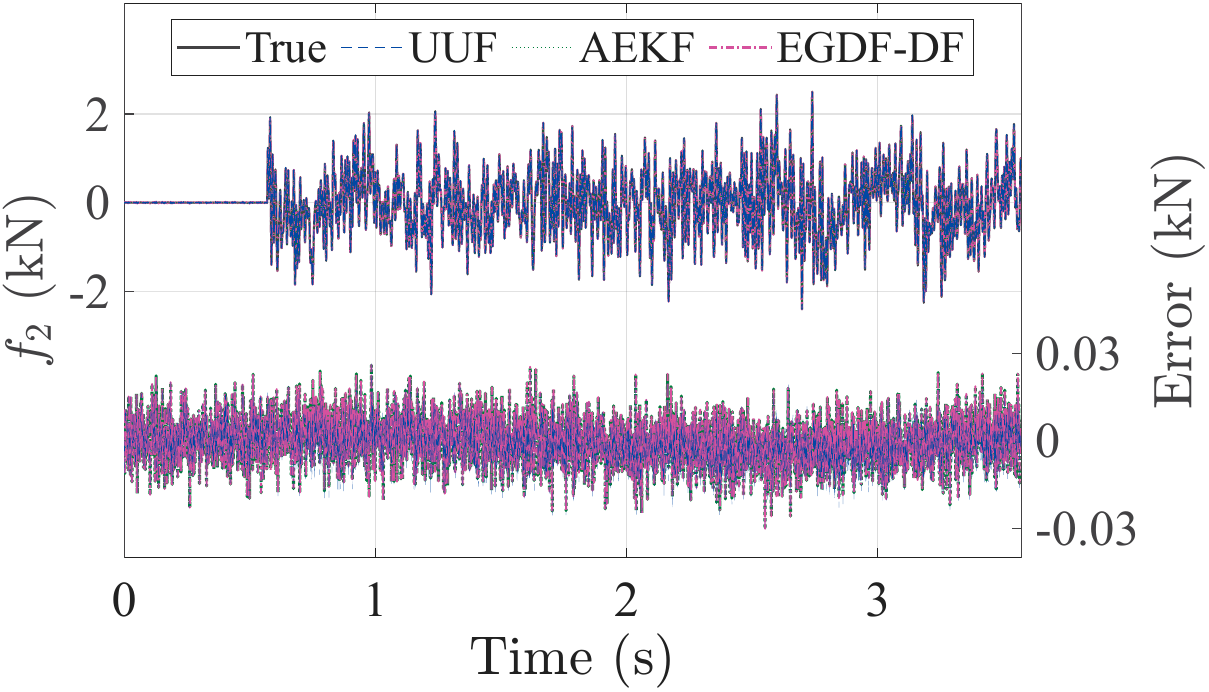}
    \end{subfigure}
    \caption{Rear axle contact force estimates by the UUF against (a) the state-of-the-art and (b) the classical methods, with element 2 damaged.}
    \label{fig:cs1_loc_input}
\end{figure}

Figs.~\ref{fig:cs1_loc_param} to~\ref{fig:cs1_loc_input} show the estimates when element 2 is damaged, for which the true stiffness is $3.41 \times 10^{10}$~$\mathrm{N \cdot m^2}$ and the bridge sensors are at node 3. In Figs.~\ref{fig:cs1_loc_state} and~\ref{fig:cs1_loc_input}, the estimation error is drawn against the right axis. As shown in Fig.~\ref{fig:cs1_loc_param_a}, the developed UUF converges to the true stiffness within about $1$~s and remains close to it until the rear axle leaves the beam at $3.57$~s. In contrast, the DKF-UKF and the IPS-UKF fluctuate around the true value until about $1.5$~s and $2$~s, respectively. In Fig.~\ref{fig:cs1_loc_param_b}, the AEKF and the EGDF-DF reach the true value at about $0.5$~s and match it until about $2.5$~s. However, both deviate considerably once the front axle leaves the beam at $3$~s, whereas the estimate of the UUF remains close to the true value.

\clearpage
Fig.~\ref{fig:cs1_loc_state} shows the displacement and the acceleration at the mid-span node 6, where neither quantity is measured. In Figs.~\ref{fig:cs1_loc_state_a} and~\ref{fig:cs1_loc_state_b}, every method matches the true displacement closely. The error of the DKF-UKF oscillates with the largest amplitude, whereas that of the UUF and IPS-UKF stays close to zero after the first $0.5$~s. The difference between the methods is again larger for the acceleration. In Fig.~\ref{fig:cs1_loc_state_c}, the errors of the DKF-UKF and the IPS-UKF grow between $1$~s and $2.5$~s to a level comparable to the true acceleration, whereas the error of the UUF remains smaller throughout. In Fig.~\ref{fig:cs1_loc_state_d}, the estimates of the AEKF and the EGDF-DF are dominated by noise, whose amplitude is much larger than that of the true acceleration.

Fig.~\ref{fig:cs1_loc_input} shows the contact force at the rear axle, which enters the beam at $0.57$~s. In Fig.~\ref{fig:cs1_loc_input_a}, the UUF and the IPS-UKF match the true force, and their errors are not visible at the scale of the figure. In contrast, the error of the DKF-UKF is comparable to the true force with a slowly varying bias. In Fig.~\ref{fig:cs1_loc_input_b}, the AEKF and the EGDF-DF also match the true force, with errors that are negligible compared to the force.

\subsection{Performance under different road classes and noise levels} \label{sec:cs1_class_noise}

\noindent The third study examines the estimation performance against the road roughness class and the measurement noise level. The road classes A, B and C of Fig.~\ref{fig:cs1_layout_b} are combined with noise levels of $1\%$, $2\%$ and $5\%$ of the RMS value of each response, giving nine scenarios. The monitored parameter is the flexural stiffness of element 4, which is reduced by $30\%$ before the vehicle enters the beam. The measurements are those of Section~\ref{sec:cs1_severity}.

\begin{figure}[ht]
    \centering
    \begin{subfigure}{0.32\textwidth}
        \caption{} \label{fig:cs1_cn_errors_a}
        \includegraphics[width=\textwidth]{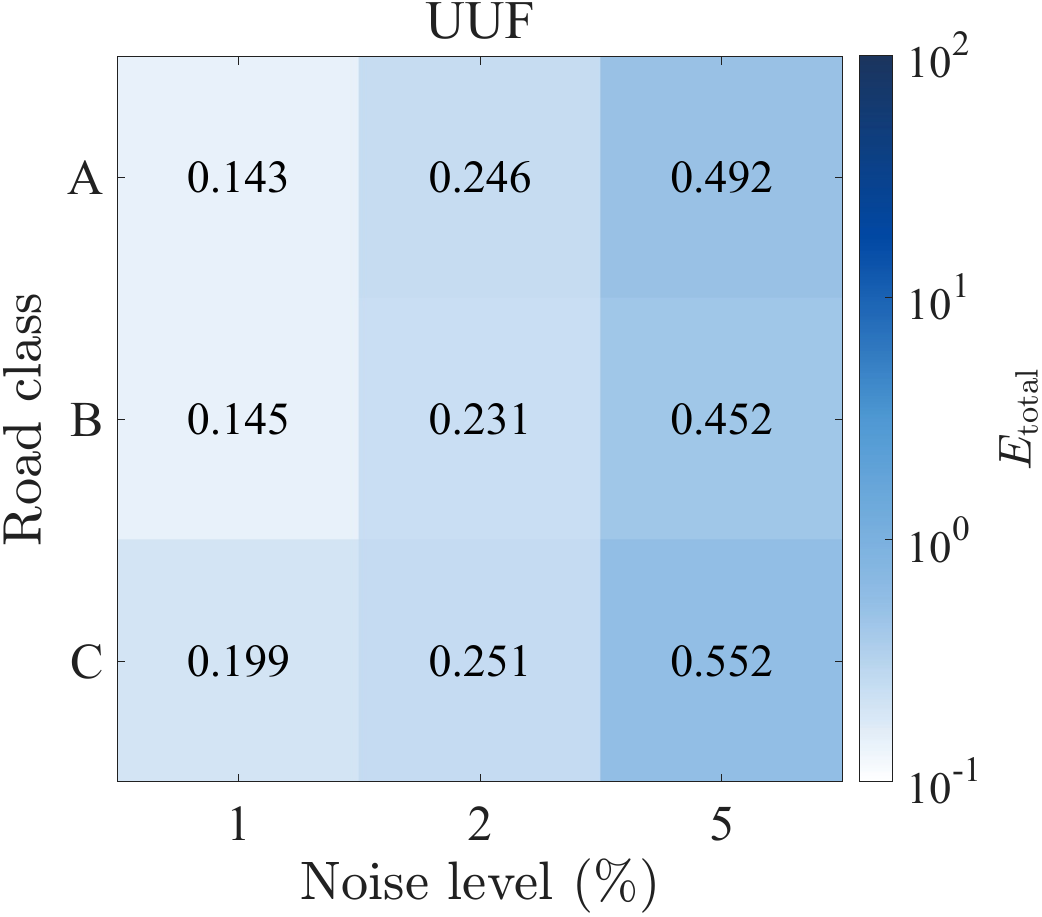}
    \end{subfigure}
    \begin{subfigure}{0.32\textwidth}
        \caption{} \label{fig:cs1_cn_errors_b}
        \includegraphics[width=\textwidth]{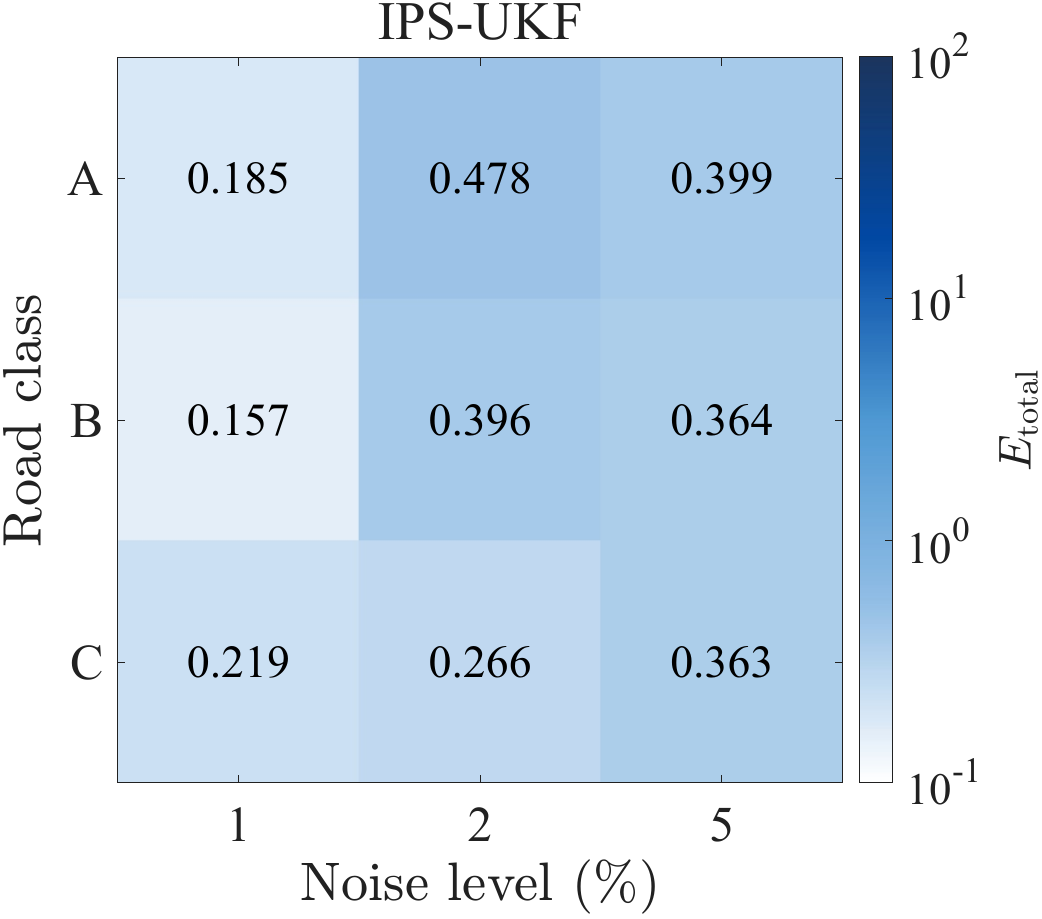}
    \end{subfigure}
    \begin{subfigure}{0.32\textwidth}
        \caption{} \label{fig:cs1_cn_errors_c}
        \includegraphics[width=\textwidth]{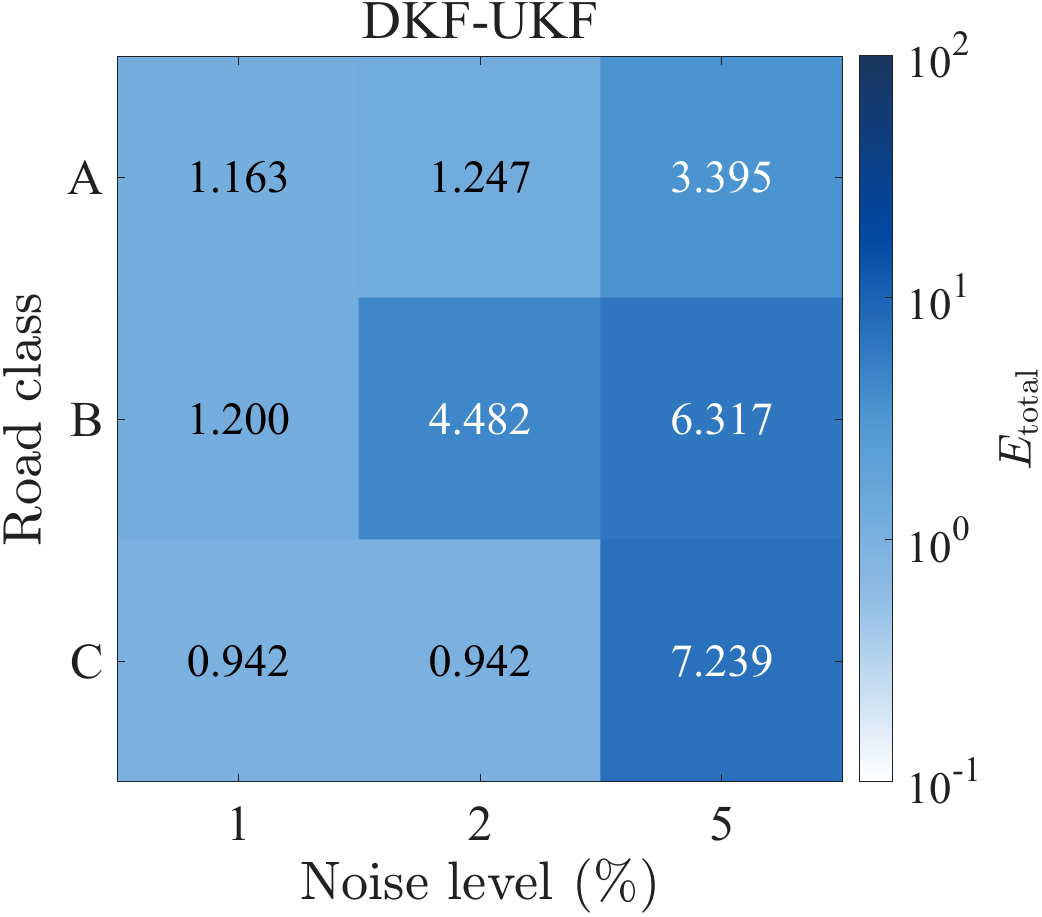}
    \end{subfigure}
    \begin{subfigure}{0.32\textwidth}
        \caption{} \label{fig:cs1_cn_errors_d}
        \includegraphics[width=\textwidth]{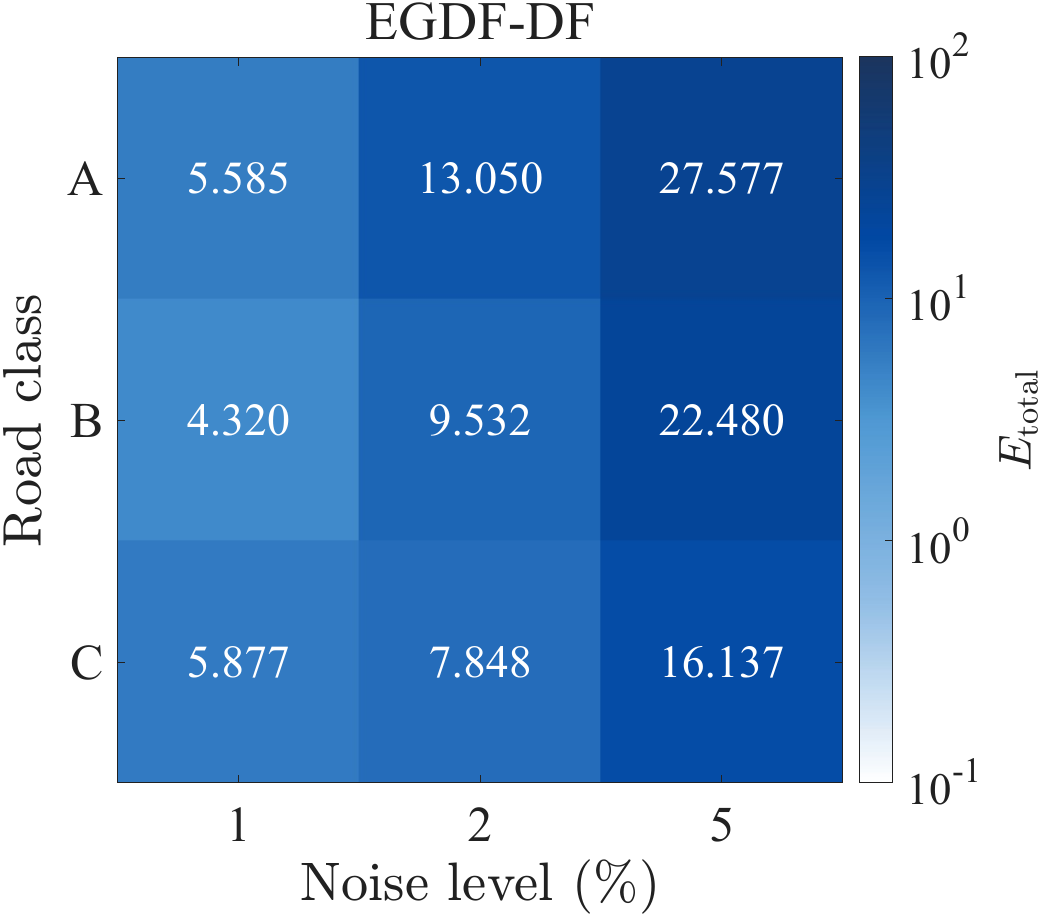}
    \end{subfigure}
    \begin{subfigure}{0.32\textwidth}
        \caption{} \label{fig:cs1_cn_errors_e}
        \includegraphics[width=\textwidth]{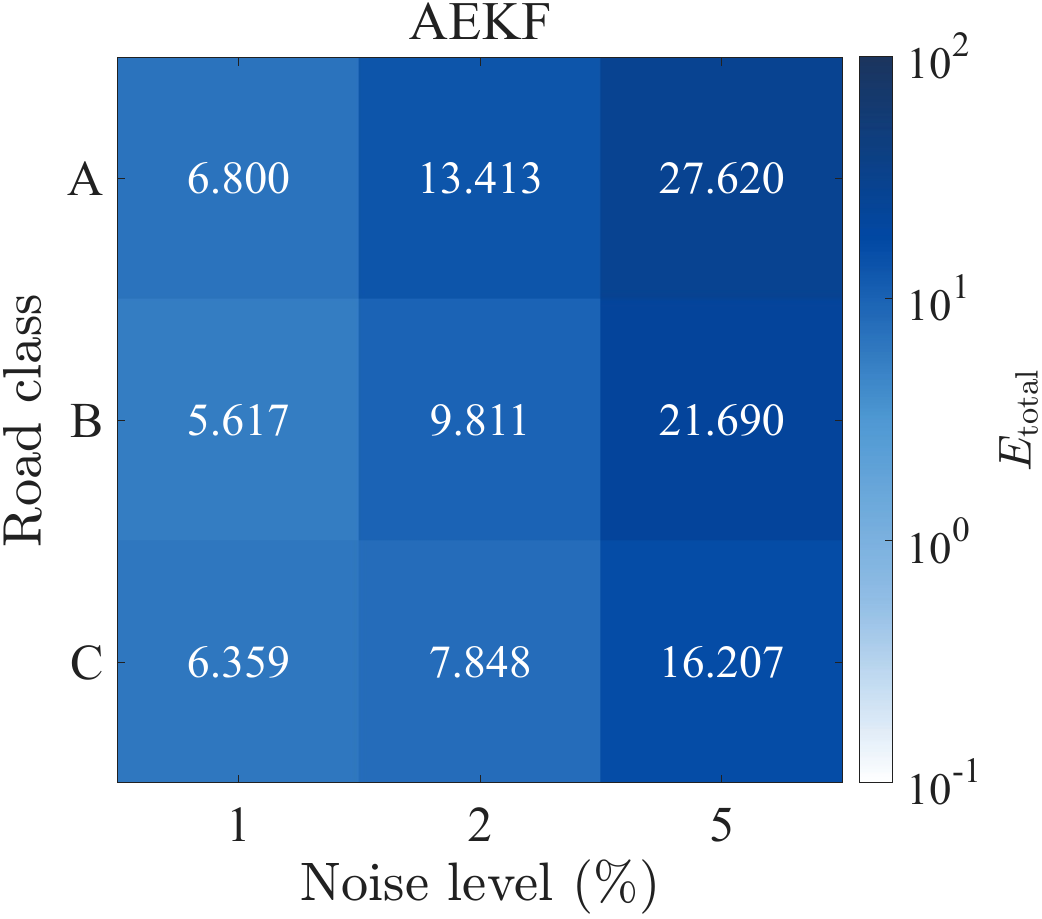}
    \end{subfigure}
    \caption{Total estimation error against the road class and the noise level: (a) UUF, (b) IPS-UKF, (c) DKF-UKF, (d) EGDF-DF and (e) AEKF.}
    \label{fig:cs1_cn_errors}
\end{figure}

Fig.~\ref{fig:cs1_cn_errors} shows the total error $E_{\mathrm{total}}$ of each method in the nine scenarios, with the value printed in each cell. The classical AEKF and EGDF-DF give total errors between $4.32$ and $27.62$, which increase considerably with the noise level. The DKF-UKF gives total errors between $0.94$ and $7.24$ and exceeds $3$ in every scenario with $5\%$ noise. In contrast, the developed UUF shows a total error between $0.14$ and $0.55$ over the nine scenarios, which is lower than those of the classical methods and the DKF-UKF in every scenario. Furthermore, the total error of the UUF varies little with the road class, i.e., from $0.14$ to $0.20$ at $1\%$ noise and from $0.23$ to $0.25$ at $2\%$ noise. This is expected, since the UUF estimates the contact forces without any model of their evolution, so a rougher road profile requires no change to the method. Compared with the IPS-UKF, the UUF gives the lower total error in all six scenarios with $1\%$ and $2\%$ noise. At $5\%$ noise, the IPS-UKF gives lower total errors, i.e., $0.36$ to $0.40$ compared to $0.45$ to $0.55$ for the UUF. Nonetheless, the total error of the UUF remains lower than those of the DKF-UKF and the classical methods at this noise level.

\section{Case Study II: the Old Ada Bridge} \label{sec:cs2}

\noindent Case Study I examined the behaviour of the methods on a simple structure with an idealised sensor configuration, across damage severities, damage locations and levels of road roughness and measurement noise. In particular, a displacement sensor was available on the bridge, which is not always possible in practice. In this case study, the bridge, the inspection vehicle and the sensor locations are taken from a field test, so that the methods are assessed under conditions close to practice. Three challenges of practice are examined. The first is that the acceleration-only configuration is often preferred, for which the auxiliary integrators of Section~\ref{sec:met_aux} are required. The second is that the suspension parameters of the vehicle are more difficult to determine than its mass properties, as noted in Section~\ref{sec:application}. The third is that the sensors may be too few for a feedforward matrix of full rank, which the UUF does not require, as stated in Section~\ref{sec:met_gains}. The set-up of the case study is presented below.

\subsection{Set-up of the case study} \label{sec:cs2_setup}

\begin{figure}[ht]
    \centering
    \begin{subfigure}[c]{0.35\textwidth}
        \caption{} \label{fig:cs2_layout_a}
        \includegraphics[width=\textwidth]{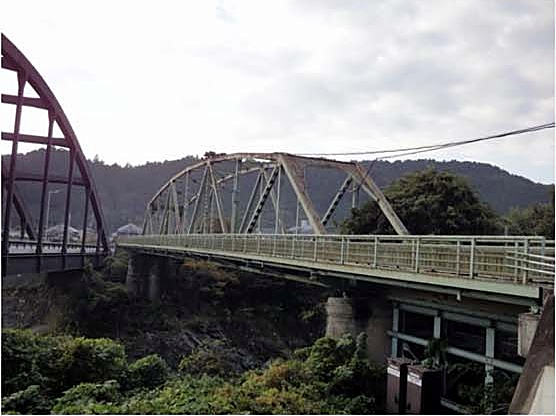}
    \end{subfigure}
    \hfill
    \begin{subfigure}[c]{0.6\textwidth}
        \caption{} \label{fig:cs2_layout_b}
        \includegraphics[width=\textwidth]{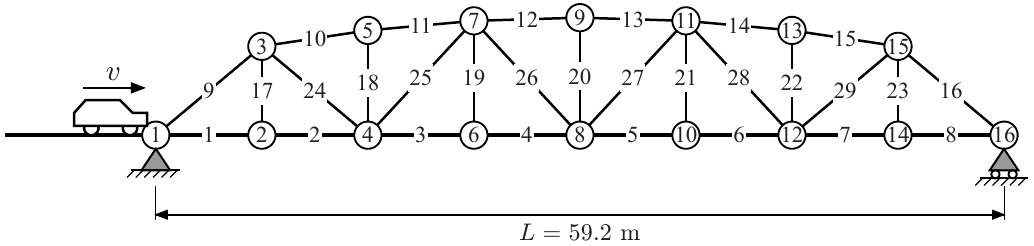}
    \end{subfigure}
    \caption{(a) Photo of the Old Ada Bridge (Photo credit:~\cite{KimAdaData2021}), and (b) the planar model with joints and members numbered.}
    \label{fig:cs2_layout}
\end{figure}

\noindent The Old Ada Bridge, shown in Fig.~\ref{fig:cs2_layout_a}, was a simply supported steel truss bridge in Nara Prefecture, Japan, with a span of $59.2$~m, built in 1959 and removed in 2012. Before its removal, a series of vibration tests was conducted on the bridge with a two-axle test vehicle~\cite{KimAdaData2021}.

Fig.~\ref{fig:cs2_layout_b} shows the planar model of the bridge, with 16 joints and 29 members. The member properties adopted in the model are those calibrated and validated in~\cite{Hamedani2025} by Bayesian model updating. The first three natural frequencies of the model are $2.98$, $6.89$ and $9.58$~Hz, within $1\%$ of the values identified from the field test. The responses of this case study are generated by this model, which is also adopted in every method compared. The zero-order hold of Eq.~\eqref{eq:vbi_dt_ss} is assumed on the input in the UUF, whereas the other methods index the input at the previous timestep, as in their original formulations. Therefore, the responses are generated for each method with its own zero-order hold assumption. The monitored parameter is the axial stiffness $EA$ of a member, and damage is introduced as a reduction of this stiffness, present before the vehicle enters the bridge.

The test vehicle is modelled following the half-car model of Section~\ref{sec:vbi_gov}, with the properties listed in Table~\ref{tab:cs2_vehicle}. The vehicle crosses the bridge at $30$~km/h on a class B road profile, and Rayleigh damping of $2\%$ is assumed for the first two modes of the bridge. The measurements are simulated with a time interval of $\Delta t = 0.002$~s and $1\%$ white noise. All other settings of the methods follow Section~\ref{sec:cs1_setup}, except the initial standard deviation of the parameter, which is $50\%$ of its nominal value. The process noise covariances are tuned by the strategy of Section~\ref{sec:app_tuning} over the same ranges.

\begin{table}[ht]
    \centering
    \caption{Properties of the half-car model of the test vehicle.}
    \label{tab:cs2_vehicle}
    \begin{tabular}{c c}
    \toprule
    Properties & Values \\
    \midrule
    Mass $m_v$ & $2~143$~kg \\
    Moment of inertia $I_v$ & $2~471$~$\mathrm{kg \cdot m^2}$ \\
    Distances to the axles $d_1$, $d_2$ & $1.35$~m \\
    Suspension stiffness $k_{v_1}$, $k_{v_2}$ & $1.30 \times 10^{5}$~N/m \\
    Suspension damping $c_{v_1}$, $c_{v_2}$ & $1.02 \times 10^{4}$~$\mathrm{N \cdot s/m}$ \\
    \bottomrule
    \end{tabular}
\end{table}

Three studies are presented hereafter that examine estimation with acceleration-only measurements, unknown suspension parameters, and a rank-deficient feedforward matrix.

\subsection{Performance with acceleration-only measurements} \label{sec:cs2_acc}

\begin{figure}[ht]
    \centering
    \includegraphics[width=0.9\textwidth]{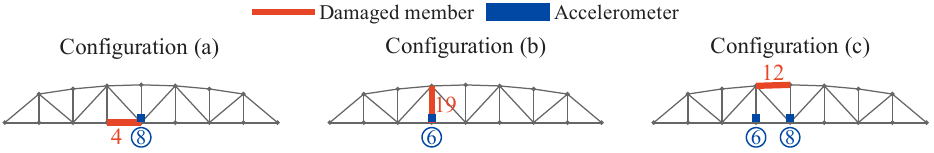}
    \caption{Damaged members and sensor locations of configurations (a) to (c).}
    \label{fig:cs2_acc_cases}
\end{figure}

\noindent The first study examines the estimation performance with acceleration-only measurements. Fig.~\ref{fig:cs2_acc_cases} shows the three configurations considered, in which the stiffness of the damaged member is reduced by $50\%$. In configuration (a), the bottom chord member 4 is monitored with an accelerometer at its end joint 8. In configuration (b), the vertical member 19 is monitored with an accelerometer at its lower joint 6. In configuration (c), the top chord member 12 is monitored with accelerometers at joints 6 and 8, neither of which is a joint of this member. This configuration is particularly challenging, since the measurements are non-collocated, i.e., the damaged member is not directly measured. In every configuration, the measurements are the two axle accelerations and the accelerations of the deck, without any displacement sensor. Therefore, the UUF is applied with the auxiliary integrators of Section~\ref{sec:met_aux}.

\begin{figure}[ht]
    \centering
    \begin{subfigure}{0.4\textwidth}
        \caption{} \label{fig:cs2_acc_c1_param_a}
        \includegraphics[width=\textwidth]{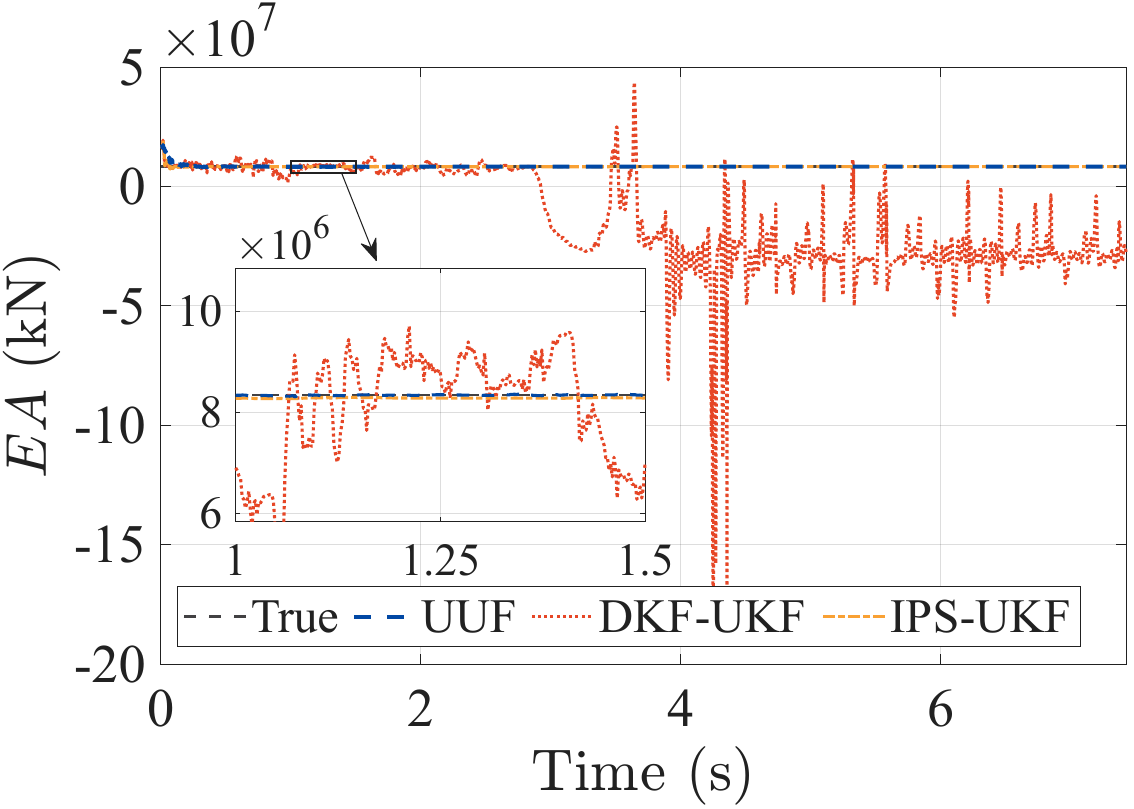}
    \end{subfigure}
    \hspace{1em}
    \begin{subfigure}{0.4\textwidth}
        \caption{} \label{fig:cs2_acc_c1_param_b}
        \includegraphics[width=\textwidth]{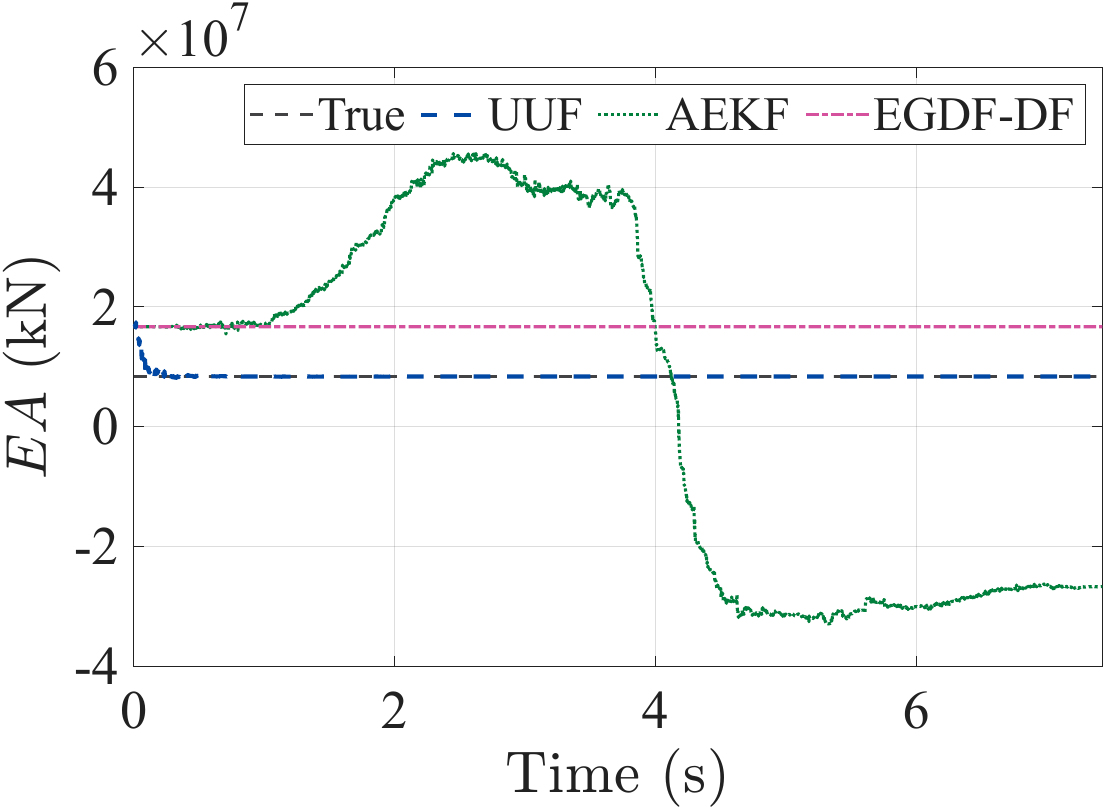}
    \end{subfigure}
    \caption{Axial stiffness estimates of member 4 by the UUF against (a) the state-of-the-art and (b) the classical methods.}
    \label{fig:cs2_acc_c1_param}
\end{figure}

\begin{figure}[ht]
    \centering
    \begin{subfigure}{0.45\textwidth}
        \caption{} \label{fig:cs2_acc_c1_state_a}
        \includegraphics[width=\textwidth]{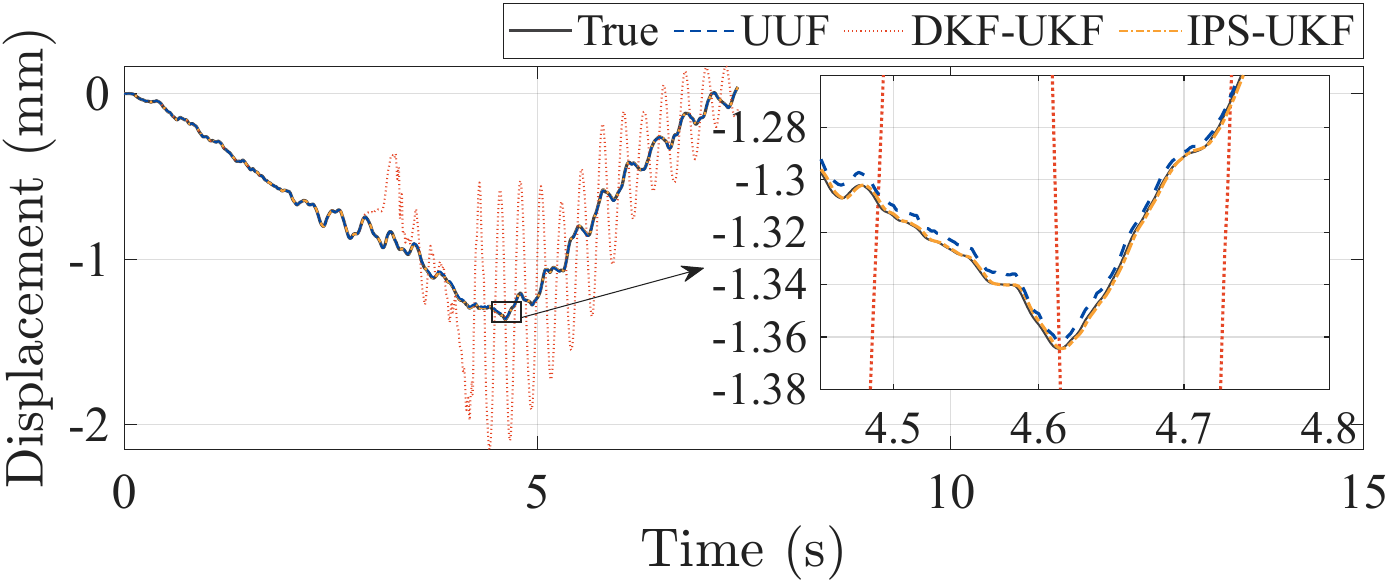}
    \end{subfigure}
    \hspace{1em}
    \begin{subfigure}{0.45\textwidth}
        \caption{} \label{fig:cs2_acc_c1_state_b}
        \includegraphics[width=\textwidth]{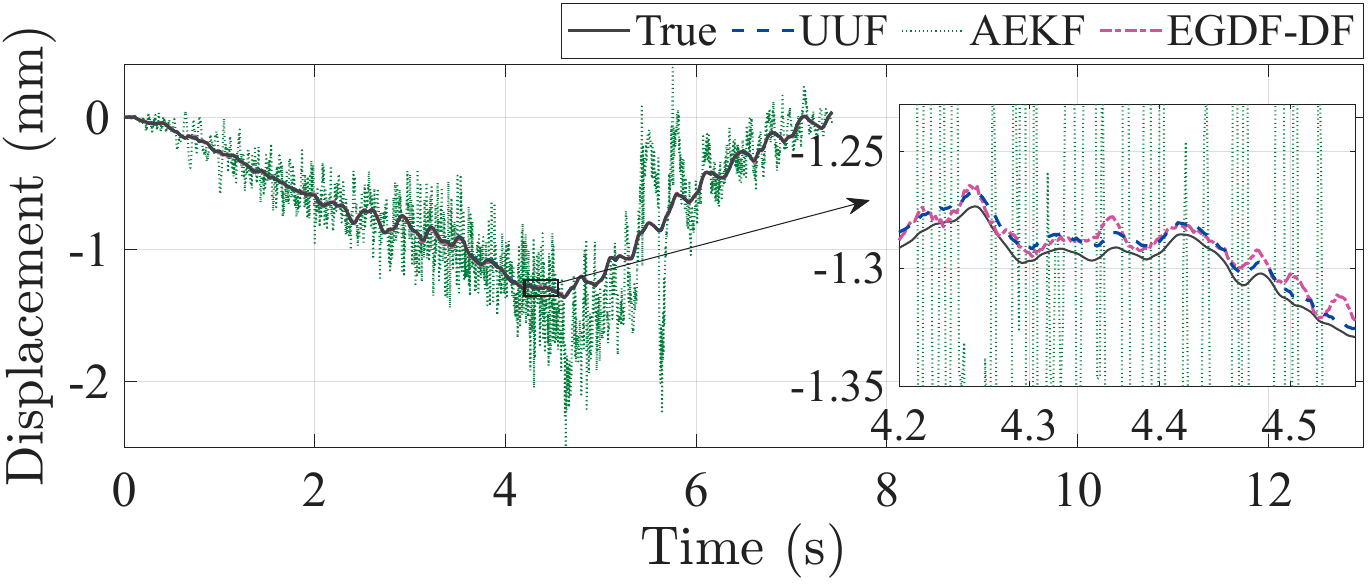}
    \end{subfigure}
    \begin{subfigure}{0.49\textwidth}
        \caption{} \label{fig:cs2_acc_c1_state_c}
        \includegraphics[width=\textwidth]{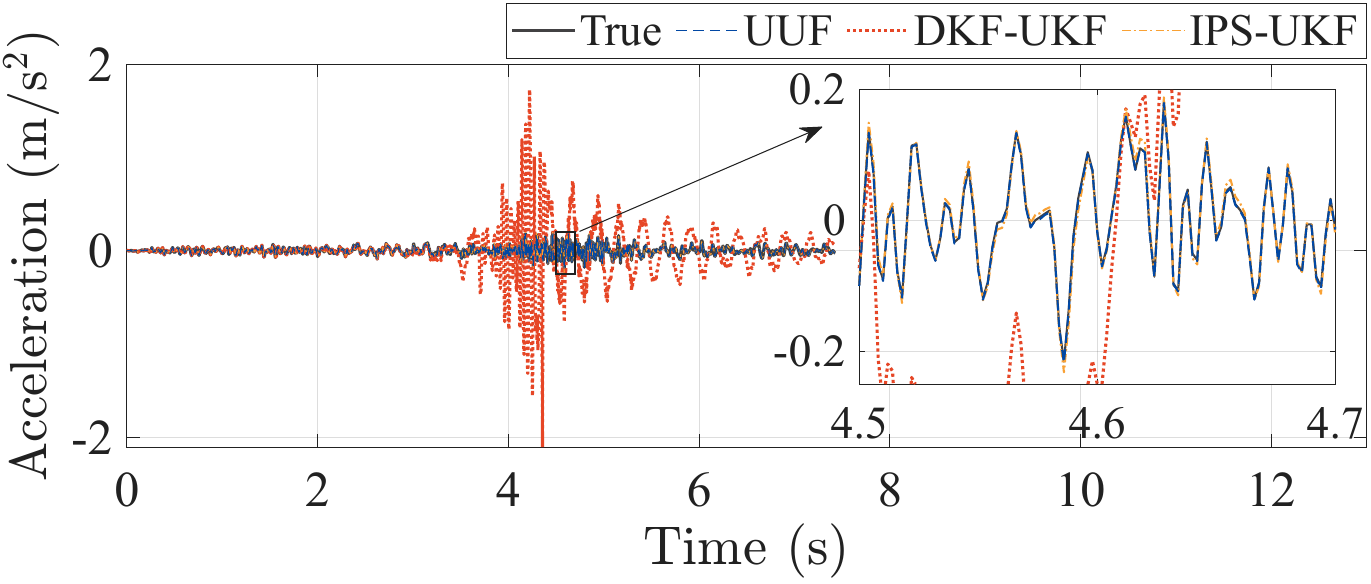}
    \end{subfigure}
    \hspace{1em}
    \begin{subfigure}{0.45\textwidth}
        \caption{} \label{fig:cs2_acc_c1_state_d}
        \includegraphics[width=\textwidth]{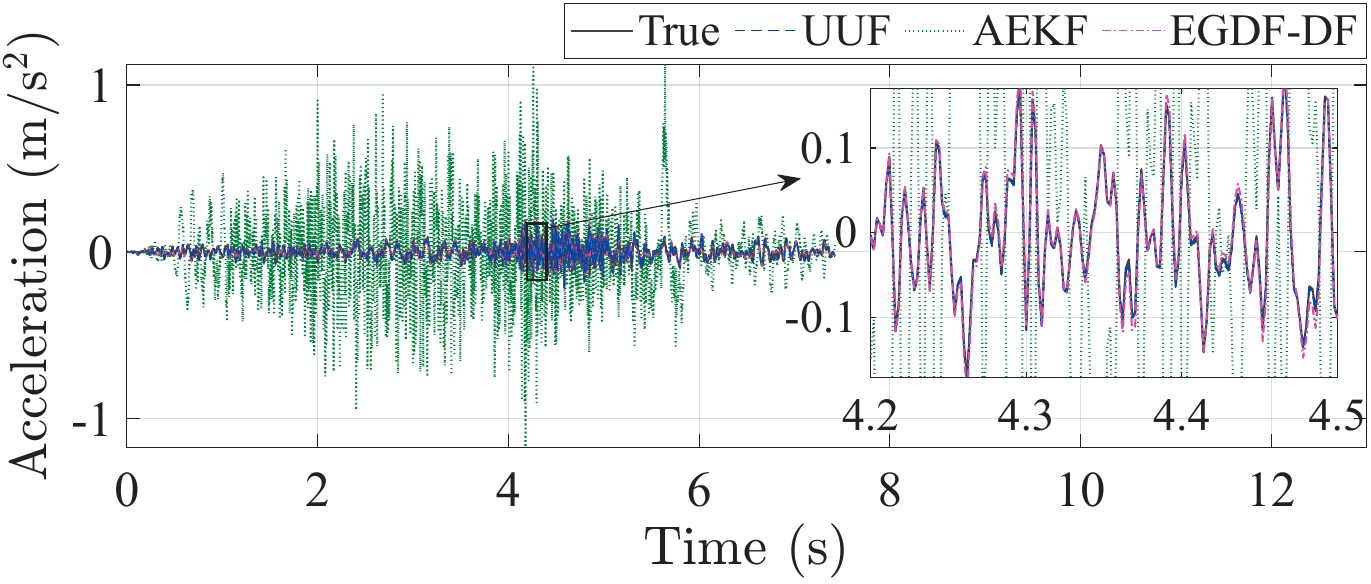}
    \end{subfigure}
    \caption{Estimates at joint 10 with member 4 damaged: (a) displacement against the state-of-the-art methods, (b) displacement against the classical methods, (c) acceleration against the state-of-the-art methods, and (d) acceleration against the classical methods.}
    \label{fig:cs2_acc_c1_state}
\end{figure}

\begin{figure}[ht]
    \centering
    \begin{subfigure}{0.4\textwidth}
        \caption{} \label{fig:cs2_acc_c1_input_a}
        \includegraphics[width=\textwidth]{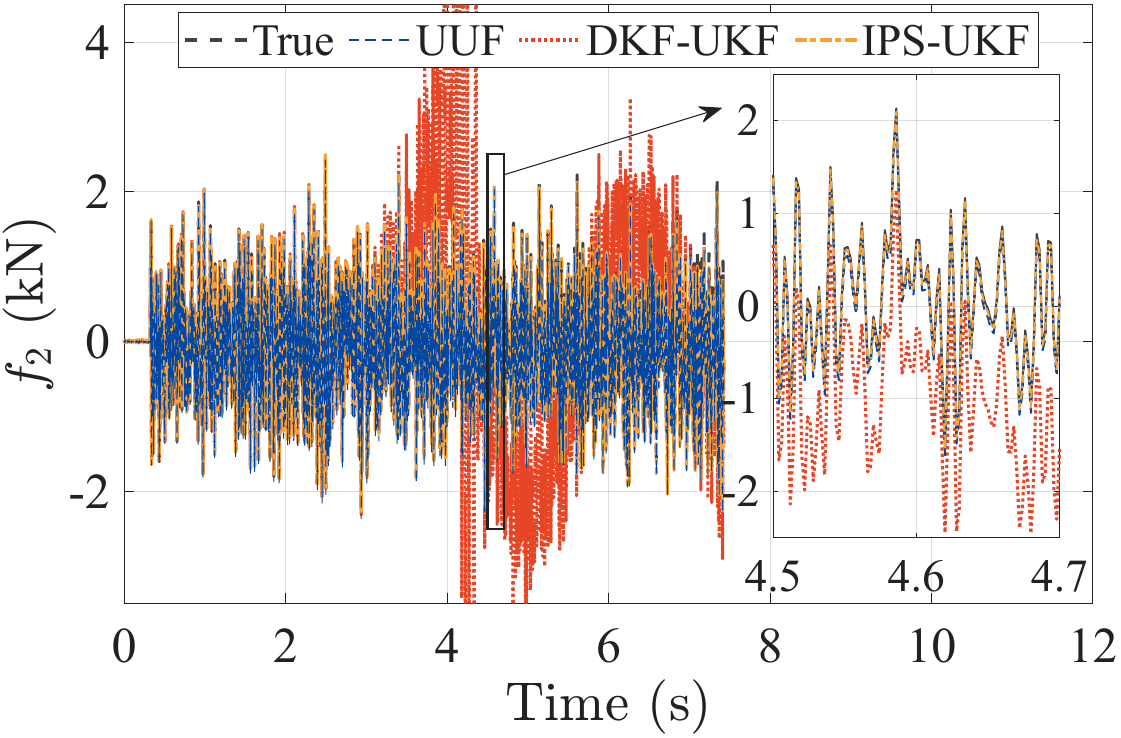}
    \end{subfigure}
    \hspace{1em}
    \begin{subfigure}{0.4\textwidth}
        \caption{} \label{fig:cs2_acc_c1_input_b}
        \includegraphics[width=\textwidth]{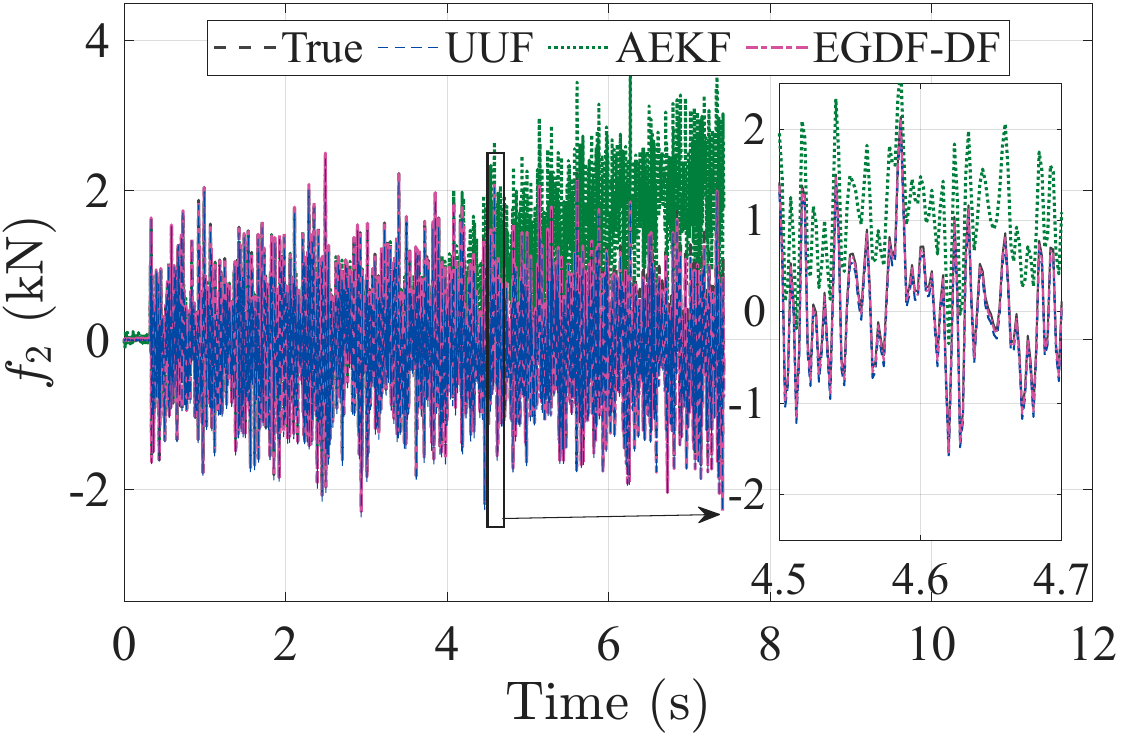}
    \end{subfigure}
    \caption{Rear axle contact force estimates by the UUF against (a) the state-of-the-art and (b) the classical methods, with member 4 damaged.}
    \label{fig:cs2_acc_c1_input}
\end{figure}

Fig.~\ref{fig:cs2_acc_c1_param} shows the stiffness estimates of damaged member 4 in configuration (a). The estimate of the developed UUF converges to the true value shortly after the vehicle enters the bridge and matches it until the vehicle leaves. The IPS-UKF performs similarly. In contrast, the estimate of the DKF-UKF diverges after about $3$~s. For the classical methods, the AEKF estimate diverges, whereas that of the EGDF-DF remains at the initial nominal value. Fig.~\ref{fig:cs2_acc_c1_state} shows the displacement and the acceleration at joint 10, where no sensor is installed. The UUF, the IPS-UKF and the EGDF-DF match the true responses, whereas the estimates of the DKF-UKF diverge, and those of the AEKF are dominated by noise. The contact force at the rear axle is shown in Fig.~\ref{fig:cs2_acc_c1_input}. The UUF, the IPS-UKF and the EGDF-DF match the true force, whereas the DKF-UKF deviates from it between about $3.5$ and $6$~s and the AEKF drifts away.

\clearpage

\begin{figure}[ht]
    \centering
    \begin{subfigure}{0.4\textwidth}
        \caption{} \label{fig:cs2_acc_c2_param_a}
        \includegraphics[width=\textwidth]{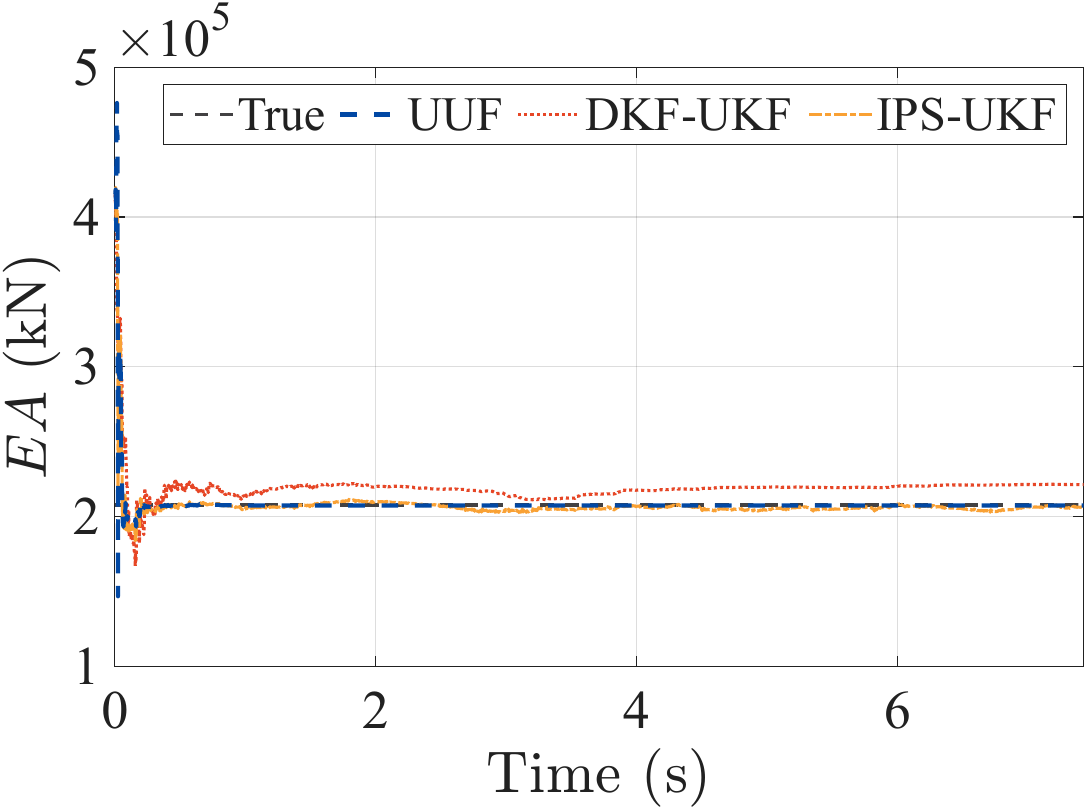}
    \end{subfigure}
    \hspace{1em}
    \begin{subfigure}{0.4\textwidth}
        \caption{} \label{fig:cs2_acc_c2_param_b}
        \includegraphics[width=\textwidth]{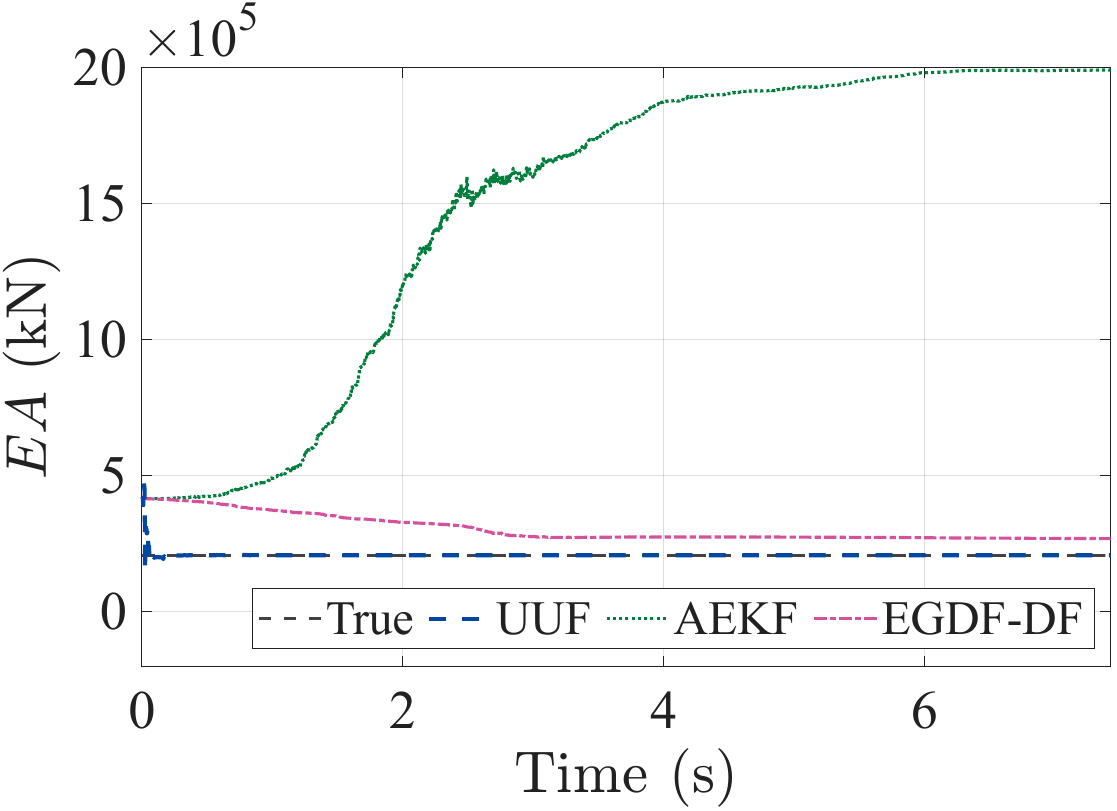}
    \end{subfigure}
    \caption{Axial stiffness estimates of member 19 by the UUF against (a) the state-of-the-art and (b) the classical methods.}
    \label{fig:cs2_acc_c2_param}
\end{figure}

\begin{figure}[ht]
    \centering
    \begin{subfigure}{0.45\textwidth}
        \caption{} \label{fig:cs2_acc_c2_state_a}
        \includegraphics[width=\textwidth]{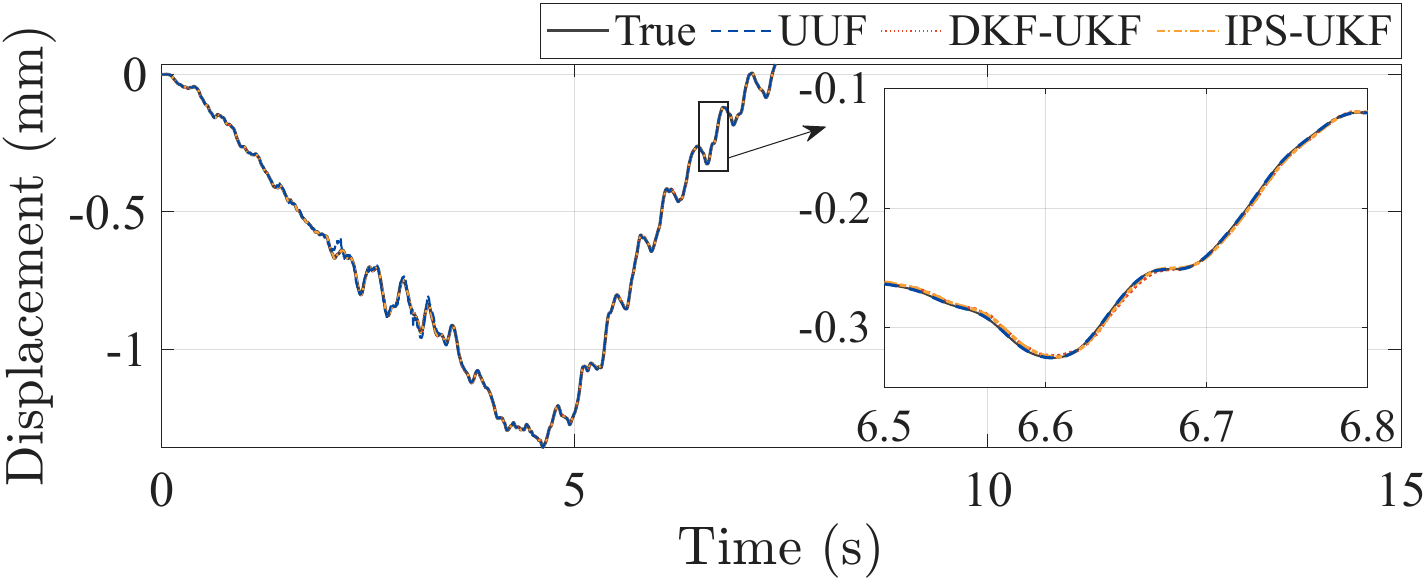}
    \end{subfigure}
    \hspace{1em}
    \begin{subfigure}{0.45\textwidth}
        \caption{} \label{fig:cs2_acc_c2_state_b}
        \includegraphics[width=\textwidth]{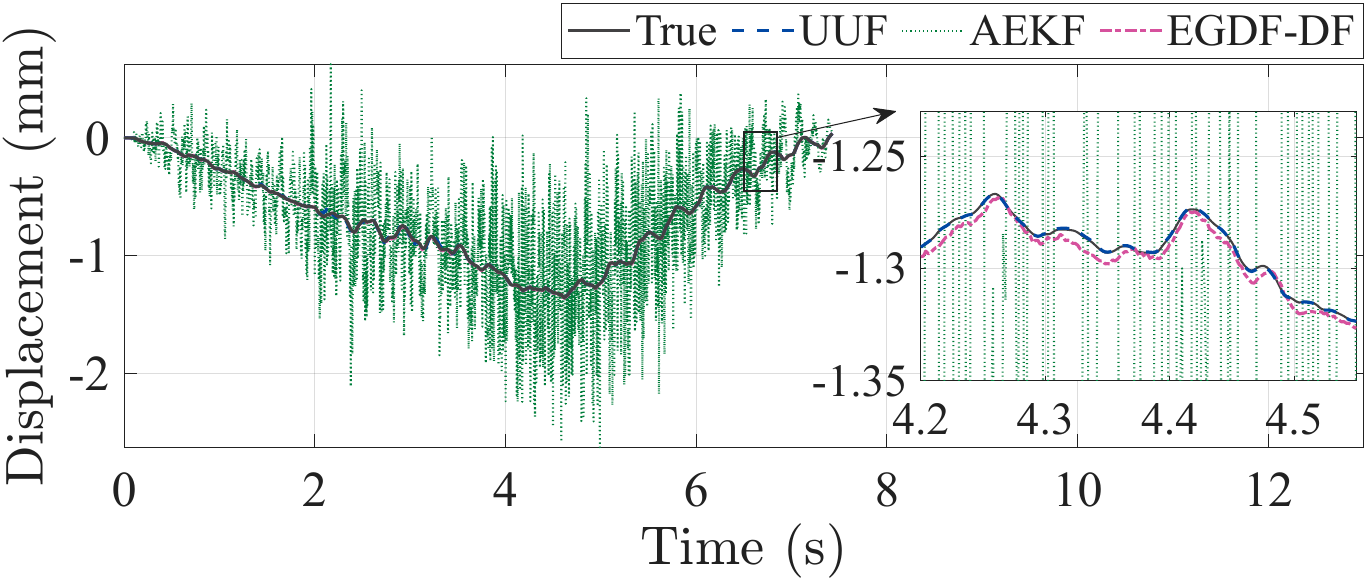}
    \end{subfigure}
    \begin{subfigure}{0.45\textwidth}
        \caption{} \label{fig:cs2_acc_c2_state_c}
        \includegraphics[width=\textwidth]{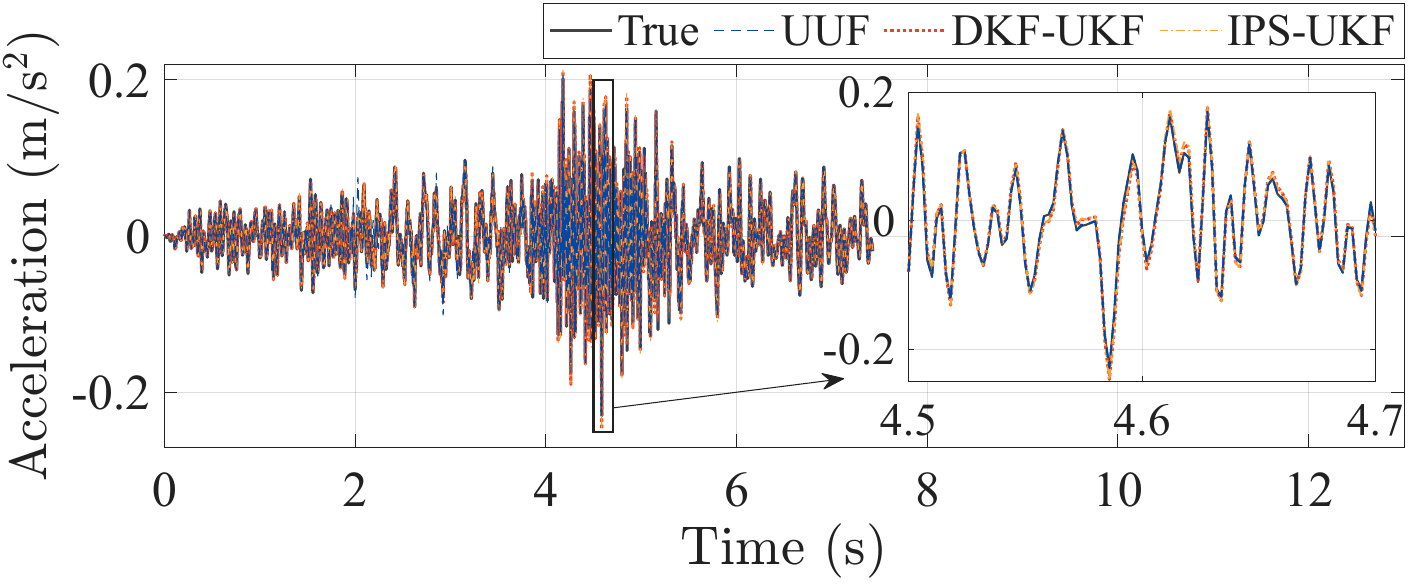}
    \end{subfigure}
    \hspace{1em}
    \begin{subfigure}{0.45\textwidth}
        \caption{} \label{fig:cs2_acc_c2_state_d}
        \includegraphics[width=\textwidth]{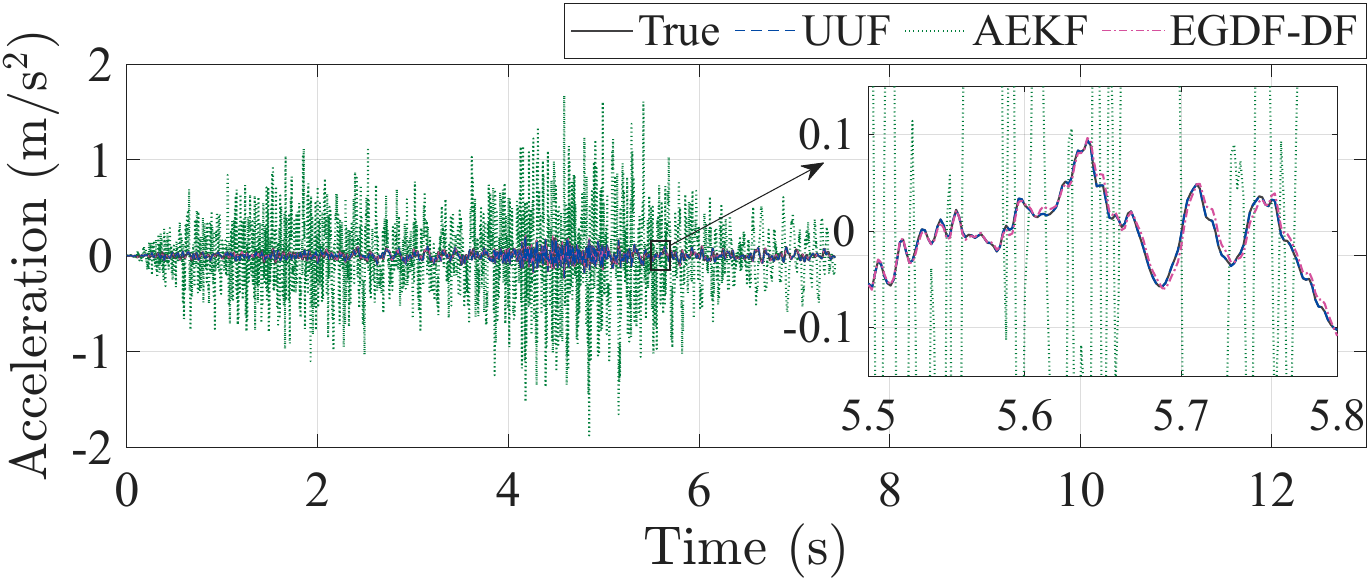}
    \end{subfigure}
    \caption{Estimates at joint 10 with member 19 damaged: (a) displacement against the state-of-the-art methods, (b) displacement against the classical methods, (c) acceleration against the state-of-the-art methods, and (d) acceleration against the classical methods.}
    \label{fig:cs2_acc_c2_state}
\end{figure}

\begin{figure}[ht]
    \centering
    \begin{subfigure}{0.4\textwidth}
        \caption{} \label{fig:cs2_acc_c2_input_a}
        \includegraphics[width=\textwidth]{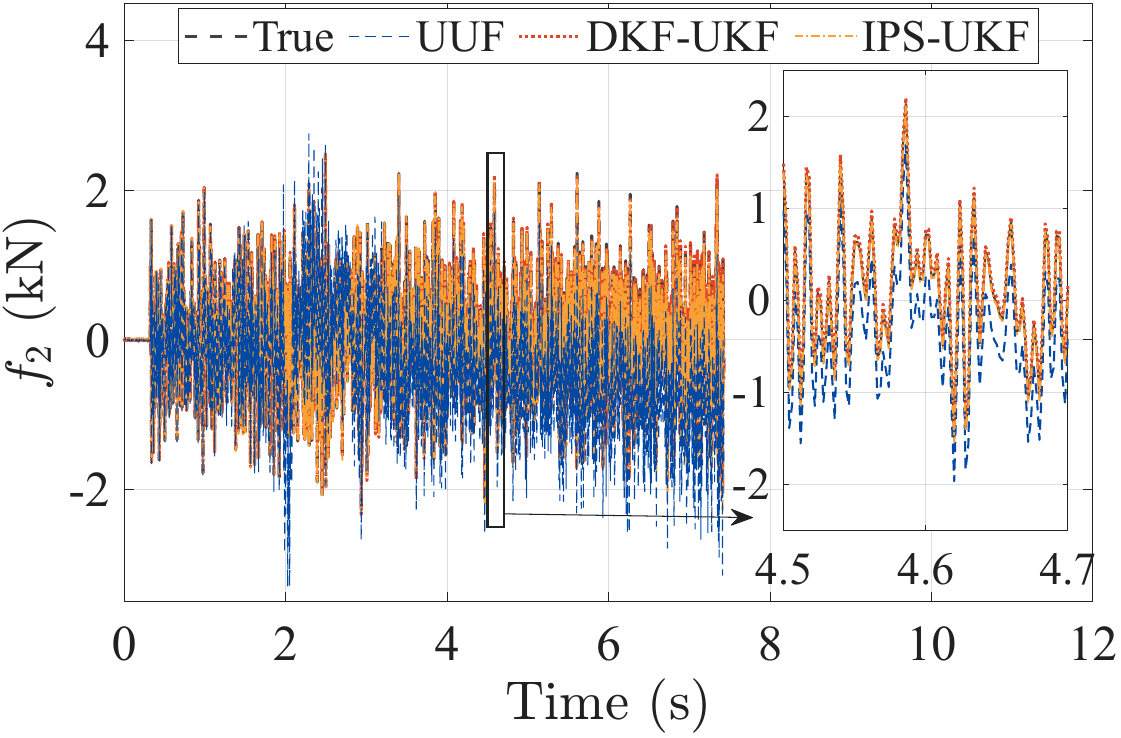}
    \end{subfigure}
    \hspace{1em}
    \begin{subfigure}{0.4\textwidth}
        \caption{} \label{fig:cs2_acc_c2_input_b}
        \includegraphics[width=\textwidth]{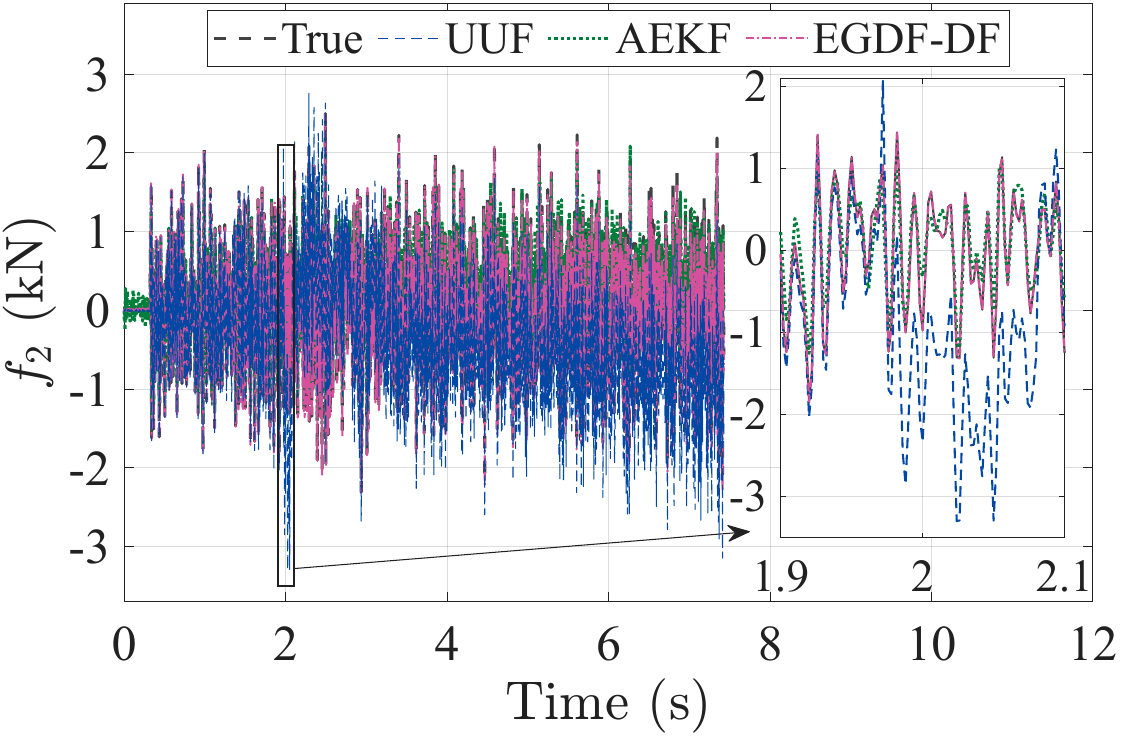}
    \end{subfigure}
    \caption{Rear axle contact force estimates by the UUF against (a) the state-of-the-art and (b) the classical methods, with member 19 damaged.}
    \label{fig:cs2_acc_c2_input}
\end{figure}

Fig.~\ref{fig:cs2_acc_c2_param} shows the stiffness estimates of damaged member 19 in configuration (b). The estimate of the developed UUF matches the true value, and that of the IPS-UKF matches it with small deviations, whereas that of the DKF-UKF remains slightly above it. The estimate of the AEKF diverges, and that of the EGDF-DF decreases towards the true value but remains above it when the vehicle leaves the bridge. In Fig.~\ref{fig:cs2_acc_c2_state}, the UUF, the state-of-the-art methods and the EGDF-DF match the true displacement and acceleration at joint 10, whereas the estimates of the AEKF are again dominated by noise. With regard to the input estimation, Fig.~\ref{fig:cs2_acc_c2_input} shows that the estimate of the UUF has a slowly varying deviation from the true force. The state-of-the-art methods and the EGDF-DF match the true force closely, and the AEKF matches it with small deviations.

\clearpage

\begin{figure}[ht]
    \centering
    \begin{subfigure}{0.4\textwidth}
        \caption{} \label{fig:cs2_acc_c3_param_a}
        \includegraphics[width=\textwidth]{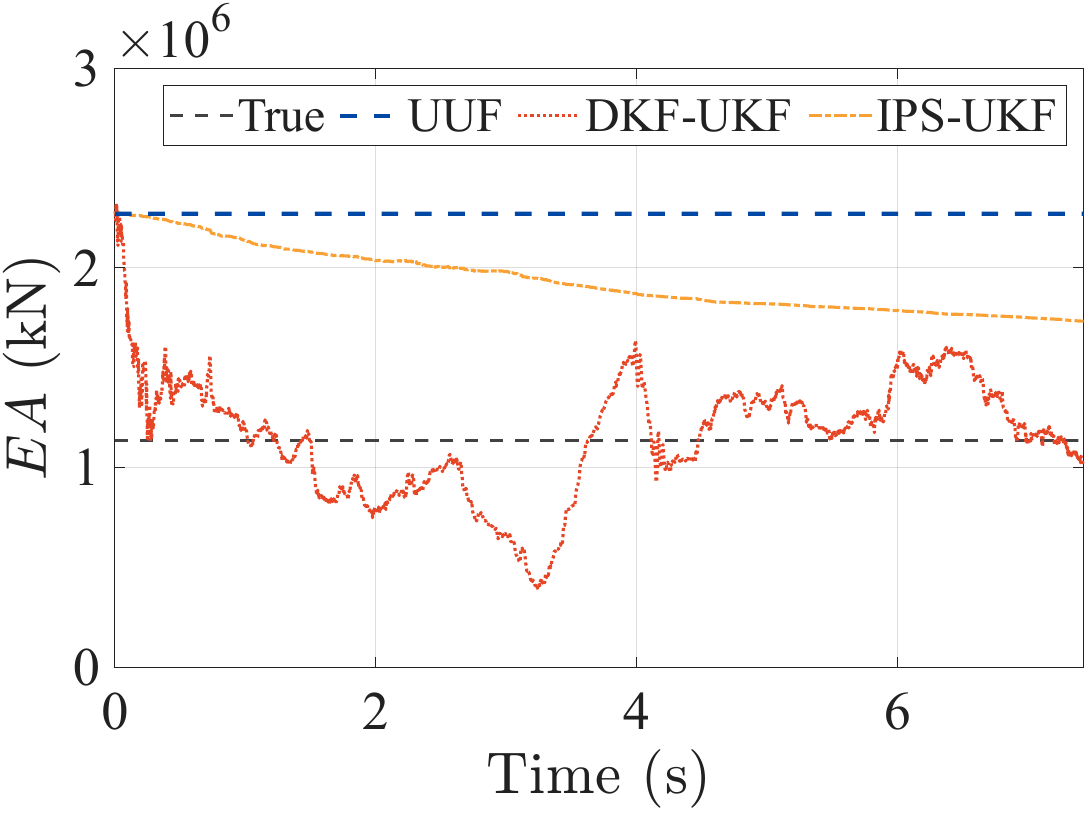}
    \end{subfigure}
    \hspace{1em}
    \begin{subfigure}{0.4\textwidth}
        \caption{} \label{fig:cs2_acc_c3_param_b}
        \includegraphics[width=\textwidth]{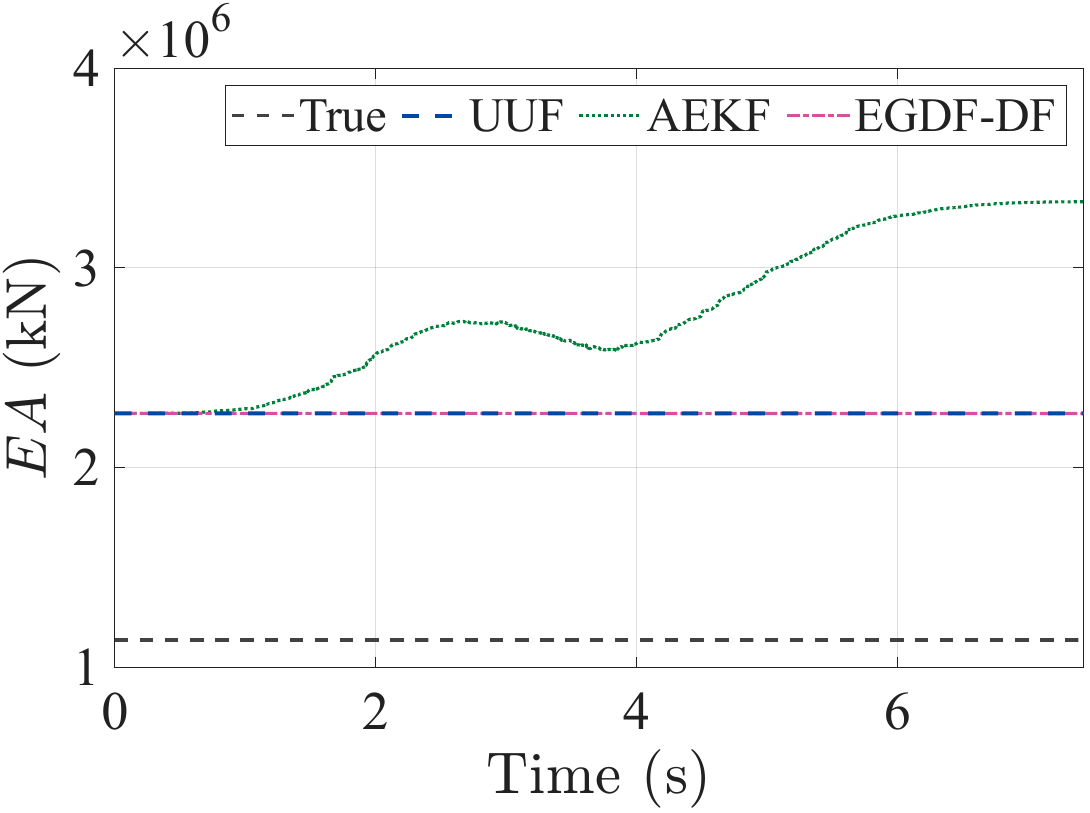}
    \end{subfigure}
    \caption{Axial stiffness estimates of member 12 by the UUF against (a) the state-of-the-art and (b) the classical methods.}
    \label{fig:cs2_acc_c3_param}
\end{figure}

\begin{figure}[ht]
    \centering
    \begin{subfigure}{0.45\textwidth}
        \caption{} \label{fig:cs2_acc_c3_state_a}
        \includegraphics[width=\textwidth]{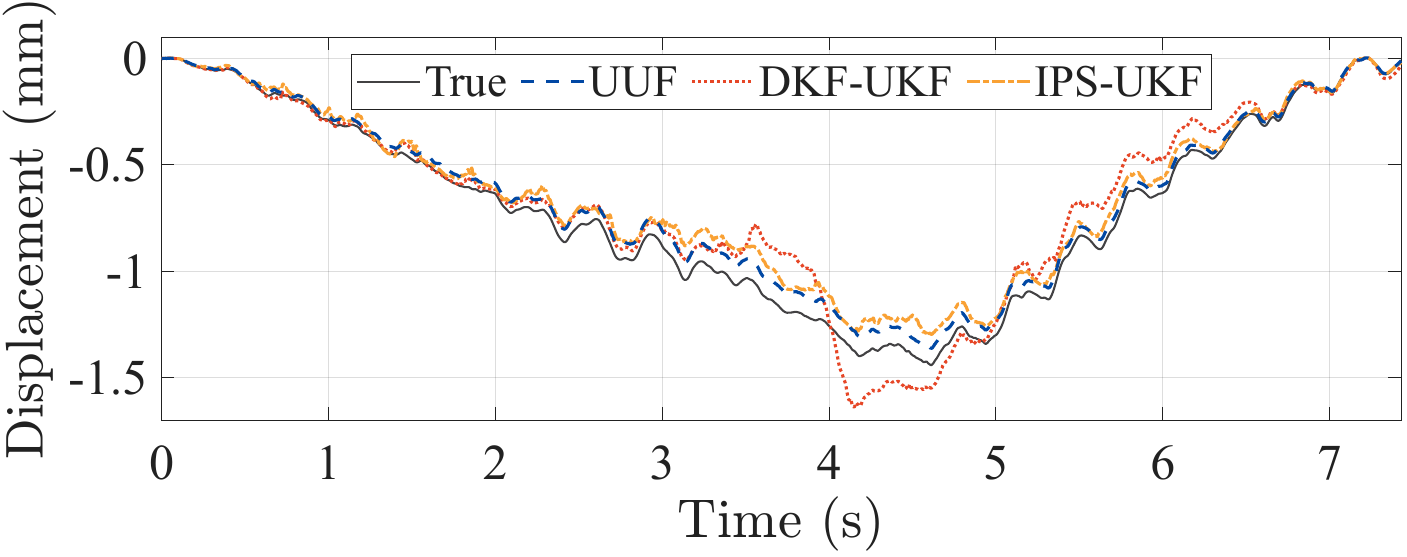}
    \end{subfigure}
    \hspace{1em}
    \begin{subfigure}{0.45\textwidth}
        \caption{} \label{fig:cs2_acc_c3_state_b}
        \includegraphics[width=\textwidth]{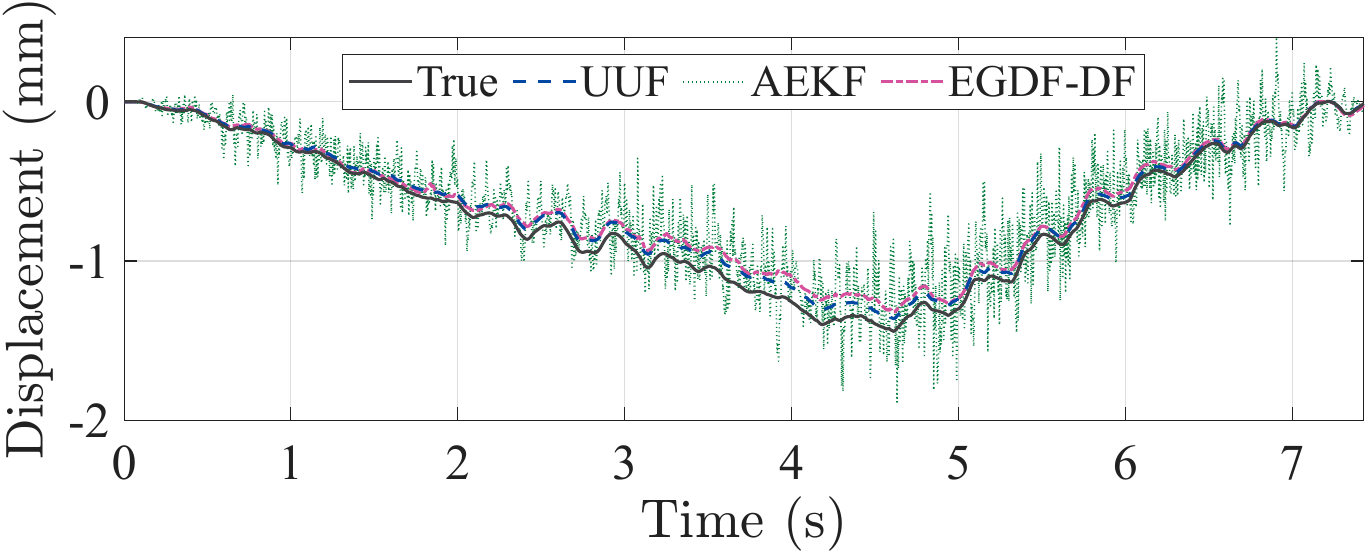}
    \end{subfigure}
    \begin{subfigure}{0.45\textwidth}
        \caption{} \label{fig:cs2_acc_c3_state_c}
        \includegraphics[width=\textwidth]{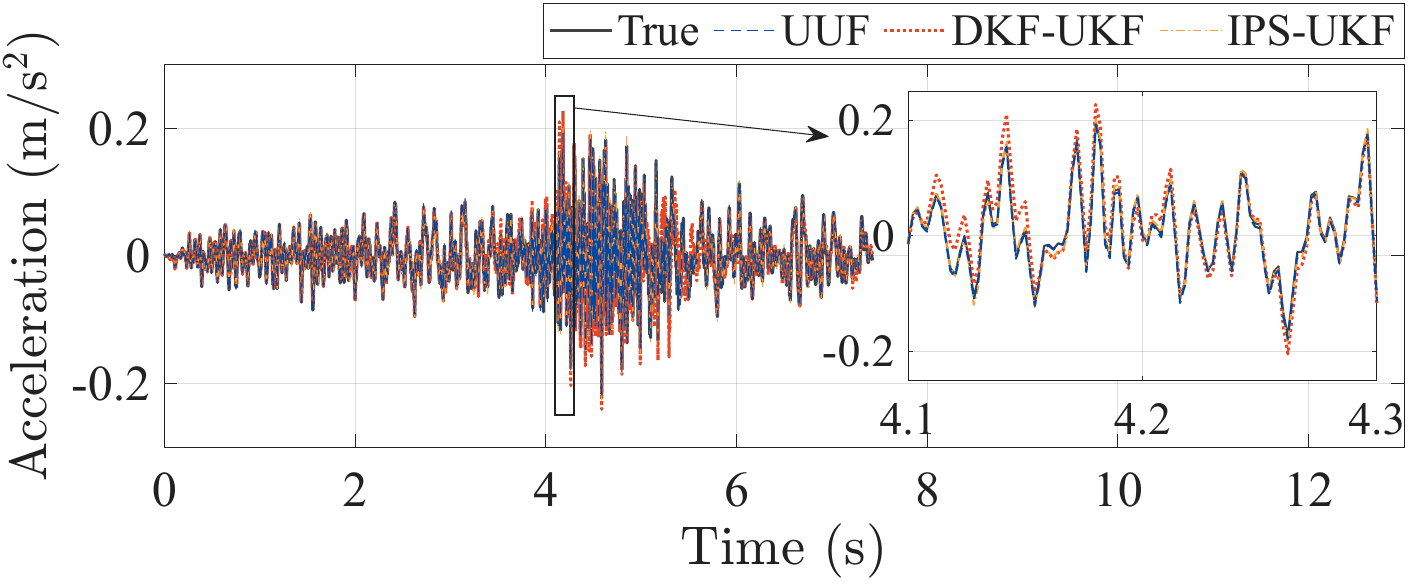}
    \end{subfigure}
    \hspace{1em}
    \begin{subfigure}{0.45\textwidth}
        \caption{} \label{fig:cs2_acc_c3_state_d}
        \includegraphics[width=\textwidth]{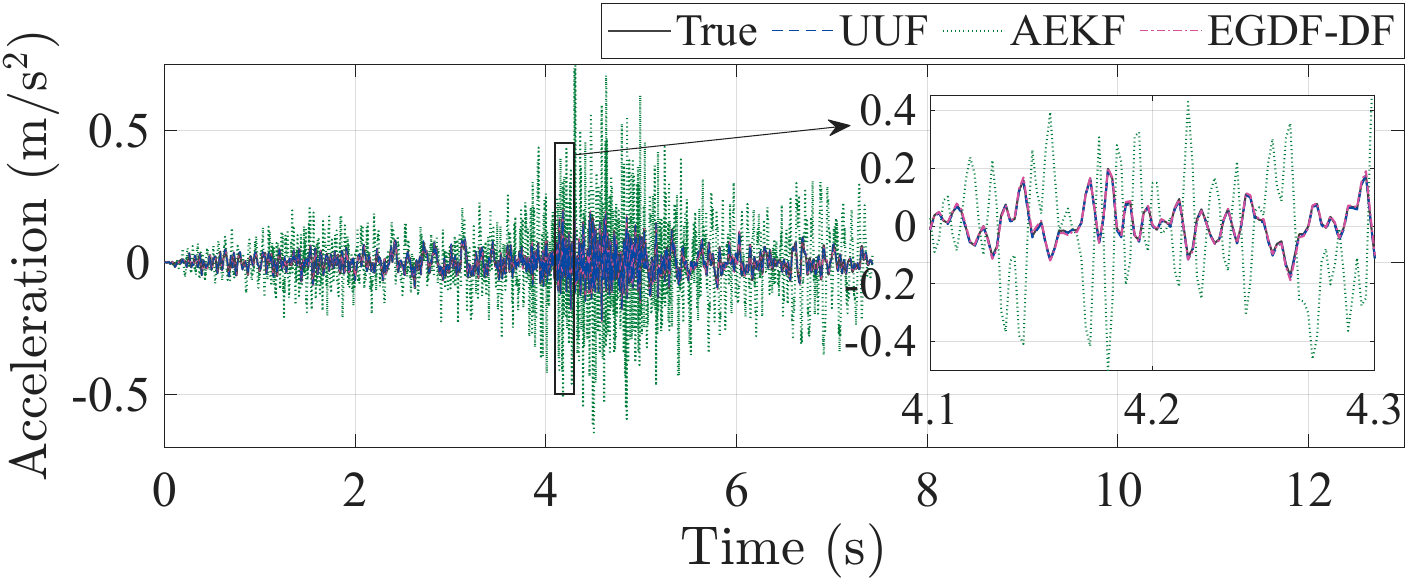}
    \end{subfigure}
    \caption{Estimates at joint 10 with member 12 damaged: (a) displacement against the state-of-the-art methods, (b) displacement against the classical methods, (c) acceleration against the state-of-the-art methods, and (d) acceleration against the classical methods.}
    \label{fig:cs2_acc_c3_state}
\end{figure}

\begin{figure}[ht]
    \centering
    \begin{subfigure}{0.4\textwidth}
        \caption{} \label{fig:cs2_acc_c3_input_a}
        \includegraphics[width=\textwidth]{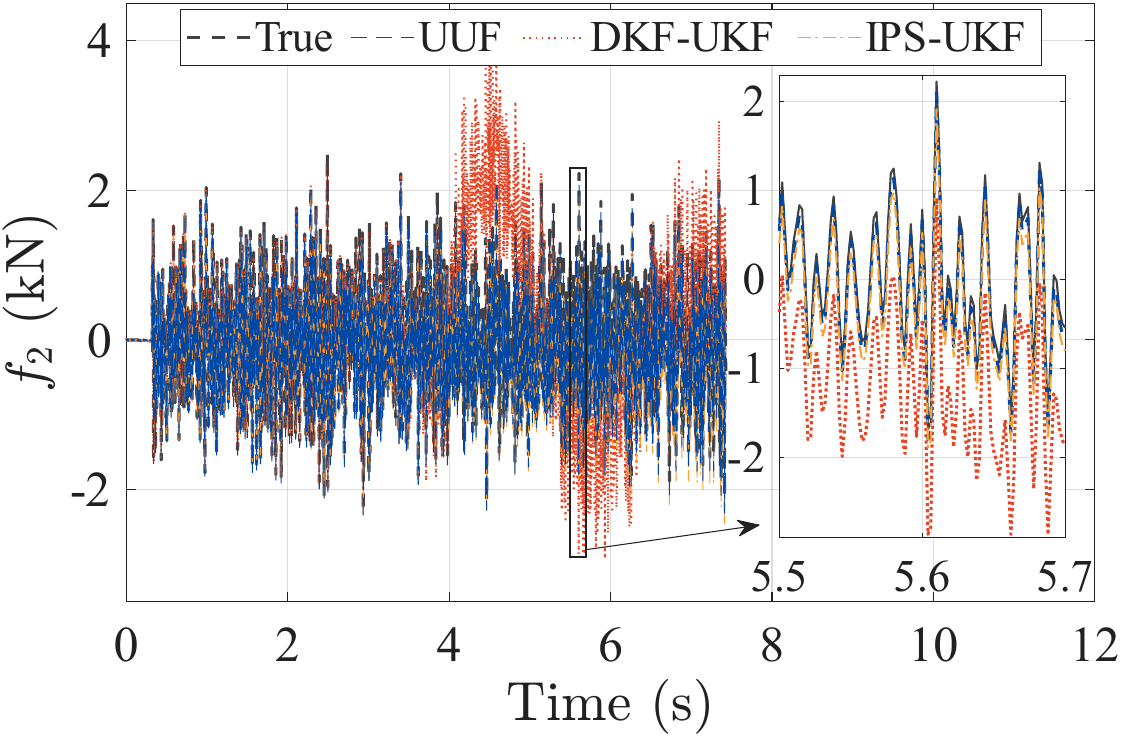}
    \end{subfigure}
    \hspace{1em}
    \begin{subfigure}{0.4\textwidth}
        \caption{} \label{fig:cs2_acc_c3_input_b}
        \includegraphics[width=\textwidth]{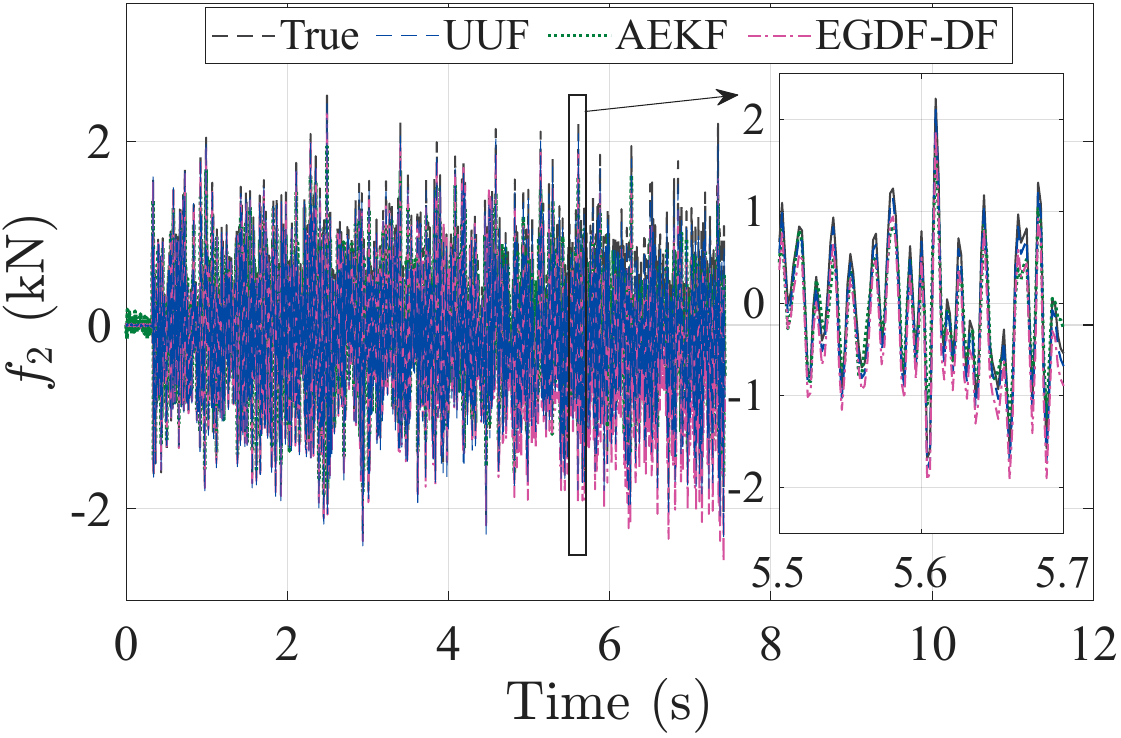}
    \end{subfigure}
    \caption{Rear axle contact force estimates by the UUF against (a) the state-of-the-art and (b) the classical methods, with member 12 damaged.}
    \label{fig:cs2_acc_c3_input}
\end{figure}

In the non-collocated configuration (c), no method identifies the stiffness of damaged member 12 from the two deck accelerations, as shown in Fig.~\ref{fig:cs2_acc_c3_param}. The estimates of the UUF and the EGDF-DF remain at their initial nominal values, while that of the IPS-UKF decreases slowly without reaching the true value. The estimate of the DKF-UKF fluctuates around the true value, and that of the AEKF diverges. In Fig.~\ref{fig:cs2_acc_c3_state}, the UUF, the IPS-UKF and the EGDF-DF match the true acceleration at joint 10, whereas their displacement estimates are smaller than the true displacement near its peak. The DKF-UKF deviates from the true displacement between about $3.5$ and $5$~s, and the estimates by the AEKF are dominated by noise. In Fig.~\ref{fig:cs2_acc_c3_input}, the UUF and the IPS-UKF match the true contact force, whereas the DKF-UKF deviates from it between about $4$ and $5$~s and after about $5.5$~s. For this type of configuration, a smoothing scheme is recommended.

Across the three configurations, the developed UUF identifies the stiffness accurately when a deck accelerometer is placed at a joint of the monitored member, for both a chord member and a vertical member. The IPS-UKF is the only method of comparable accuracy, whereas the classical AEKF and EGDF-DF do not identify the stiffness in any configuration, given the limited measurement. In the non-collocated configuration (c), the stiffness is not identified by any method.

It should be noted that the contact forces are estimated from accelerations only in this study. By Eq.~\eqref{eq:vbi_susp_force}, the total suspension forces satisfy $\mathbf{L}_1F_1(t) + \mathbf{L}_2F_2(t) = -\mathbf{M}_v\ddot{\mathbf{u}}_v(t)$, so they are fixed by the two axle accelerations at every timestep. The contact force caused by the road roughness is the remaining part of the suspension force, i.e., $f_i(t) = k_{v_i}\left[z_{a_i}(t) - \boldsymbol{\iota}_i(t)^T\mathbf{u}_b(t)\right] + c_{v_i}\left[\dot{z}_{a_i}(t) - \boldsymbol{\iota}_i(t)^T\dot{\mathbf{u}}_b(t)\right] - F_i(t)$, which depends on the displacements and velocities of the vehicle body and of the bridge deck. From accelerations, these are obtained by integration, and their low-frequency components are constrained by the pseudo-measurements only above the cut-off frequency $\omega_c$ of the auxiliary integrators. A slowly varying error in axle displacement $z_{a_i}(t)$ therefore enters $f_i(t)$ through $k_{v_i}$, which is the deviation observed in configuration (b). Since the bridge carries the total force $F_i(t)$ in Eq.~\eqref{eq:vbi_bridge_eom}, which is fixed by the measured accelerations, this deviation does not affect the responses of the bridge or the identified stiffness. The DKF-UKF shows no such deviation, since the matrix of its fictitious input model is set to $0.95\mathbf{I}_2$, as stated in Section~\ref{sec:cs1_setup}. The input predicted at each timestep is therefore its previous estimate scaled by $0.95$, which prevents a slowly varying error from accumulating, at the expense of larger errors in the state and the stiffness. In contrast, the developed UUF requires no such input model.

\subsection{Performance with unknown suspension parameters} \label{sec:cs2_susp}

\noindent The second study repeats configuration (b) of Section~\ref{sec:cs2_acc} without any knowledge of the suspension, i.e., $k_{v_1}$, $k_{v_2}$, $c_{v_1}$ and $c_{v_2}$ are unknown and set to zero, as listed in Table~\ref{tab:cs2_susp_vehicle}. The mass, the moment of inertia and the axle distances of the vehicle remain known, and the vertical member 19 is monitored with the two axle accelerations and the deck acceleration at joint 6 as before. Therefore, the unknown input of each axle represents the total force transmitted by its tyre-suspension system, of which the contact force caused by the road roughness is one part. For brevity, this total force is still denoted by $f_i$ in this study, and the input estimates are compared with it. The UUF is compared with the state-of-the-art DKF-UKF and IPS-UKF.
\begin{table}[ht]
    \centering
    \caption{Properties of the half-car model assumed by every method in the second study, with the unknown suspension parameters shaded.}
    \label{tab:cs2_susp_vehicle}
    \begin{tabular}{c c}
    \toprule
    Properties & Values \\
    \midrule
    Mass $m_v$ & $2~143$~kg \\
    Moment of inertia $I_v$ & $2~471$~$\mathrm{kg \cdot m^2}$ \\
    Distances to the axles $d_1$, $d_2$ & $1.35$~m \\
    \rowcolor{gray!25} Suspension stiffness $k_{v_1}$, $k_{v_2}$ & $0$~N/m (unknown)\\
    \rowcolor{gray!25} Suspension damping $c_{v_1}$, $c_{v_2}$ & $0$~$\mathrm{N \cdot s/m}$ (unknown) \\
    \bottomrule
    \end{tabular}
\end{table}

\begin{figure}[ht]
    \centering
    \begin{subfigure}{0.4\textwidth}
        \caption{} \label{fig:cs2_susp_param_a}
        \includegraphics[width=\textwidth]{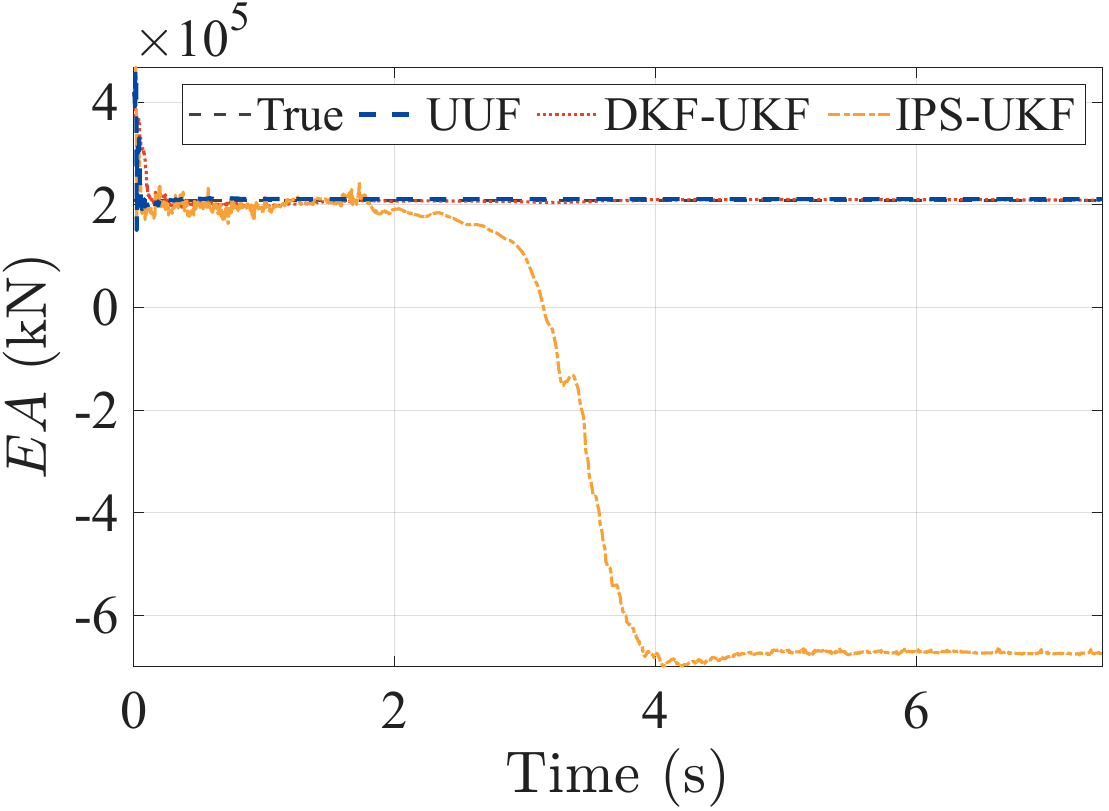}
    \end{subfigure}
    \hspace{1em}
    \begin{subfigure}{0.4\textwidth}
        \caption{} \label{fig:cs2_susp_param_b}
        \includegraphics[width=\textwidth]{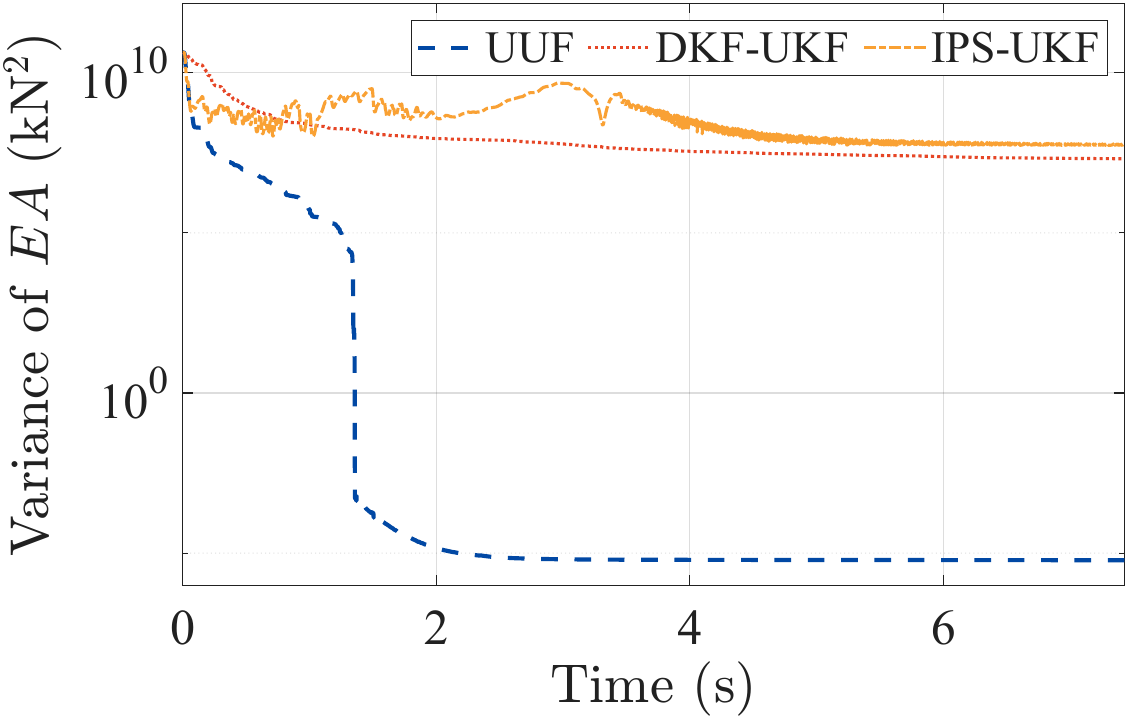}
    \end{subfigure}
    \caption{Axial stiffness of member 19: (a) estimates, and (b) variance of the estimates.}
    \label{fig:cs2_susp_param}
\end{figure}

Fig.~\ref{fig:cs2_susp_param_a} shows the stiffness estimates of damaged member 19. The estimate of the developed UUF converges to the true value in about $0.5$~s and remains close to it until the vehicle leaves the bridge. The DKF-UKF performs similarly. In contrast, the estimate of the IPS-UKF fluctuates around the true value until about $2$~s, then decreases and settles at a negative value from about $4$~s onward. Fig.~\ref{fig:cs2_susp_param_b} shows the variance of the stiffness estimate given by each method. The variance of the UUF decreases sharply once the estimate has converged and remains nearly constant thereafter, whereas those of the DKF-UKF and the IPS-UKF decrease slowly over the whole time history.

\begin{figure}[ht]
    \centering
    \begin{subfigure}{0.75\textwidth}
        \caption{} \label{fig:cs2_susp_state_a}
        \includegraphics[width=\textwidth]{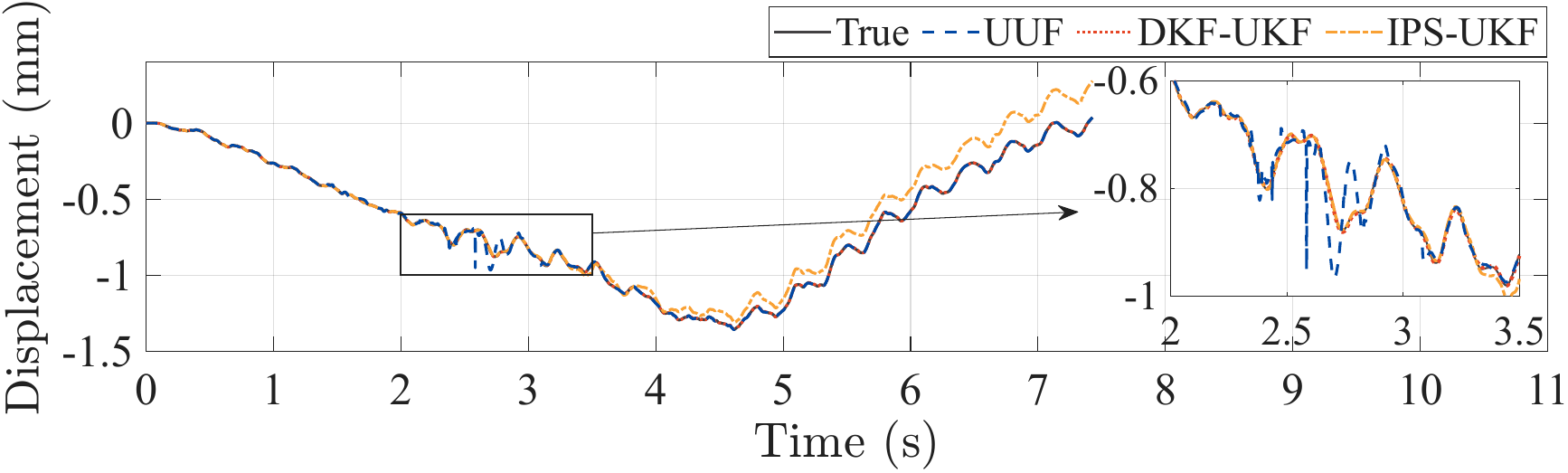}
    \end{subfigure}
    \hspace{1em}
    \begin{subfigure}{0.75\textwidth}
        \caption{} \label{fig:cs2_susp_state_b}
        \includegraphics[width=\textwidth]{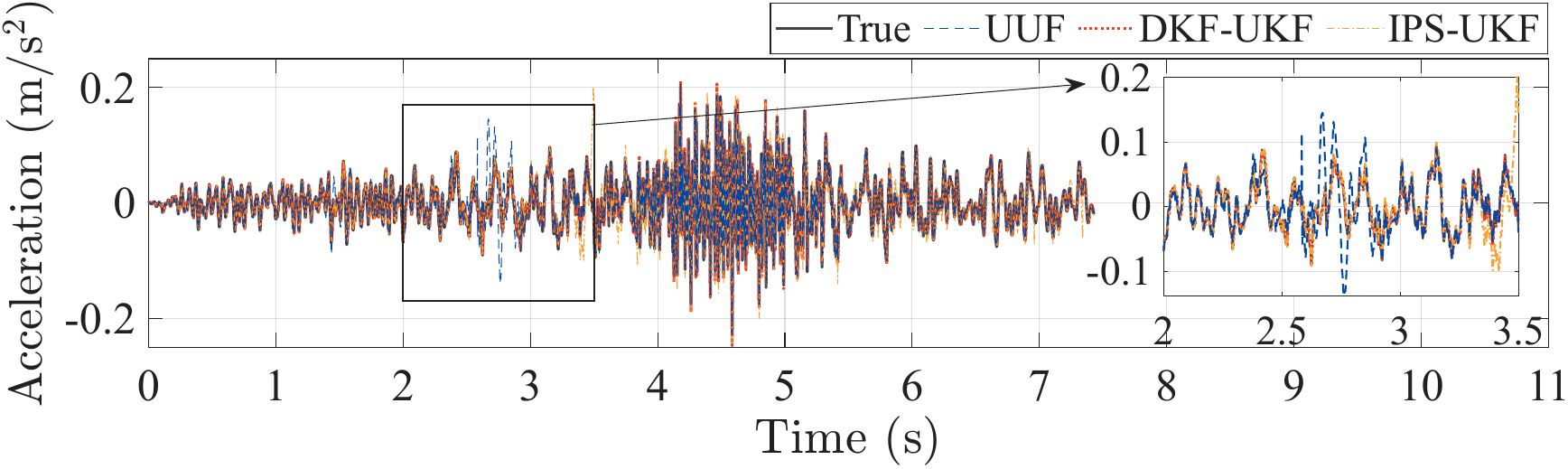}
    \end{subfigure}
    \caption{Estimates at joint 10: (a) displacement, and (b) acceleration.}
    \label{fig:cs2_susp_state}
\end{figure}

\begin{figure}[ht]
    \centering
    \begin{subfigure}{0.4\textwidth}
        \caption{} \label{fig:cs2_susp_input_a}
        \includegraphics[width=\textwidth]{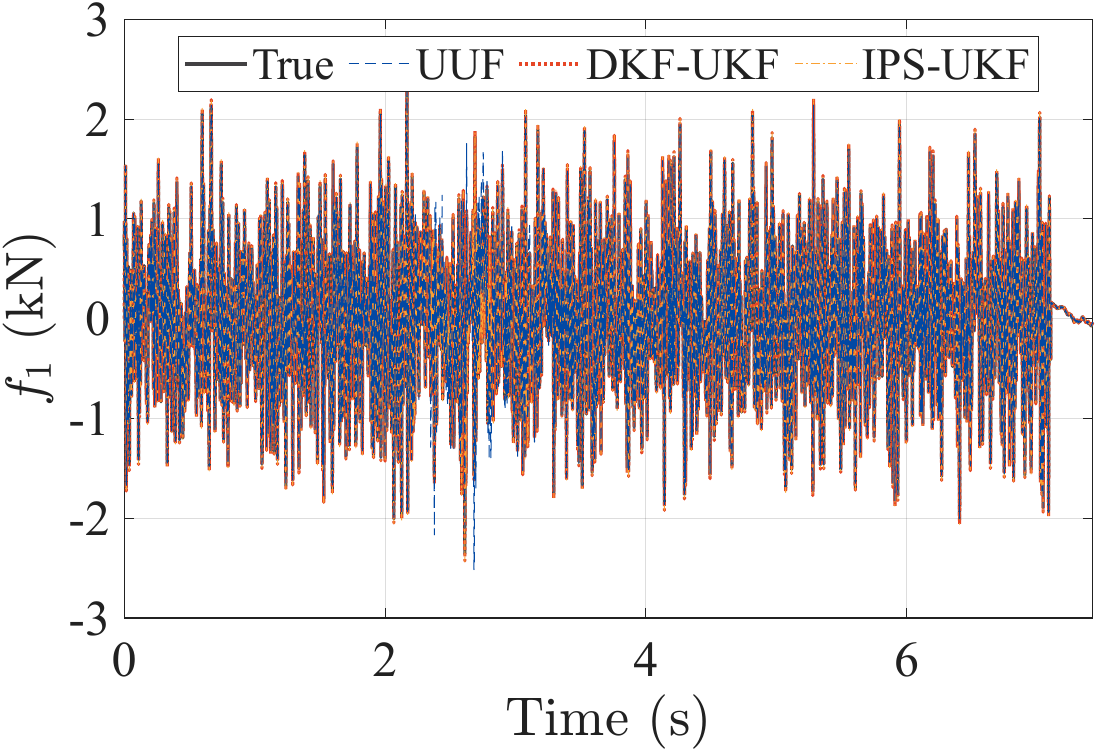}
    \end{subfigure}
    \hspace{1em}
    \begin{subfigure}{0.4\textwidth}
        \caption{} \label{fig:cs2_susp_input_b}
        \includegraphics[width=\textwidth]{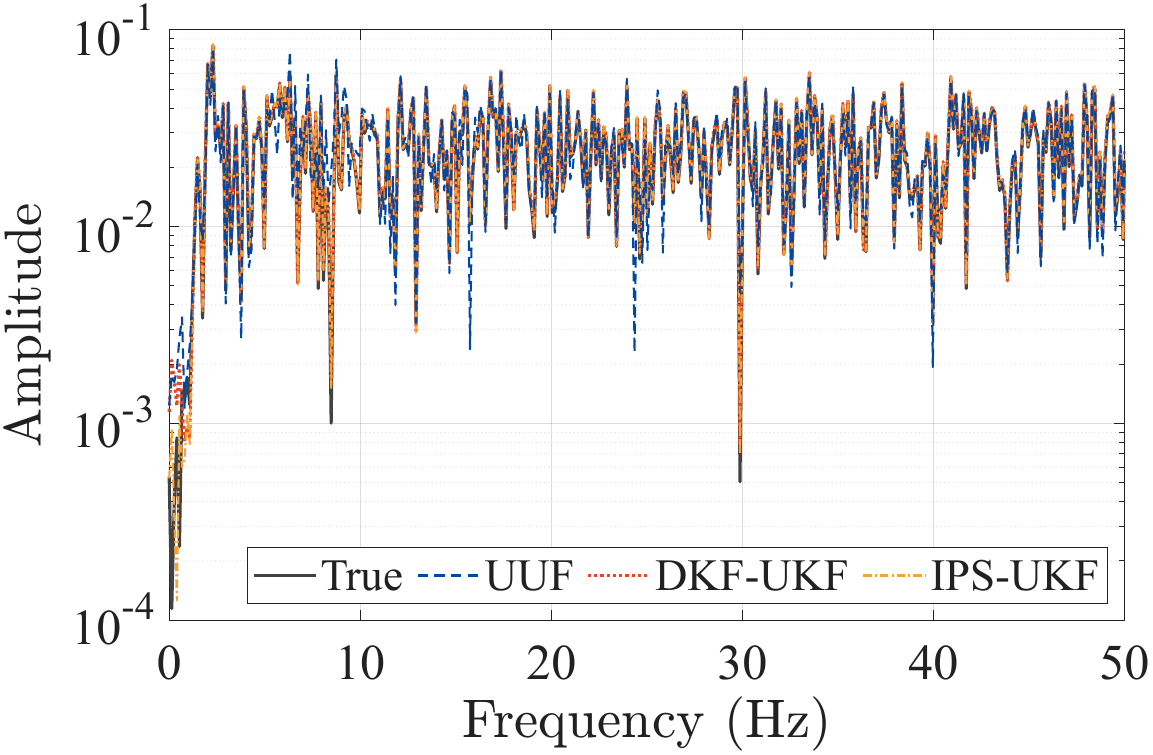}
    \end{subfigure}
    \begin{subfigure}{0.4\textwidth}
        \caption{} \label{fig:cs2_susp_input_c}
        \includegraphics[width=\textwidth]{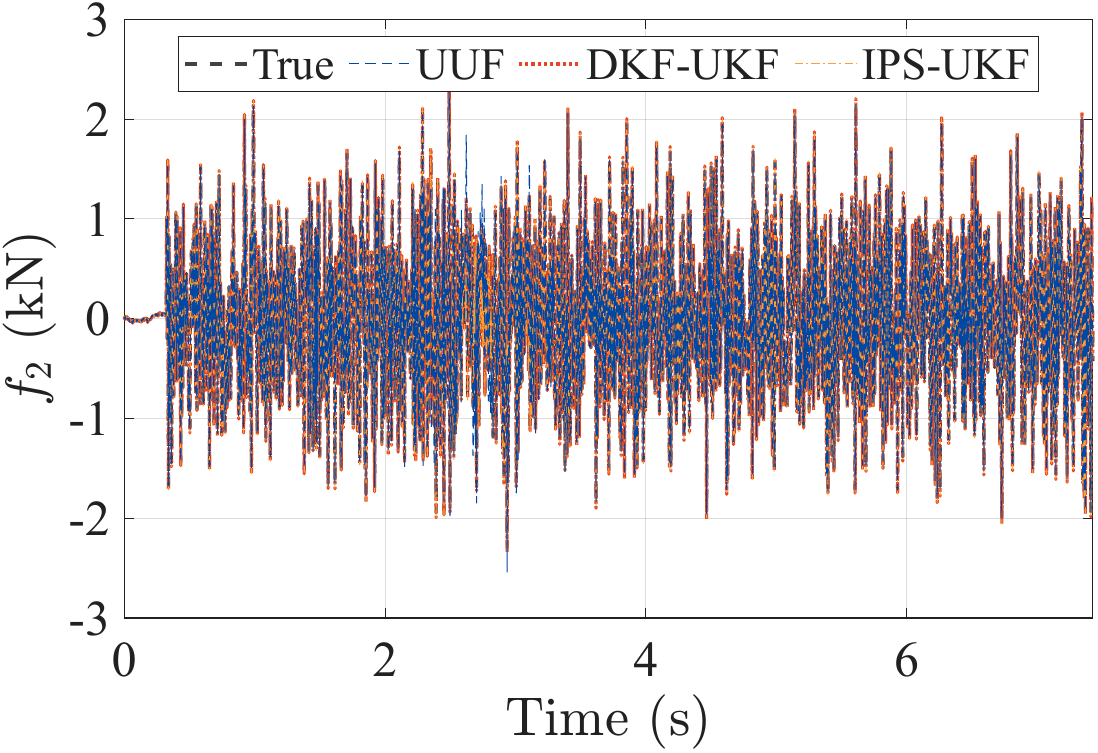}
    \end{subfigure}
    \hspace{1em}
    \begin{subfigure}{0.4\textwidth}
        \caption{} \label{fig:cs2_susp_input_d}
        \includegraphics[width=\textwidth]{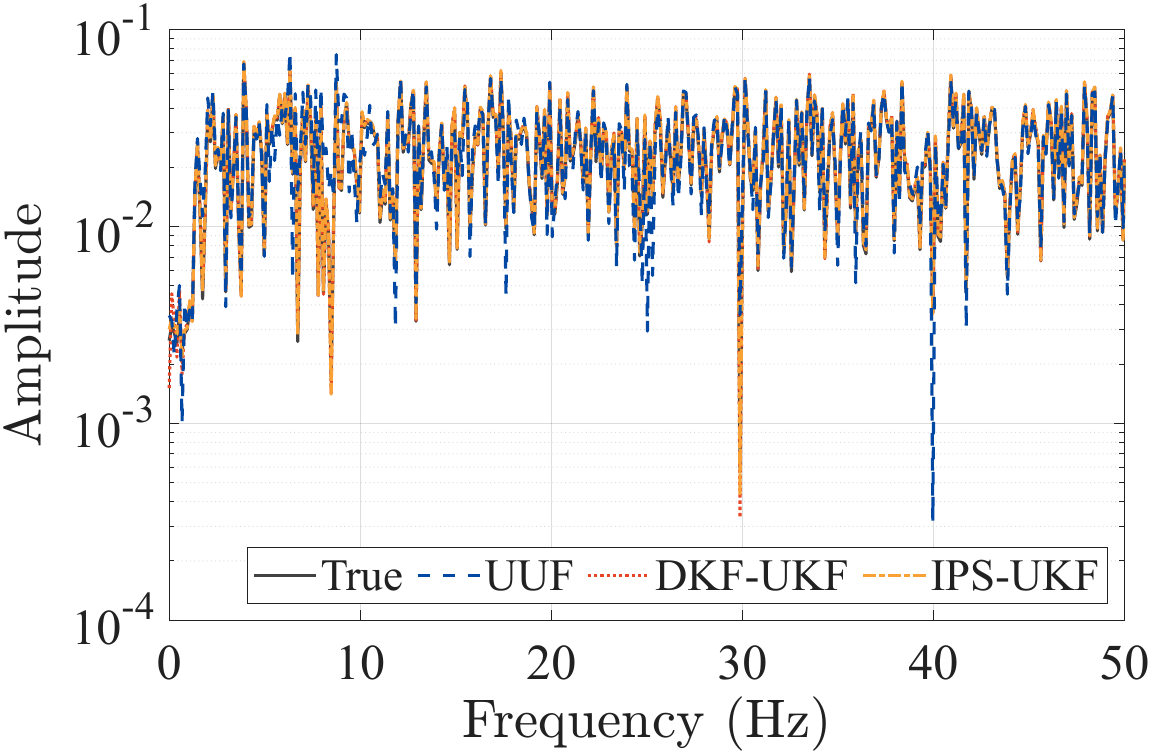}
    \end{subfigure}
    \caption{Force estimates at the front axle in (a) the time and (b) the frequency domains, and at the rear axle in (c) the time and (d) the frequency domains.}
    \label{fig:cs2_susp_input}
\end{figure}

The displacement and the acceleration at joint 10 are shown in Fig.~\ref{fig:cs2_susp_state}. The UUF and the DKF-UKF match the true responses, except for short deviations of the UUF around $2.5$~s. In contrast, the displacement estimated by the IPS-UKF drifts from the true displacement from about $4$~s. With regard to the input estimation, Fig.~\ref{fig:cs2_susp_input} shows the forces at the two axles in the time and frequency domains. The three methods match the true forces, with short deviations of the UUF around $2.5$~s, and their spectra coincide with the true ones except at the lowest frequencies. Overall, the UUF identifies the stiffness without the suspension parameters. Therefore, the bridge can be assessed with the mass properties of the vehicle only, which are easier to determine in practice than the suspension parameters, as noted in Section~\ref{sec:application}.

\clearpage
Over the first two studies, the methods are compared without the non-collocated configuration (c), in which no method identifies the stiffness. Neither of the state-of-the-art methods identifies the stiffness in all the remaining scenarios. The IPS-UKF identifies the stiffness in configurations (a) and (b) of Section~\ref{sec:cs2_acc}, whereas its estimate diverges when the suspension parameters are unknown. The DKF-UKF identifies the stiffness in configuration (b) and with unknown suspension parameters, whereas its estimate diverges in configuration (a). With regard to the contact forces, the DKF-UKF shows no slowly varying deviation under acceleration-only measurements, at the expense of larger errors in the state and the stiffness. In contrast, the developed UUF identifies the stiffness in all of these scenarios, with the same tuning strategy and without any model of the input.

\subsection{Performance with a rank-deficient feedforward matrix} \label{sec:cs2_rank}

\begin{figure}[ht]
    \centering
    \includegraphics[width=0.9\textwidth]{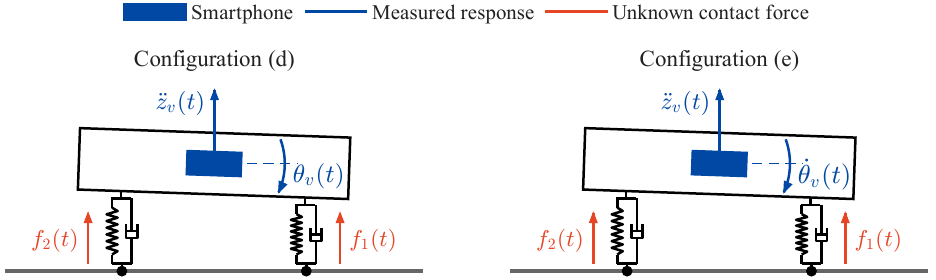}
    \caption{Sensor configurations of the smartphone: vertical acceleration and pitch angle in configuration (d), and vertical acceleration and pitch rate in configuration (e).}
    \label{fig:cs2_rank_config}
\end{figure}

\noindent In the third study, the vehicle carries a single sensor unit at its centre of gravity, e.g., a smartphone, and the bridge is instrumented with one displacement sensor at joint 6. The vertical member 19 is monitored as in configuration (b) of Section~\ref{sec:cs2_acc}, with its stiffness reduced by $50\%$. Two sensor configurations are considered, as shown in Fig.~\ref{fig:cs2_rank_config}. In configuration (d), the smartphone measures the vertical acceleration and the pitch angle of the vehicle body. In configuration (e), the pitch angle is replaced by the pitch rate.

In both configurations, only one measurement is an acceleration, so the feedforward matrix $\mathbf{D}_k$ has a rank of one, whereas two contact forces are unknown. The rank condition of the EGDF-DF is therefore not satisfied, and neither is that of the DKF-UKF, which requires an accelerometer at the position of each unknown force~\cite{Dertimanis2019}. The IPS-UKF is not applicable either, since it estimates the input from the equation of motion with the measured accelerations~\cite{ImpraimakisSmyth2022}, whereas the pitch acceleration is not measured. In contrast, the rank condition of the developed UUF is placed on the matrix $\mathbcal{G}_k$ of Section~\ref{sec:met_gains}, which has a rank of two in both configurations. The performance of the UUF is therefore examined in the two configurations.

\begin{figure}[ht]
    \centering
    \includegraphics[width=0.4\textwidth]{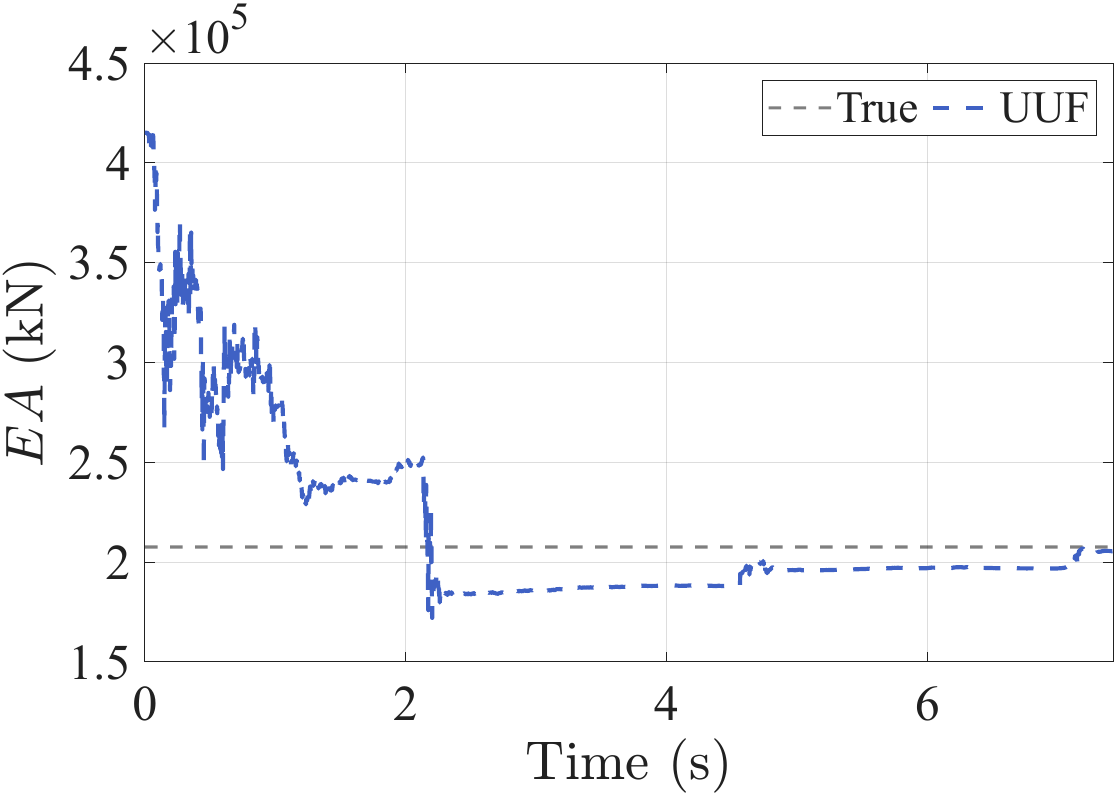}
    \caption{Axial stiffness estimates of member 19 in configuration (d).}
    \label{fig:cs2_rank_d_param}
\end{figure}

\begin{figure}[ht]
    \centering
    \begin{subfigure}{0.75\textwidth}
        \caption{} \label{fig:cs2_rank_d_state_a}
        \includegraphics[width=\textwidth]{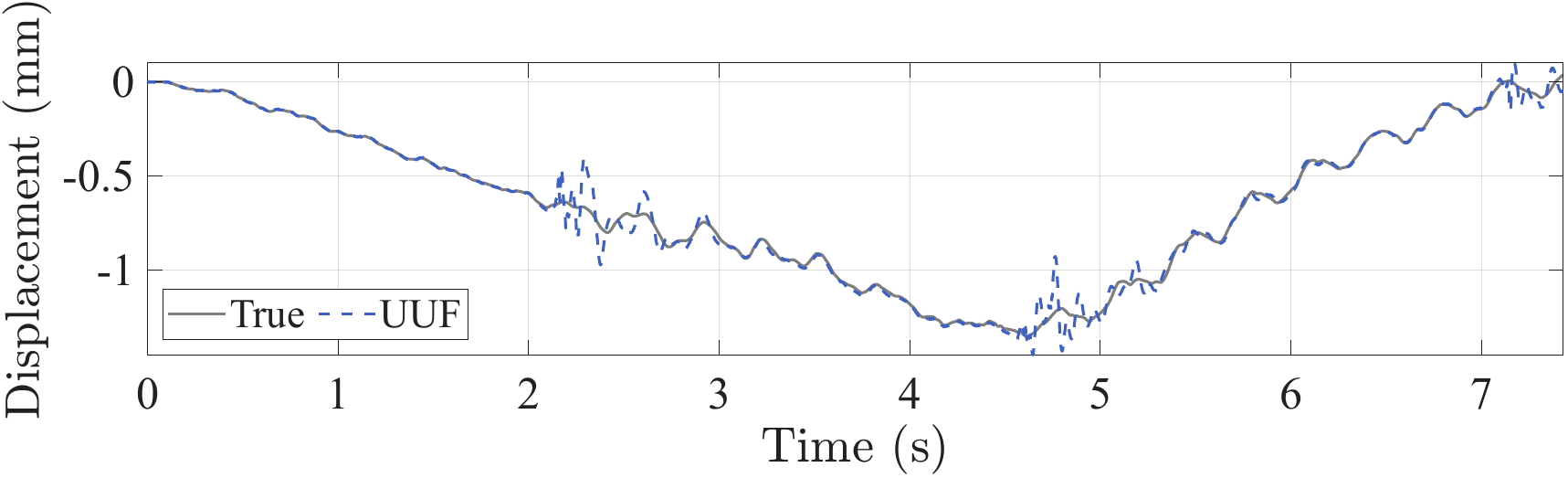}
    \end{subfigure}
    \hspace{1em}
    \begin{subfigure}{0.75\textwidth}
        \caption{} \label{fig:cs2_rank_d_state_b}
        \includegraphics[width=\textwidth]{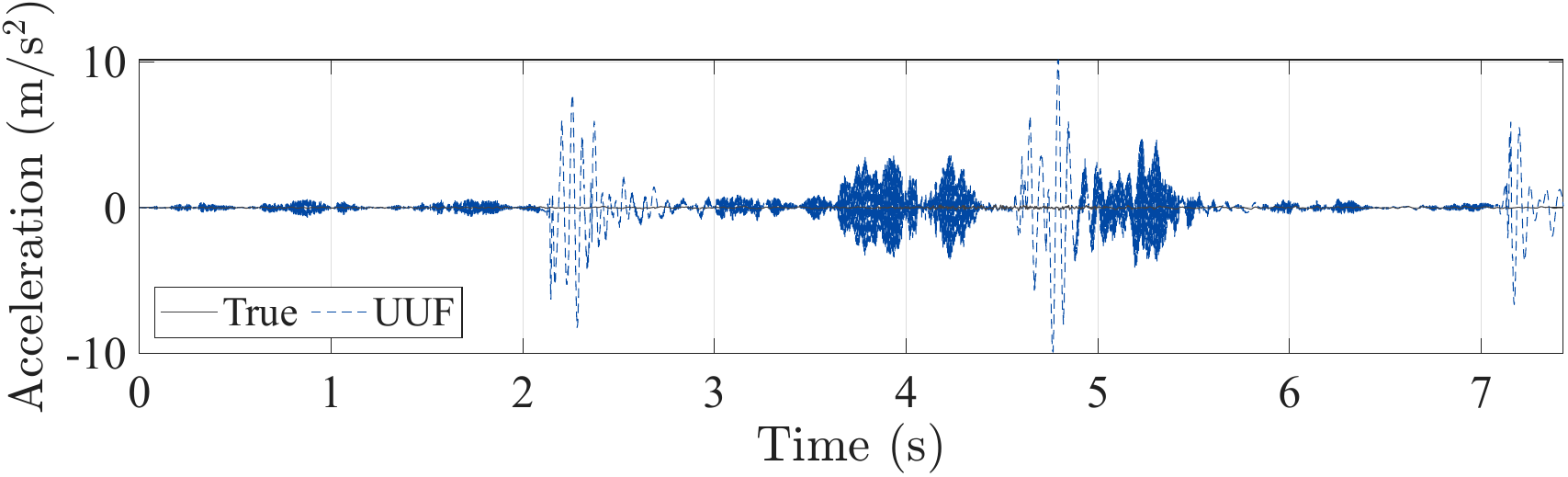}
    \end{subfigure}
    \caption{Estimates at joint 10 in configuration (d): (a) displacement, and (b) acceleration.}
    \label{fig:cs2_rank_d_state}
\end{figure}

The results of configuration (d) are presented first. Fig.~\ref{fig:cs2_rank_d_param} shows the stiffness estimates of damaged member 19. The estimate of the developed UUF decreases from its initial nominal value and remains below the true value from about $2$~s onward, approaching it only when the vehicle leaves the bridge. Fig.~\ref{fig:cs2_rank_d_state} shows the displacement and the acceleration at joint 10, where no sensor is installed. The displacement estimate matches the true response, except near $2.5$ and $5$~s, and when the vehicle left the bridge. However, the acceleration estimate is dominated by noise.

\begin{figure}[ht]
    \centering
    \begin{subfigure}{0.4\textwidth}
        \caption{} \label{fig:cs2_rank_d_input_a}
        \includegraphics[width=\textwidth]{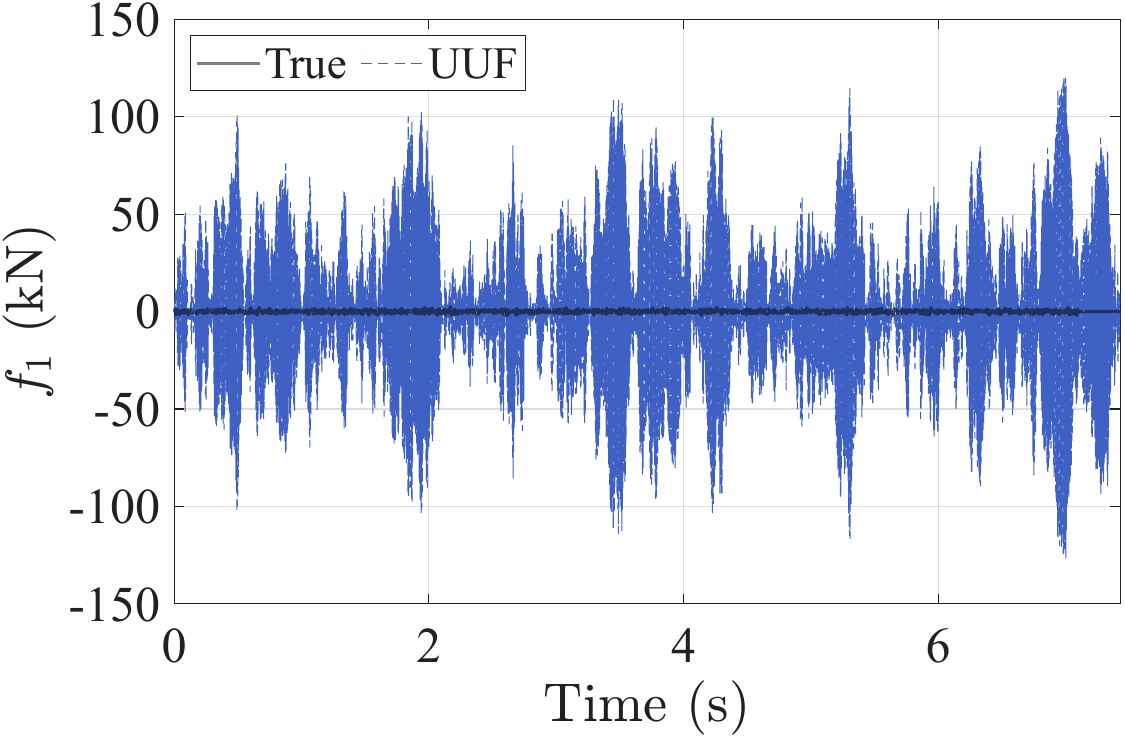}
    \end{subfigure}
    \hspace{1em}
    \begin{subfigure}{0.4\textwidth}
        \caption{} \label{fig:cs2_rank_d_input_b}
        \includegraphics[width=\textwidth]{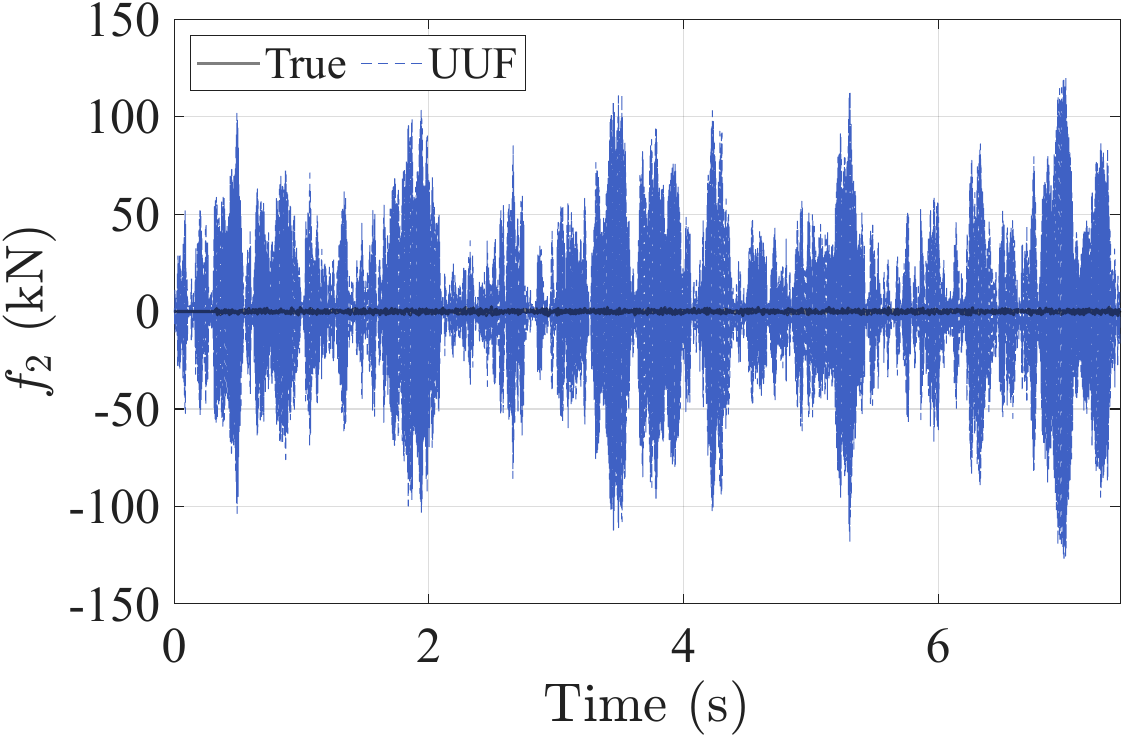}
    \end{subfigure}
    \begin{subfigure}{0.4\textwidth}
        \caption{} \label{fig:cs2_rank_d_input_c}
        \includegraphics[width=\textwidth]{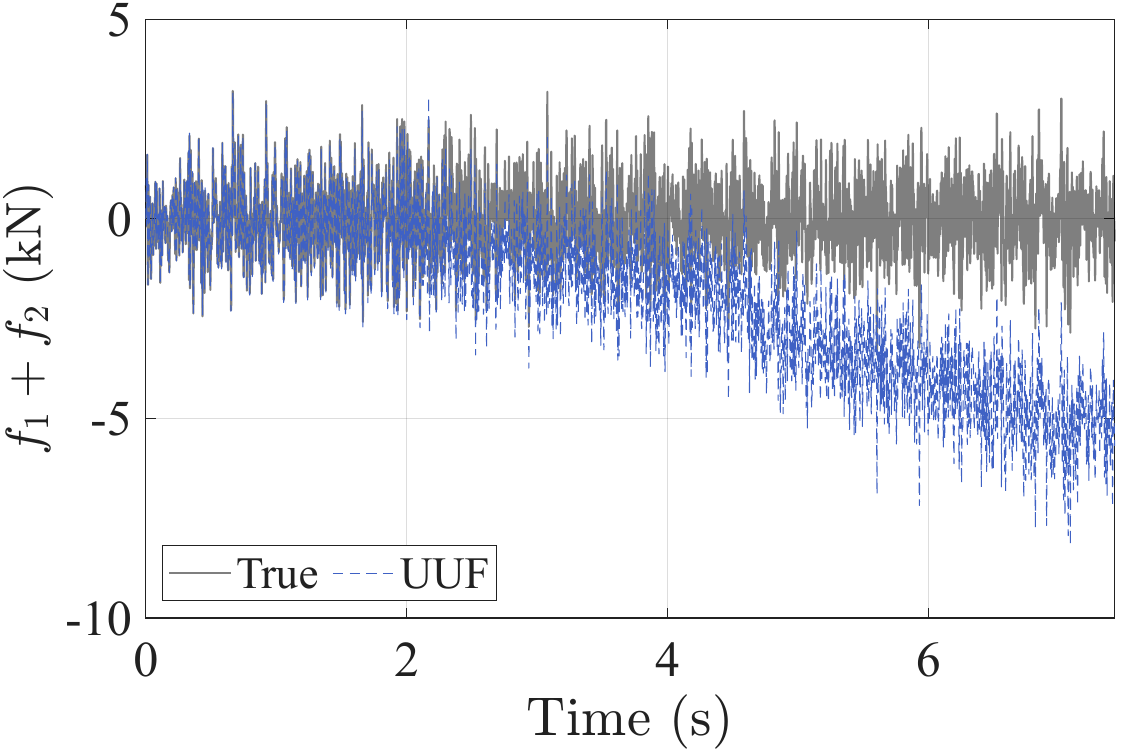}
    \end{subfigure}
    \caption{Contact force estimates in configuration (d): (a) front axle, (b) rear axle, and (c) sum of the two forces.}
    \label{fig:cs2_rank_d_input}
\end{figure}

With regard to the input estimation, Figs.~\ref{fig:cs2_rank_d_input_a} and~\ref{fig:cs2_rank_d_input_b} show that the contact force estimates at both the front and rear axles are dominated by noise with amplitudes far larger than the true forces. These oscillations cancel in the sum of the two forces, as shown in Fig.~\ref{fig:cs2_rank_d_input_c}. The estimated sum matches the true sum in the first $2$~s, and then drifts slowly from it. These results follow from the way the two contact forces act on the vehicle body. By Eqs.~\eqref{eq:vbi_vehicle_eom} and~\eqref{eq:vbi_axle_disp}, with $d_1 = d_2$ for this vehicle, they contribute $\left[f_1(t) + f_2(t)\right]/m_v$ to the vertical acceleration $\ddot{z}_v(t)$ and $d_1\left[f_1(t) - f_2(t)\right]/I_v$ to the pitch acceleration $\ddot{\theta}_v(t)$. The sum of the two forces therefore acts on a measured quantity and is determined by the vertical acceleration. Its slow drift arises from the low-frequency components of the vehicle body motion, which are not determined from an acceleration.

In contrast, the two forces are separated only by the pitch measurement, since the pitch acceleration is not measured. Over one time interval, a difference $f_1 - f_2$ changes the pitch rate by $\Delta t\, d_1\left[f_1 - f_2\right]/I_v$ and the pitch angle by $\Delta t^2 d_1\left[f_1 - f_2\right]/\left(2I_v\right)$. With $\Delta t = 0.002$~s, the pitch angle is far less sensitive to this difference than the pitch rate. Furthermore, an error in the estimated pitch rate is not corrected by the pitch angle measurement. It is carried to the following timesteps, where it is compensated by the two force estimates. As a result, the two force estimates are not accurate in configuration (d). Therefore, the pitch angle is replaced by the pitch rate in configuration (e).

\begin{figure}[ht]
    \centering
    \includegraphics[width=0.4\textwidth]{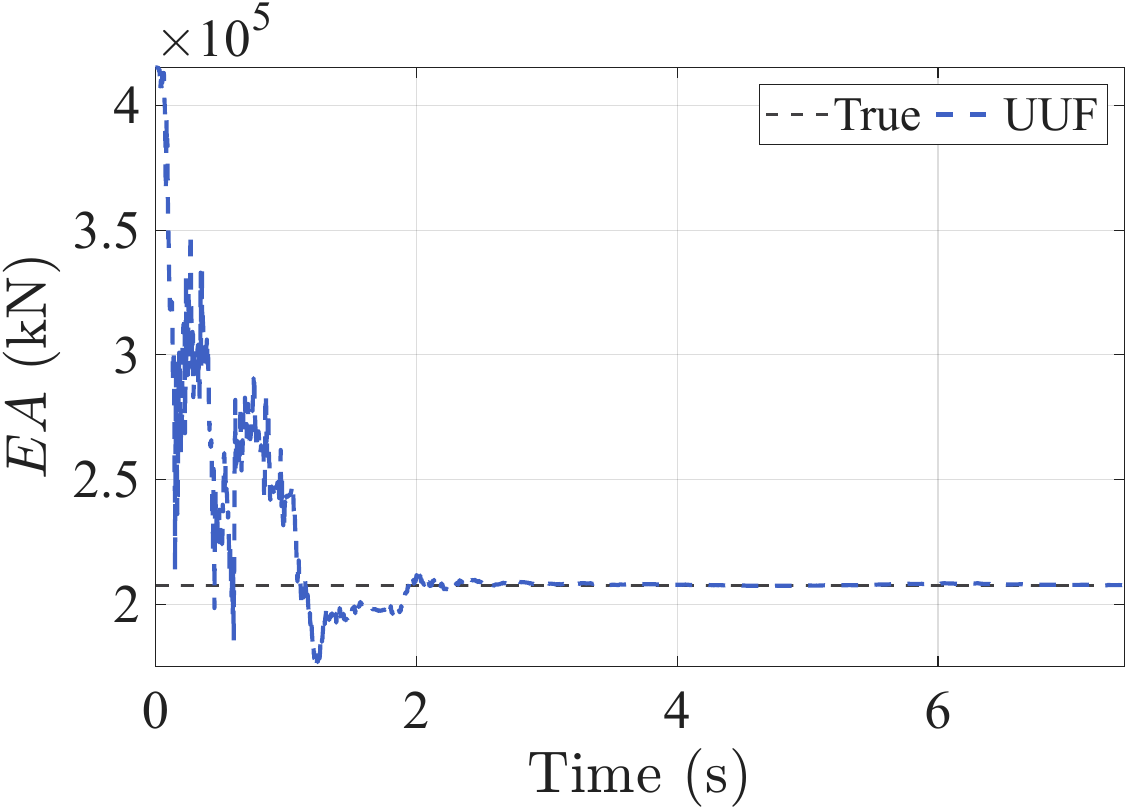}
    \caption{Axial stiffness estimates of member 19 in configuration (e).}
    \label{fig:cs2_rank_e_param}
\end{figure}

\begin{figure}[ht]
    \centering
    \begin{subfigure}{0.75\textwidth}
        \caption{} \label{fig:cs2_rank_e_state_a}
        \includegraphics[width=\textwidth]{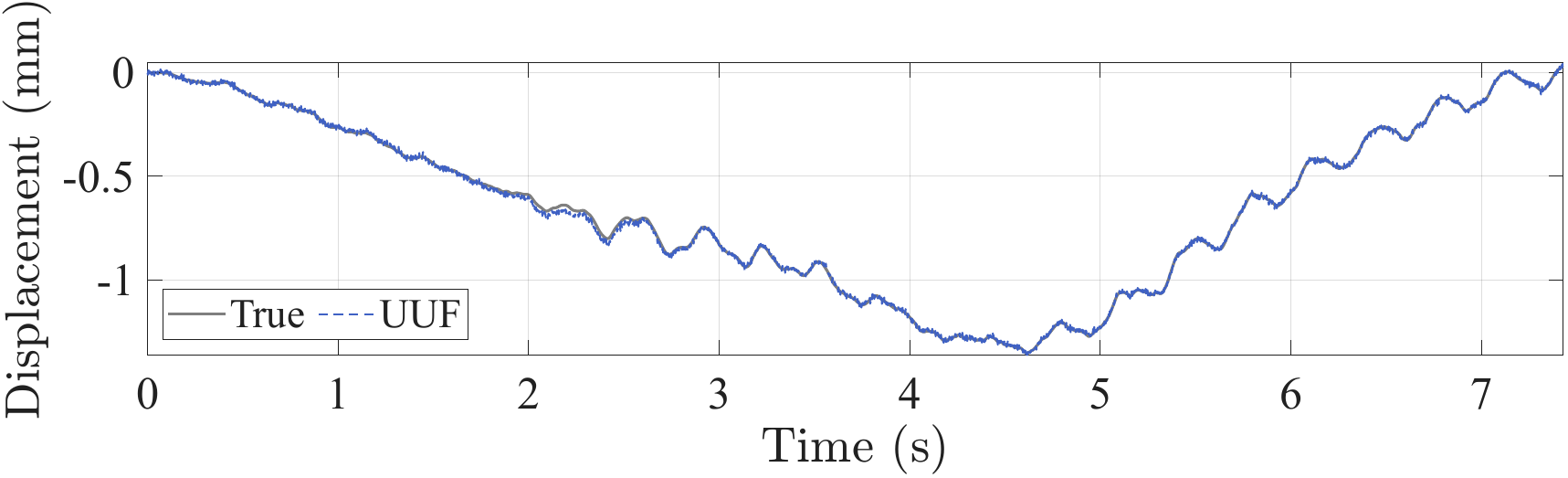}
    \end{subfigure}
    \hspace{1em}
    \begin{subfigure}{0.75\textwidth}
        \caption{} \label{fig:cs2_rank_e_state_b}
        \includegraphics[width=\textwidth]{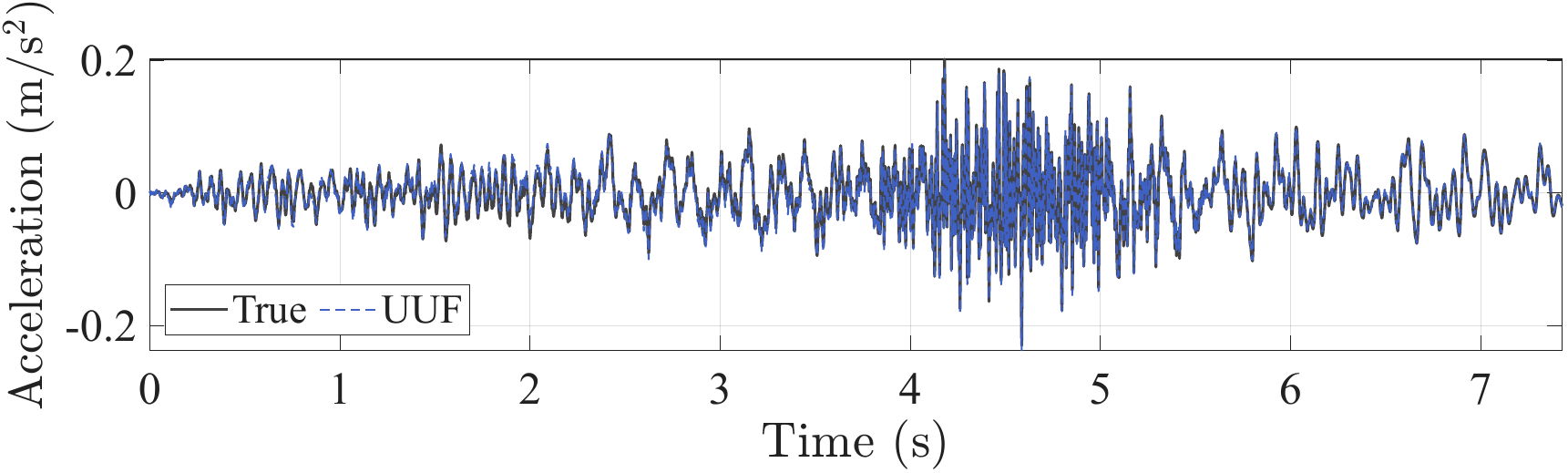}
    \end{subfigure}
    \caption{Estimates at joint 10 in configuration (e): (a) displacement, and (b) acceleration.}
    \label{fig:cs2_rank_e_state}
\end{figure}

\begin{figure}[ht]
    \centering
    \begin{subfigure}{0.4\textwidth}
        \caption{} \label{fig:cs2_rank_e_input_a}
        \includegraphics[width=\textwidth]{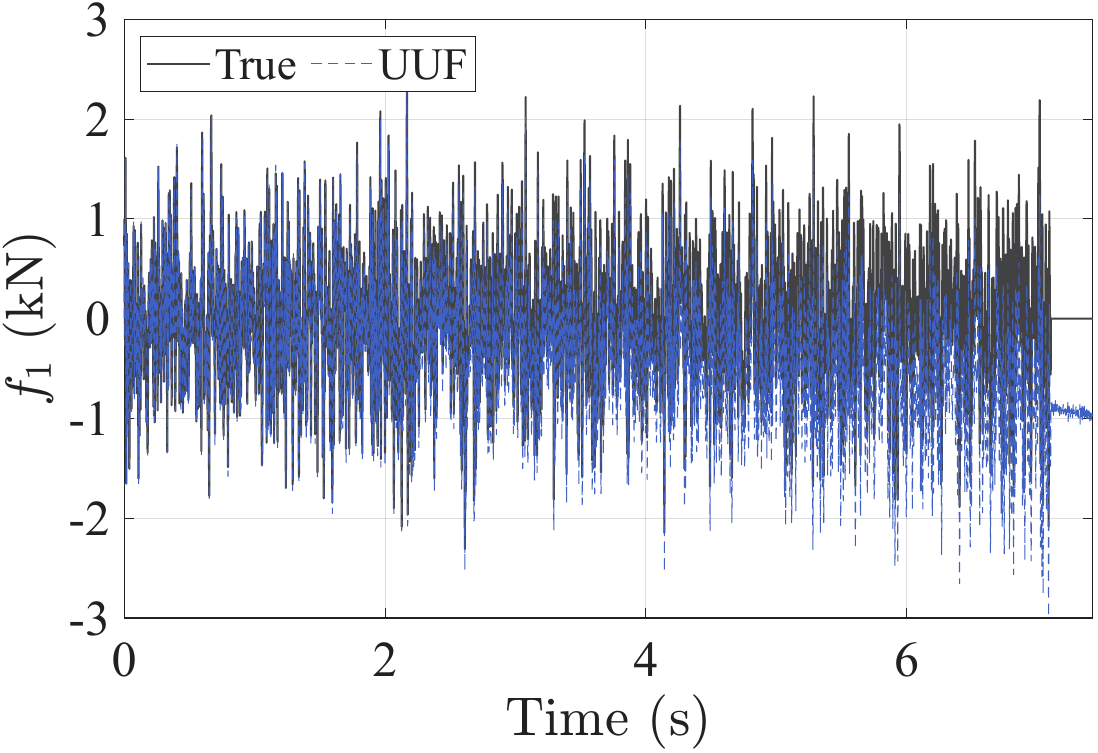}
    \end{subfigure}
    \hspace{1em}
    \begin{subfigure}{0.4\textwidth}
        \caption{} \label{fig:cs2_rank_e_input_b}
        \includegraphics[width=\textwidth]{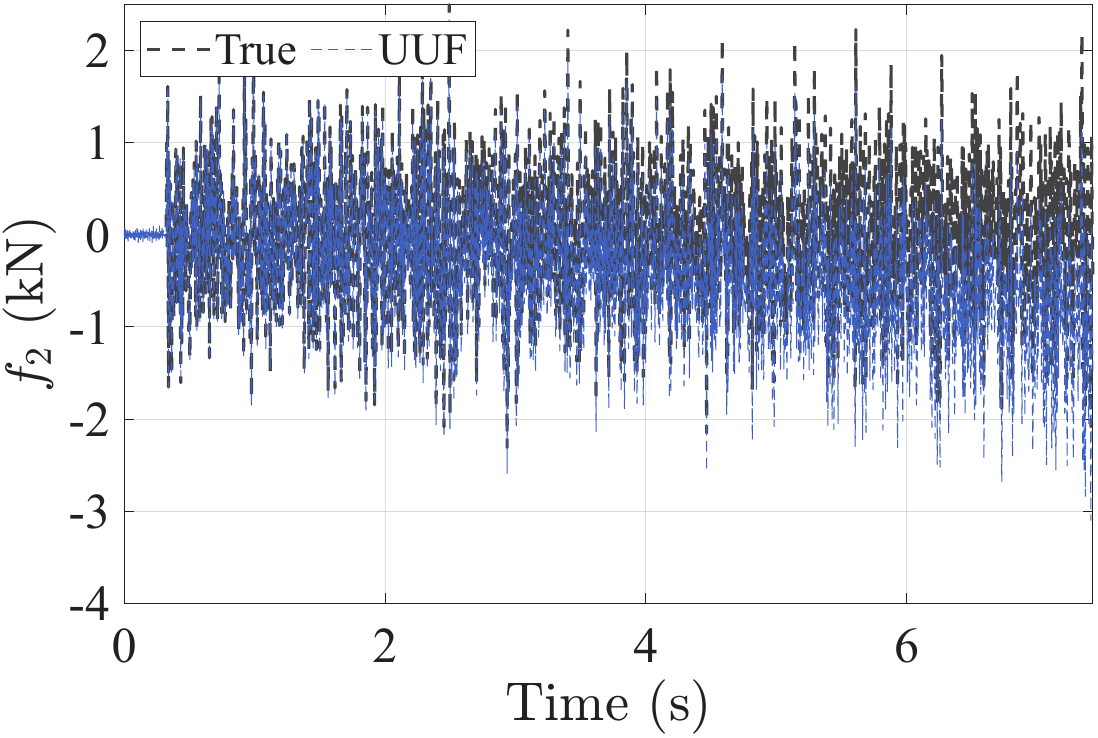}
    \end{subfigure}
    \begin{subfigure}{0.4\textwidth}
        \caption{} \label{fig:cs2_rank_e_input_c}
        \includegraphics[width=\textwidth]{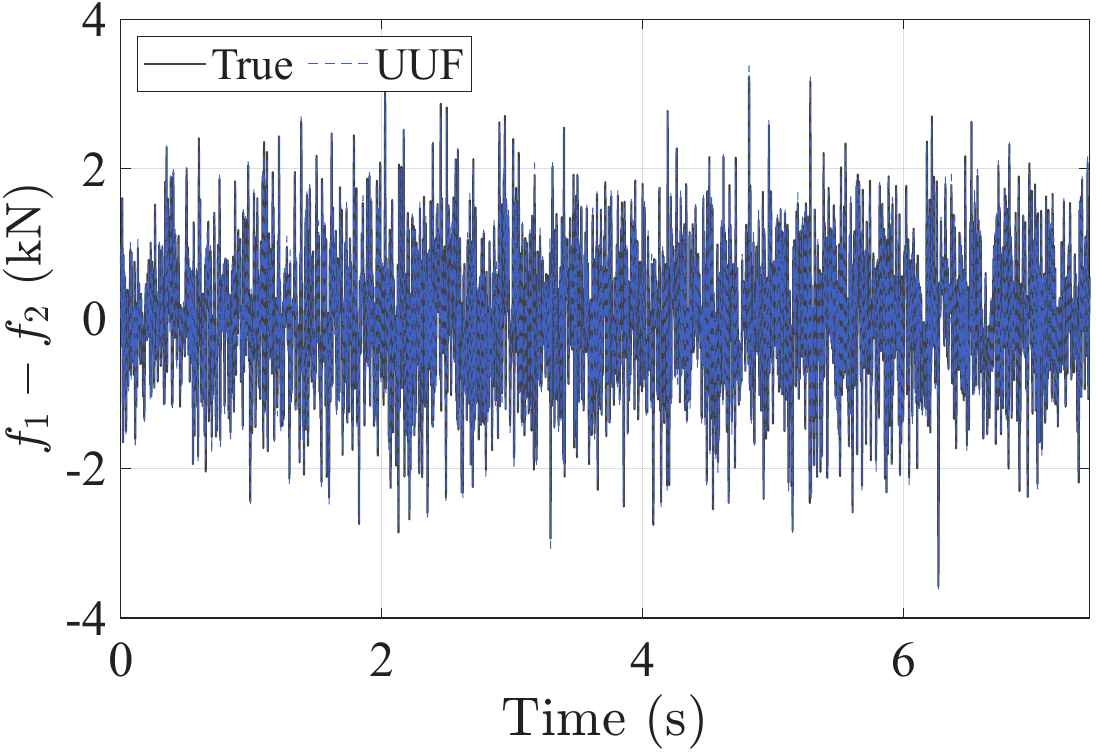}
    \end{subfigure}
    \caption{Contact force estimates in configuration (e): (a) front axle, (b) rear axle, and (c) difference of the two forces.}
    \label{fig:cs2_rank_e_input}
\end{figure}

Fig.~\ref{fig:cs2_rank_e_param} shows that the stiffness estimate in configuration (e) fluctuates at first, converges to the true value at about $2$~s, and matches it until the vehicle leaves the bridge. In Fig.~\ref{fig:cs2_rank_e_state}, the UUF matches the true displacement and acceleration at joint 10, with small deviations in the displacement around $2$~s. With regard to the input estimation, Fig.~\ref{fig:cs2_rank_e_input_c} shows that the UUF matches the true difference of the two forces closely. The contact forces at the front and rear axles are shown in Figs.~\ref{fig:cs2_rank_e_input_a} and~\ref{fig:cs2_rank_e_input_b}, respectively. The estimates exhibit a slowly varying deviation from the true forces that grows until the vehicle leaves the bridge. This deviation is common to both forces and therefore cancels out in their difference. As in configuration (d), it arises from the low-frequency components of the vehicle body motion, which are not determined from the vertical acceleration. It should be noted that this deviation is not reflected in the stiffness and the responses of the bridge.

Therefore, the pitch rate is recommended when the vehicle is instrumented at its centre of gravity only. Overall, the developed UUF performs joint input-state-parameter estimation with a rank-deficient feedforward matrix, for which the EGDF-DF, DKF-UKF, and IPS-UKF are not applicable.

\clearpage
\section{Conclusions} \label{sec:conclusions}

\noindent A novel nonlinear filter in the Universal Filter family, designated as the Unscented Universal Filter (UUF), was proposed for joint input-state-parameter estimation of structural systems with unknown inputs. The unknown input was estimated by weighted least squares from the innovation, and the state gain was obtained by minimising the trace of the state estimation error covariance. The nonlinear propagation was linearised by the scaled unscented transform, and the residuals of the linearisation were propagated into the error covariances. For the acceleration-only configuration, auxiliary integrators were introduced to form pseudo-measurements from the acceleration measurements themselves.

Unlike Kalman-type methods, such as the AEKF and the DKF-UKF, the proposed method requires no evolution model or statistics for the unknown input. In contrast with the EGDF-DF, no rank condition is imposed on the feedforward matrix, so that systems with a rank-deficient feedforward matrix are covered. In addition, no Jacobian matrix is required.

The UUF was applied to drive-by structural health monitoring on the coupled VBI system. Since the road profile acts on the system through two quantities only, the unknown input was re-parameterised as the two contact forces. The performance of the UUF was assessed in two numerical case studies against classical and state-of-the-art methods, all tuned by the same practical strategy that requires no true response. The main conclusions are drawn below:

\begin{itemize}
    \item On the simply supported beam, the UUF gave the lowest total error in most of the scenarios examined, and its total error varied little with the damage severity, the damage location and the road class. The IPS-UKF was the only method of comparable accuracy, and it gave lower total errors at the highest noise level. The total errors of the AEKF, the EGDF-DF and the DKF-UKF were clearly larger.

    \item On the Old Ada Bridge, the UUF with the auxiliary integrators identified the stiffness of the damaged member from accelerations only, provided that an accelerometer was placed at a joint of this member. The contact forces showed a slowly varying deviation, which did not affect the identified stiffness.

    \item Without any knowledge of the suspension parameters, the UUF identified the stiffness within $2\%$ of the true value. Therefore, the bridge can be assessed with the mass properties of the vehicle only. Over these two studies, neither state-of-the-art method identified the stiffness throughout.

    \item With a single sensor unit at the centre of gravity of the vehicle, the feedforward matrix is rank-deficient, and the EGDF-DF, the DKF-UKF and the IPS-UKF are not applicable. With the pitch rate, the UUF could still estimate the stiffness, the bridge responses and the difference of the two contact forces. With the pitch angle, the stiffness estimate was less accurate and the two contact forces were not estimated accurately. Therefore, the pitch rate is recommended for such a sensor unit.
\end{itemize}

The numerical case studies highlighted the flexibility of the UUF across sensor configurations, suggesting its potential for drive-by structural health monitoring with limited instrumentation. It is worth mentioning that this paper focused on a numerical study. Future research with field validation is recommended. In addition, for non-collocated configurations, in which the damaged member is not directly measured, no method identified the stiffness, and a smoothing extension of the UUF will be investigated.

\section*{Acknowledgements}
The authors gratefully acknowledge the support of the Australian Research Council (ARC) through the Discovery Early Career Researcher Award (DECRA) scheme under grant DE210101625. The authors also acknowledge the support provided by the Australia-Japan Innovation Fund (AJIF).

\bibliography{ref}

\end{document}